\documentclass[]{aa}
\usepackage{natbib}
\usepackage{makecell}
\usepackage{graphicx}
\usepackage{txfonts}
\usepackage{pdflscape}
\usepackage{multirow}

\mathchardef\mhyphen="2D

\newcommand{\Gaia}{{\sl Gaia}} 

\newcommand{\deltanu}{\mbox{$\langle \Delta\nu \rangle$}}
\newcommand{\numax}{\mbox{$\nu_{\rm max}$}}

\newcommand{\Kepler}{\textit{Kepler}}
\def\Teff{$T_{\rm eff}$}
\def\teff{$T_{\rm eff}$}
\def\logg{$\log g$}

\def\msun{M$_\odot$}
\newcommand{\DM}{\mbox{$\langle \Delta M \rangle$}}

\usepackage{pdfpages}

\begin{document}

\title{
Age dissection of the Milky Way discs: Red giants in the \Kepler\ field\thanks{Table \ref{tab:datamodel} is only available in electronic form at the CDS via anonymous ftp to {\tt cdsarc.u-strasbg.fr (130.79.128.5)} or via {\tt http://cdsweb.u-strasbg.fr/cgi-bin/qcat?J/A+A/}}}

\titlerunning{Age dissection of the Milky Way discs}
   \author{A.~Miglio\inst{\ref{inst1},\ref{inst3}}\and
   C.~Chiappini\inst{\ref{inst2}}\and
   J.~T.~Mackereth\inst{\ref{inst1}}\and 
   G.~R.~Davies\inst{\ref{inst1},\ref{inst3}}\and
   K.~Brogaard\inst{\ref{inst3},\ref{inst4}}\and 
   L.~Casagrande\inst{\ref{inst15}, \ref{inst13}}\and
   W.~J.~Chaplin\inst{\ref{inst1},\ref{inst3}}\and
   L.~Girardi\inst{\ref{inst14}}\and 
   D.~Kawata\inst{\ref{inst5}}\and 
   S.~Khan\inst{\ref{inst1}}\and 
   R.~Izzard\inst{\ref{inst6}}\and  
   J.~Montalb\'an\inst{\ref{inst1}}\and 
   B.~Mosser\inst{\ref{inst7}}\and  
   F.~Vincenzo\inst{\ref{inst8},\ref{inst9},\ref{inst1}}\and 
   D.~Bossini\inst{\ref{inst12}}\and 
   A.~Noels\inst{\ref{inst10}}\and 
   T.~Rodrigues\inst{\ref{inst14}}\and 
   M.~Valentini\inst{\ref{inst2}}\and
   I.~Mandel \inst{\ref{inst11},\ref{inst16},\ref{inst13},\ref{inst1}}
}

 \institute{
 {School of Physics and Astronomy, University of Birmingham, Birmingham B15 2TT, United Kingdom}\label{inst1} \and
 {Stellar Astrophysics Centre (SAC), Department of Physics and Astronomy, Aarhus University, Ny Munkegade 120, DK-8000 Aarhus C, Denmark}\label{inst3} \and
 {Leibniz-Institut für Astrophysik Potsdam (AIP), An der Sternwarte 16, 14482 Potsdam, Germany}\label{inst2}\and
 {Institute of Theoretical Physics and Astronomy, Vilnius University, Sauletekio av. 3, 10257 Vilnius, Lithuania} \label{inst4} \and
 {Research School of Astronomy and Astrophysics, Mount Stromlo Observatory, The Australian National University, ACT 2611, Australia}\label{inst15} \and
{ARC Centre of Excellence for All Sky Astrophysics in 3 Dimensions (ASTRO 3D), Australia}\label{inst13} \and
 {Osservatorio Astronomico di Padova -- INAF, Vicolo dell'Osservatorio 5, I-35122 Padova, Italy}\label{inst14} \and
{Mullard Space Science Laboratory, University College London,	Holmbury St. Mary, Dorking, Surrey, RH5 6NT, United Kingdom}\label{inst5}\and
 {Astrophysics Research Group, University of Surrey, Guildford, Surrey, GU2 7XH, United Kingdom}\label{inst6}\and 
 {LESIA, Observatoire de Paris, Universit\'e PSL, CNRS, Sorbonne Universit\'e, Universit\'e de Paris, 5 place Jules Janssen, 92195 Meudon, France}\label{inst7} \and
 {Center for Cosmology and AstroParticle Physics, Ohio State University, Columbus, OH 43210, USA}\label{inst8} \and
 {Department of Astronomy, Ohio State University, Columbus, OH 43210, USA}\label{inst9} \and
{Instituto de Astrof\'isica e Ci\^encias do Espa\c{c}o, Universidade do Porto, CAUP, Rua das Estrelas, 4150-762 Porto, Portugal}\label{inst12} \and
{Space sciences, Technologies and Astrophysics Research (STAR) Institute, Université de Liège, 19C Allée du 6 Août, B-4000 Liège, Belgium}\label{inst10} \and
{Monash Centre for Astrophysics, School of Physics and Astronomy, Monash University, Clayton, Victoria 3800, Australia}\label{inst11} \and
{ARC Centre of Excellence for Gravitational Wave Discovery (OzGrav), Australia}\label{inst16}
}

\authorrunning{A.~Miglio, et al.}

   \date{Received \today; accepted \today}

  \abstract{
Ensemble studies of red-giant stars with exquisite asteroseismic (\Kepler), spectroscopic (APOGEE), and astrometric (\Gaia) constraints offer a novel opportunity to recast and address long-standing questions concerning the evolution of stars and of the Galaxy.
{Here, we infer masses and  ages  for  nearly 5400 giants with available \Kepler\ light curves and APOGEE spectra  using  the  code {\sc param}, and discuss some of the systematics that may affect the accuracy of the inferred stellar properties. We then present patterns in mass, evolutionary state, age, chemical abundance, and orbital parameters that we deem robust against the systematic uncertainties explored.}
{
First, we look at age-chemical-abundances ([Fe/H] and $\mathrm{[\alpha/Fe]}$) relations. We find a dearth of young,  metal-rich ($\mathrm{[Fe/H] > 0.2)}$ stars, and the existence of a significant population of old (8-9 Gyr), low-$\mathrm{[\alpha/Fe]}$, super-solar metallicity stars, reminiscent of the age and metallicity of the well-studied open cluster NGC6791. The age-chemo-kinematic properties of these stars indicate that efficient radial migration happens in the thin disc.
We find that ages and masses of the nearly 400 $\alpha$-element-rich red-giant-branch (RGB) stars in our sample are compatible with those of an old ($\sim 11$ Gyr), nearly coeval,  chemical-thick disc population.  Using a statistical model, we show that the width of the observed age distribution is dominated by the random uncertainties on age, and that the spread of the inferred intrinsic age distribution is such that 95\% of the population was born within $\sim 1.5$ Gyr. Moreover, we find a difference in the vertical velocity dispersion between low- and high-$\mathrm{[\alpha/Fe]}$ populations. This discontinuity, together with the chemical one in the $\mathrm{[\alpha/Fe]}$ versus [Fe/H] diagram, and with the inferred age distributions, not only confirms the different chemo-dynamical histories of the chemical-thick and thin discs, but it is also suggestive of a halt in the star formation (quenching) after the formation of the chemical-thick disc.
We then exploit the almost coeval $\alpha$-rich population  to  gain  insight into  processes  that  may  have  altered  the mass of a star along its evolution, which are key to improving the mapping of the current, observed, stellar mass to the initial mass and thus to the age. Comparing the mass distribution of stars on the lower RGB ($R< 11\, \mathrm{R_\odot}$) with those in the red clump (RC), we find evidence for a mean integrated RGB mass loss $\DM\ = 0.10\pm 0.02$ M$_\odot$. Finally, we find that the occurrence of massive ($M\gtrsim 1.1\, {\rm M}_\odot$) $\alpha$-rich stars is of the order of 5\% on the RGB, and significantly higher in the RC, supporting the scenario in which most of these stars had undergone an interaction with a companion. 
}
}
   \maketitle
%

\section{Introduction and motivation}
Asteroseismic constraints, coupled with information on photospheric chemical abundances and temperature, have given us the ability to measure the masses of tens of thousands of red giant stars.
Precise masses of red-giant stars enable a robust inference of their ages, given the strong relation between the initial mass of a star and the duration of the main-sequence phase of evolution and hence its age on the red-giant branch (RGB).

Thanks to these unprecedented constraints on mass and age, ensemble studies of solar-like oscillating red-giant stars allow significant progress to be made, both in our understanding of the Milky Way (MW) and of stellar structure and evolution.
We can now start exploiting the orthogonal constraints offered by age, chemistry, and dynamics to infer the Milky Way's formation and evolution (e.g. \citealt{Anders2017} using CoRoT and APOGEE { (CoRoGEE)}, \citealt{SilvaAguirre2018} using {\it Kepler} and APOGEE, \citealt{Rendle2019b} using K2, APOGEE and Gaia-ESO,  \citealt{Valentini2019} using K2 and RAVE). 
Moreover, large datasets of red-giant stars spanning wide mass and metallicity ranges can be used to revisit long-standing questions in stellar physics leading to improved stellar models, hence to more reliable inferences on stellar ages, which are inherently model-dependent. Such questions concern, for example, the boundary mixing of convective envelopes \citep{Khan2018}, the near-core structure of helium-burning stars \citep[e.g.][]{Vrard2016, Bossini2017}, and the efficiency of angular momentum transport \citep{Gehan2018, Eggenberger2019}. Furthermore, they  potentially indirectly constrain  stellar outer-boundary conditions \citep{Tayar2017, Salaris2018} for stars of different mass and metallicity.

{The aim of the present paper is to use the $\sim 5400$ red giants with available \Kepler\ light curves and APOGEE spectra to: a) identify the main properties and correlations of their age-mass-metallicity ([Fe/H] and [$\alpha$/Fe]) distributions, enabling inferences about the age of the thick disc (here defined as the high-$\mathrm{[\alpha/Fe]}$ sequence), as well as checking for evidence of radial migration in the thin disc, which are both key constraints to the Milky Way evolution, and b) gain insight into processes that may have altered the mass of a star along its evolution (e.g. mass loss during the RGB phase).}
The results presented here illustrate the impact precise ages can have on our understanding of the dominant events shaping our Galaxy. Theoretical work on Galactic archaeology has shown that on top of processes such as gas accretion (infall) and inside-out disc formation (e.g. \citealt{Chiappini1997}, \citealt{Chiappini2001}, \citealt{Brook2006}, and, more recently, \citealt{Grisoni2017}, \citealt{Noguchi2018}, \citealt{Grand2018}, \citealt{Mackereth2018a}, \citealt{Spitoni2019}, \citealt{Nuza2019}), additional secular processes, such as radial migration (e.g. \citealt{Wielen1977}, \citealt{Sellwood2002}, \citealt{Roskar2008}, \citealt{Minchev2010}) and non-secular processes such as mergers (e.g. \citealt{Abadi2003}, \citealt{Bird2013}, \citealt{Villalobos2008}), are responsible for moving stars away from their birthplace. Luckily, the chemical (\citealt{Minchev2013,Minchev2014}, \citealt{Bergemann2018}) and dynamical (e.g. \citealt{Antoja2018}, \citealt{Bland-Hawthorn2019}) signatures of these processes can be extracted from the exquisite datasets presently available for the MW, especially when precise ages are known in a wide age range (see discussion in \citealt{Miglio2017}). 

In particular, as it became clear after \Gaia\ DR2, the Milky Way suffered a large collision with another dwarf Galaxy, the  so-called Gaia-Enceladus \citep{Helmi2018} or Gaia-Sausage \citep{Belokurov2018}, roughly estimated to have happened around 10 Gyrs ago, contributing to the halo and/or thick disc population observed today \citep{Haywood2018, Sahlholdt2019, diMatteo2019, Mackereth2019a, Deason2019, Mackereth2020}.  However, many questions remain  open, namely: What was the state of the Milky Way when these mergers occurred? Were the thick disc, bulge and an in-situ halo in place? What fraction of the halo observed today is actually made of stars from Gaia-Enceladus?  Was the thick disc a result of MW-Enceladus collision or was the disc already forming when the impact occurred, and continued to be formed afterwards, as claimed by some authors \citep[e.g.][]{Grand2020}? {From the colour-magnitude diagram (CMD) analysis of stars within $\mathrm{2~kpc}$ from the Sun, \citet{Gallart2019} suggested that the components of the double sequence observed in the CMD of a sample of kinematically-defined halo stars are coeval but have different metallicity, with the bluer sequence being associated with the accretion event (e.g.
\citealt{Helmi2018, Belokurov2018, Haywood2018}).} The ages inferred by comparing stellar-model tracks to observed quantities in the CMD suggest a sharper halo age cut around 10 Gyr upon the major accretion event to the MW, while these authors also suggested the thick disc component to be younger. However, finding robust answers to these questions require precise ages. {Asteroseismology is starting to provide relevant constraints also in the metal-poor regime \cite[e.g., see][]{Valentini2019}, and to provide high-precision ages for stars that were likely born in-situ \citep{Chaplin2020} or accreted from Gaia-Enceladus \citep{Montalban2020}.} 

We devote a significant part of this work to exploring and characterising the uncertainty on our mass and age estimates.
We then present general results and trends which we find to be robust against such uncertainties. 
Independent measurements of  masses and radii were shown to be within a few percent those of obtained from asteroseismology (based, e.g., on observations of eclipsing binaries, stellar clusters, and stars with precise distances, see \citealt{Stello2016, Miglio2016, Handberg2017, Brogaard2018, Buldgen2019, Zinn2019, Khan2019, Hall2019, Joergensen2020}). 
Provided that the inferred {mass} is accurate, one can show that the age of such a star is largely related to its main-sequence lifetime. The latter is primarily determined by the star's luminosity, and hence the mass of the star, by its chemical composition (both {heavy-elements and helium mass fraction}), and affected by additional uncertainties related to the modelling of stars (e.g. nuclear reaction rates, occurrence of diffusion, convective boundary mixing in stars that develop a convective core). One additional limitation to using measured mass as an age proxy of giant stars is the possible difference between the current and initial stellar mass (e.g. due to { mass loss} along the red-giant branch or by the occurrence of {mass exchange and coalescence} in binary systems, see e.g. \citealt{DeMarco2017} for a review).

The paper is organised as follows.
In Sec. \ref{sec:data_models} we present the target sample and the stellar models used in this work. In Sec. \ref{sec:method} we describe the methods used to infer stellar properties.
The main results on ensemble age-chemistry studies are reported in Sec. \ref{sec:results}, where particular care is taken to infer properties of stars in the $\alpha$-rich sequence. 
We then devote Section \ref{sec:stevo} to investigating the properties and occurrence of stars that have likely lost or accreted mass during their evolution.
As discussed above, a significant component to current uncertainties on mass and age is systematic and stems from either model parameters which are poorly constrained (e.g. initial helium mass fraction), or fundamental uncertainties in stellar models, or systematic uncertainties in the observational constraints (e.g. effective temperatures). We explore some of these effects in Appendix \ref{sec:systematics}, where we include tests of seismically inferred properties based on independent information from \Gaia\ DR2 parallaxes. The outcome of these tests is used to inform our conclusions about the robustness of the trends evinced from the ensemble of stars studied here. Finally, a summary of the results and our conclusions are reported in Sec. \ref{sec:summary}.

\section{Observational constraints and stellar models}
\label{sec:data_models}
{ We  consider \Kepler\ solar-like oscillating giants whose spectroscopic parameters (\teff, [Fe/H] and [$\alpha$/M]) are available from SDSS APOGEE  DR14 \citep{DR14}. The list of targets with detected oscillations are taken either from \citet{Pinsonneault2018} or from the wider sample presented in \citet{Yu2018}. We have not excluded any stars based on orbital parameters. All stars with \texttt{STAR\_BAD} or \texttt{STAR\_WARN} flags from APOGEE were removed.} The total number of stars with such constraints is nearly 5400.

{As mentioned previously, one of our aims is to explore the effect on the inferred masses and ages of potential biases in the data (e.g. \Teff\ and metallicity scales, definition of seismic average parameters) and of using different grids of stellar models. The various assumptions taken in each modelling run are reported in the last column of Table \ref{tab:runs}.}
 
 \begin{sidewaystable*}

  \renewcommand{\arraystretch}{1.6}

  \begin{tabular}{llcccccc|p{3cm}ccm{3cm}}
     Run &  &  $\mu_{\rm Age}$ & $\delta_{\rm Age}$&  $\mu_{\rm Mass}$ & $\delta_{\rm Mass}$ & $\epsilon_{\rm Mass}$ & \DM &\multicolumn{4}{c}{Notes} \\
     & & [Gyr] & [Gyr] &  [M$_{\odot}$]& [M$_{\odot}$] & & [M$_{\odot}$] & \deltanu&\numax&Model grid& Other\\
\hline
\hline
 R1
 &    RGB       & $    10.98^{+0.13}_{-0.14}$ & $     0.76^{+0.27}_{-0.23}$ & $    0.972^{+0.004}_{-0.004}$ & $    0.050^{+0.008}_{-0.008}$ & $     0.06^{+0.02}_{-0.01}$  & &  Yu et al.,&  Mosser & Diffusion (G2)& \\ 
 &  RC &       &       & $    0.872^{+0.005}_{-0.005}$ & $    0.082^{+0.012}_{-0.010}$ & $     0.18^{+0.03}_{-0.03}$  & $0.10^{+0.01}_{-0.01}$  &\\ 
 \hline
 R2 &    RGB       & $    11.06^{+0.16}_{-0.17}$ & $     0.96^{+0.35}_{-0.30}$ & $    0.978^{+0.004}_{-0.004}$ & $    0.038^{+0.008}_{-0.006}$ & $     0.05^{+0.01}_{-0.01}$  & &  Yu et al.,&  Mosser &  No diffusion (G1)& \\  
 & RC &       &       & $    0.877^{+0.006}_{-0.006}$ & $    0.114^{+0.012}_{-0.012}$ & $     0.21^{+0.03}_{-0.03}$  & $0.10^{+0.01}_{-0.01}$\\
 \hline
 
 R3&     RGB  & $    10.70^{+0.16}_{-0.15}$ & $     0.83^{+0.29}_{-0.26}$ & $    0.977^{+0.004}_{-0.004}$ & $    0.046^{+0.008}_{-0.008}$ & $     0.05^{+0.01}_{-0.01}$  & &  individual radial-mode frequencies $\nu_{i}$ & Mosser & Diffusion (G2) & [$\alpha$/Fe]>0.1\\ 
 &  RC&       &       & $    0.887^{+0.006}_{-0.006}$ & $    0.084^{+0.014}_{-0.012}$ & $     0.18^{+0.03}_{-0.03}$  &$0.09^{+0.01}_{-0.01}$ \\ 
 
 \hline
  R4&          RGB & $    11.05^{+0.21}_{-0.20}$ & $     1.07^{+0.39}_{-0.32}$ & $    0.979^{+0.004}_{-0.004}$ & $    0.044^{+0.008}_{-0.008}$ & $     0.05^{+0.01}_{-0.01}$  &  &  $\nu_{i}$ & Mosser &  No diffusion (G1) & [$\alpha$/Fe]>0.1\\ 
 & RC &       &       & $    0.881^{+0.007}_{-0.006}$ & $    0.118^{+0.014}_{-0.012}$ & $     0.19^{+0.04}_{-0.03}$  & $0.10^{+0.01}_{-0.01}$\\ 
 \hline
 R5&          RGB & $     9.12^{+0.16}_{-0.14}$ & $     0.74^{+0.27}_{-0.24}$ & $    1.013^{+0.004}_{-0.004}$ & $    0.052^{+0.008}_{-0.008}$ & $     0.05^{+0.01}_{-0.01}$  & &  $\nu_{i}$ & Mosser &  $\Delta$Y/$\Delta$Z=2 (G3) & [$\alpha$/Fe]>0.1\\ 
 \hline
 
  R6&      RGB & $    10.70^{+0.15}_{-0.15}$ & $     0.85^{+0.28}_{-0.25}$ & $    0.978^{+0.004}_{-0.004}$ & $    0.052^{+0.010}_{-0.008}$ & $     0.05^{+0.02}_{-0.01}$  & & $\nu_{i}$ & Mosser & Diffusion (G2) & [$\alpha$/Fe]>0.1, \newline \teff+100 K \\ 
 & RC &       &       & $    0.882^{+0.006}_{-0.006}$ & $    0.100^{+0.014}_{-0.014}$ & $     0.17^{+0.04}_{-0.03}$  &$0.10^{+0.01}_{-0.01}$ \\ 
 \hline
  R7&      RGB & $    10.03^{+0.16}_{-0.15}$ & $     0.87^{+0.31}_{-0.27}$ & $    0.983^{+0.004}_{-0.004}$ & $    0.060^{+0.010}_{-0.010}$ & $     0.05^{+0.02}_{-0.01}$  &  & $\nu_{i}$ & Mosser & Diffusion (G2) & [$\alpha$/Fe]>0.1, [Fe/H]-0.1\\ 
 & RC &       &       & $    0.892^{+0.007}_{-0.006}$ & $    0.102^{+0.014}_{-0.014}$ & $     0.17^{+0.03}_{-0.03}$  & $0.09^{+0.01}_{-0.01}$\\ 
 \hline
  R8&      RGB & $    11.49^{+0.17}_{-0.17}$ & $     0.83^{+0.32}_{-0.26}$ & $    0.968^{+0.004}_{-0.004}$ & $    0.050^{+0.008}_{-0.008}$ & $     0.05^{+0.02}_{-0.01}$  & & $\nu_{i}$ &  Mosser & Diffusion (G2) & [$\alpha$/Fe]>0.1,   [m/H]=[Fe/H]+[$\alpha$/Fe] \\ 
 & RC &       &       & $    0.878^{+0.006}_{-0.006}$ & $    0.098^{+0.014}_{-0.014}$ & $     0.17^{+0.03}_{-0.03}$  & $0.09^{+0.01}_{-0.01}$\\ 
 \hline
 R9&      RGB & $    12.01^{+0.17}_{-0.16}$ & $     0.86^{+0.31}_{-0.25}$ & $    0.947^{+0.004}_{-0.004}$ & $    0.042^{+0.008}_{-0.008}$ & $     0.05^{+0.01}_{-0.01}$  & & $\nu_{i}$ & Mosser & Diffusion (G2)& [$\alpha$/Fe]>0.1, \numax\ increased by 1\% \\ 
 & RC &       &       & $    0.871^{+0.005}_{-0.005}$ & $    0.082^{+0.012}_{-0.012}$ & $     0.16^{+0.03}_{-0.03}$  &$0.08^{+0.01}_{-0.01}$ \\ 
 
 \hline


 
  R10&          RGB & $    12.62^{+0.21}_{-0.21}$ & $     0.76^{+0.36}_{-0.20}$ & $    0.945^{+0.004}_{-0.004}$ & $    0.038^{+0.008}_{-0.006}$ & $     0.05^{+0.01}_{-0.01}$  & &  Mosser &  Mosser & No diffusion (G1) &\\ 
 & RC &       &       & $    0.891^{+0.006}_{-0.006}$ & $    0.110^{+0.012}_{-0.012}$ & $     0.20^{+0.04}_{-0.03}$  &$0.06^{+0.01}_{-0.01}$ \\ 
 \hline
 R11&          RGB & $    12.56^{+0.17}_{-0.16}$ & $     0.86^{+0.32}_{-0.27}$ & $    0.937^{+0.004}_{-0.004}$ & $    0.044^{+0.008}_{-0.008}$ & $     0.06^{+0.01}_{-0.01}$  & & Mosser & Mosser & Diffusion (G2) &\\ 
& RC &       &       & $    0.892^{+0.005}_{-0.005}$ & $    0.076^{+0.010}_{-0.012}$ & $     0.15^{+0.03}_{-0.03}$  & $0.05^{+0.01}_{-0.01}$\\ 
 \hline

 \end{tabular}
 \vspace{0.5cm}
 \caption{Description of the various assumptions taken while running {\sc param} (see Sec. \ref{sec:data_models} for details).  Median age and mass ($\mu_{\rm Age}$, $\mu_{\rm Mass}$), and intrinsic age and mass spread ($\delta_{\rm Age}$, $\delta_{\rm Mass}$) of the   $\mathrm{[\alpha/Fe]>0.1}$ population are reported together with their uncertainties (based on the 14th and 86th percentiles of the distribution). $\delta_{\rm Mass}$ is defined as twice the standard deviation of the Gaussian describing the intrinsic mass distribution, and $\delta_{\rm Age}$ as the age range between $\mu-\sigma$ and $\mu+\sigma$, with $\mu$ and $\sigma$ the mean and standard deviation of the Gaussian in $\log_{10}(\mathrm{age})$ (see Appendix \ref{sec:HBM}).  $\epsilon$ describes the contaminant fraction as inferred from the mass distribution of both RGB and RC stars.}
\label{tab:runs}

 \end{sidewaystable*}

\subsection{Asteroseismic constraints}
\label{sec:asterodata}
As asteroseismic constraints we use the average indices \numax\ and \deltanu.
The former is determined using the method described in  \citet{Mosser2011}.
We use two different measurements of the average large frequency separation \deltanu: from \citet{Mosser2011} and from \citet{Yu2018}.
Moreover, in a sub-sample of stars ($\alpha$-rich stars, as defined in Sec. \ref{sec:rgbmass_age}), we  measure  \deltanu\,  also by fitting individual radial-mode frequencies (also known as `peakbagging') following the approach presented in \citet{Davies2016}. The latter approach allows for a definition of \deltanu\ closer to that adopted in our models, as described in Sections \ref{sec:models} and \ref{sec:method} (see also \citealt{Khan2019} and \citealt{Viani2019}). { Since measuring individual-mode frequencies is not a fully automated procedure yet, we apply this approach to the limited set of stars which we then use to make detailed, precise inferences on mass loss and age.}

Moreover, as presented in \citet{Khan2019}, \deltanu\ from \citet{Yu2018} is close  to that derived from individual-mode frequencies, while the \deltanu\ as determined by \citet{Mosser2011} has a different definition, closer to the analytical asymptotic relation, and its value for RGB stars is systematically larger by $\simeq 1\%$  compared to the one from individual mode frequencies. This difference stems from the acoustic glitches due the second ionisation of Helium \citep{Vrard2015}.

Since in the modelling code used here ({\sc param}) we define \deltanu\ from the individual radial-mode frequencies (\citealt{Rodrigues2017}), our preferred choice for \deltanu\ is that from peakbagging, or from \citet{Yu2018} when the former is not available (i.e. for the stars in the low-$\alpha$ sequence). The comparison of observed and modelled parameters defined in similar manners ensures that artefacts, such as the glitches mentioned before, do not affect the analysis. {Our approach is therefore different, for example, to \citet{Pinsonneault2018}, where empirical calibrations are used to rescale and combine asteroseismic results from different pipelines.}

Core-He burning stars in the sample are identified using the properties of their mixed-modes frequency spectra \citep{Bedding2011, Mosser2011, Vrard2016, Elsworth2017}, and specifically the `consensus evolutionary state' described in \citet{Elsworth2019}.

\subsection{Orbital parameters}
\label{sec:orbitalparam}
We measure the orbital parameters for the stars in question by taking 100 samples of the covariance matrix formed from the reported observed RA, DEC, proper motion in RA and DEC, distance and radial velocity, and their uncertainties and correlation coefficients. Distance, { as determined using the code {\sc param} (see Sec. \ref{sec:method})} and radial velocity are uncorrelated with each other and the measures from \Gaia\ DR2 \citep{GaiaCollaboration2018}. 
We then estimate the orbital parameters of each of these samples using the fast orbit-estimation method of \citet{Mackereth2018} implemented in \texttt{galpy} \citep{Bovy2015}. We assume the simple \texttt{MWPotential2014} potential. We assume the position of the Sun to be $R_0 = 8.125$ kpc \citep{Gravity2018}, and $z_0 = 0.02$ kpc \citep{Bennett2019}, and its velocity to be $\vec{v}_0=[U,V,W]=[-11.1,245.6,7.25]\ \mathrm{km\ s^{-1}}$, based on the SGR A* proper motion \citep{Gravity2018} and the solar motion derived by \citet{Schoenrich2010}. We estimate pericentre ($R_{\rm peri}$) and apocentre radii ($R_{\rm apo}$), orbital eccentricity and the maximum vertical excursion ($z_{\rm max}$), their uncertainties and correlation coefficients for each star.

\subsection{Spectroscopic constraints}
Spectroscopic parameters are taken from SDSS-IV/APOGEE DR14 \citep{DR14}.
 APOGEE data products used in this paper are those output by the standard data analysis pipeline, the APOGEE Stellar Parameters and Chemical Abundances Pipeline (ASPCAP, \citealt{GarciaPerez2016}), which uses a precomputed spectral library \citep{Zamora2015}, synthesised using a customised H-band linelist \citep{Shetrone2015}, to measure stellar parameters and element abundances. A full description and examination of the analysis pipeline is given in \citet{Holtzman2018}. { For a comprehensive review of the APOGEE survey, see \citet{Majewski2017}}.  As in \citet{Pinsonneault2018}, we use ASPCAP's [$\alpha$/M] as a proxy for [$\alpha$/Fe].  In our sample, the difference between [$\alpha$/M] and an average  [$\alpha$/Fe] defined using O, Mg, Si, S, and Ca over Fe are negligible (with a mean offset equal to $\mathrm{-0.01\ dex}$ and a standard deviation of $\mathrm{0.02\ dex}$).

The median uncertainty on \Teff\ is $\sigma_{\rm T_{\rm eff}}=75$ K, while $\sigma_{\rm  [Fe/H]}=0.030$ dex,  and $\sigma_{\rm [\alpha/Fe]}= 0.012$ dex.  In the modelling runs, we increase the uncertainties on spectroscopic parameters by a factor 2 as the quoted uncertainties are internal errors  only, and cross-validation against other surveys shows larger differences \citep[e.g., see][]{Rendle2019b, Hekker2019, Anguiano2018}. 
Moreover, model-predicted $T_{\rm eff}$ suffer from large uncertainties associated with the modelling of outer boundary conditions and near-surface convection, hence we prefer to downplay the role of $T_{\rm eff}$.

\begin{figure}
   \resizebox{\hsize}{!}{\includegraphics{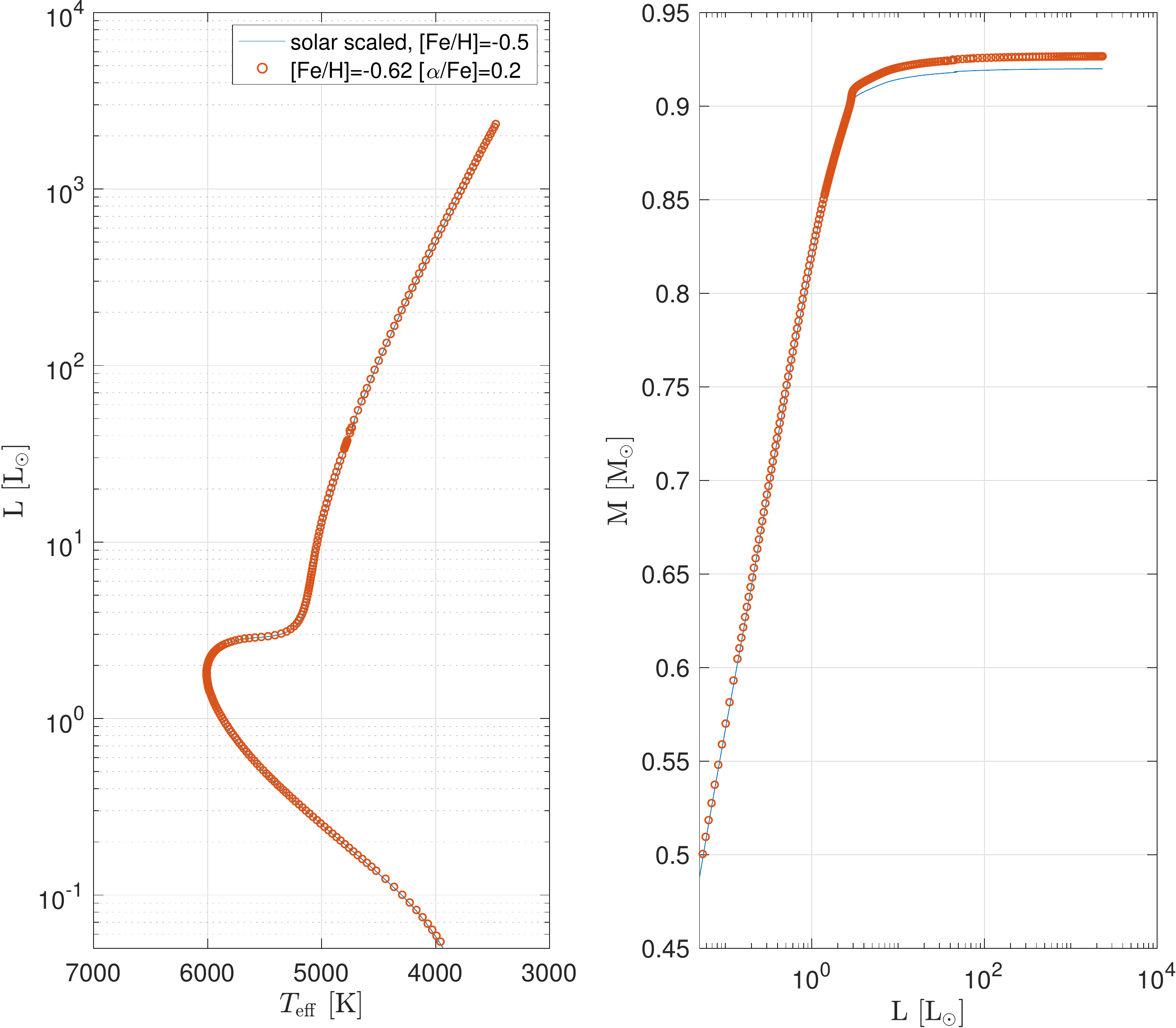}}
   \caption{Comparison of DSEP \citep{Dotter2008} 11-Gyr isochrones computed with $\alpha$-enrichment ([Fe/H]=-0.62 and [$\alpha$/Fe]=0.2, circles) and with a solar-scaled metallicity (line), following the formula of \citet{Salaris1993}.}
 \label{fig:alphatrack}
 \end{figure}

For stars showing enhancement in the $\alpha$ elements we apply the prescription described by \citet{Salaris1993}, updated to use of  \citet{Grevesse93} solar abundances \citep[see also][]{Valentini2019}.  To check whether  adopting the \citet{Salaris1993} rescaling introduces significant biases in the analysis,  we consider models based on the DSEP code\footnote{\texttt{http://stellar.dartmouth.edu/models/isolf\_new.html}} \citep{Dotter2008} for a chemical composition which corresponds to an extreme $\alpha$ enrichment for the stars  in our sample. Models computed  both with solar-scaled (following \citealt{Salaris1993}) and with $\alpha$-rich abundances are compared  in  Fig. \ref{fig:alphatrack}. At a given age (11 Gyr), the differences on the HR diagram and in the mass of RGB stars between the two sets of tracks appears to be negligible ($\lesssim 1\%$). Despite this, to estimate the relevance of accounting for $\alpha$ enrichment in our sample  we explore the effect, for instance, of overestimating such a correction in one of our modelling runs (R8, see Table \ref{tab:runs}). Also, the effect of potential systematic effects in the overall $\mathrm{[Fe/H]}$ scale of the order of $\mathrm{0.1 dex}$ are considered in the modelling run R7.

\subsection{Stellar models}
\label{sec:models}
To explore how sensitive our results are to the input models, we have considered three sets of  of evolutionary tracks. 

{The first set (G1) is described in \citet{Rodrigues2017}. These tracks are computed using {\sc mesa} \citep{Paxton2011, Paxton2013, Paxton2015} assuming a solar metal-mixture and no diffusion. The evolution is followed from the pre-main sequence up to the first thermal pulse. We refer to \citet{Rodrigues2017} for more information about the choice of the micro- and macro-physics adopted in the models.}

A second set of isochrones (G2) was computed including microscopic diffusion in the models. In {\sc mesa}, we adopt the implementation of microscopic diffusion described in \citet{Choi2016},  but do not turn off diffusion in the post-main-sequence phase, which is expected to have, however, limited impact on the stars of interest here. A non-negligible effect on the model properties, on the other hand, is that in a grid computed with diffusion, the calibrated solar model has a different mixing-length parameter ($\alpha_{\rm MLT}=2.12$), and a different initial helium and heavy-element mass fractions ($Y_0=0.274$, $Z_0=0.0197$) compared to the solar model without diffusion ($\alpha_{\rm MLT}=1.96$, $Y_0=0.266$, $Z_0=0.0176$).

As  in most grids of stellar models \citep[see e.g.][]{Pietrinferni2004,Bressan2012, Choi2016}, when computing models at different $Z$  we assume a linear relation between $Y$ and $Z$, using as calibrating points the Sun and  the primordial helium abundance ($Y_{\rm P} = 0.2485$, \citealt{Aver2013}).
This assumption leads to differences in $Y$ which may become substantial ($\gtrsim 0.01$) at solar and super-solar metallicity. In the grid computed with diffusion, for instance, models with twice the solar metallicity reach $Y\simeq 0.30$, which is compatible with the helium abundance estimated in the open cluster NGC6791 \citep{Brogaard2012}, while lower values are assumed for the grid without diffusion ($Y\simeq0.28$, see Fig. \ref{fig:DYDZ}).

The hypothesis of a linear relation between $Y$ and $Z$ is of course a simplification, as shown, for instance, by  helium abundance variations within globular clusters and by the debated complex chemical enrichment of the bulge \citep[e.g. see][and references therein]{Nataf2016, Milone2018}. However, in most disc stars, a nearly linear relation between $Y$ and $Z$ is expected from chemical evolution models \citep[e.g.][]{Chiosi1982, Vincenzo2019} as well as empirical determinations (e.g., \citealt{Ribas2000, Casagrande2007}).

Helium-enrichment relations significantly and systematically affect ages given the precision enabled by asteroseismic constraints \citep[e.g. see][]{Lebreton2014}, hence we argue that one should at the very least estimate the systematic uncertainties related to such an assumption. For this reason we compute an additional grid (G3) where we consider a helium-metallicity enrichment relation which is twice that of the original grid of \citet{Rodrigues2017}. A more detailed discussion on the effect of  different helium enrichment relations on the models is presented in Sec. \ref{sec:he_enrich}.

\section{Method}
\label{sec:method}
Masses, ages, radii and distances are inferred using the code {\sc param} \citep{daSilva2006, Rodrigues2017}.  Asteroseismic constraints are included in a self-consistent manner, whereby \deltanu\  is computed using the radial-mode frequencies of the models in the grid, not added as an a-posteriori correction to the scaling relation between \deltanu\ and the square root of the stellar mean density. This approach has limitations, primarily related to the accuracy of model predictions,  but reduces the additional uncertainty on how to apply the corrections to the \deltanu\ scaling relation  (see e.g. \citealt{Brogaard2018}).  At this time this approach has yielded masses and radii which show no systematic deviations to within few percent of independent estimates (see e.g. \citealt{Miglio2016}, \citealt{Rodrigues2017}, \citealt{Handberg2017},  \citealt{Brogaard2018}, who partially revisited the work by \citealt{Gaulme2016}, and \citealt{Buldgen2019}).
One should, however, be aware that such tests have been carried out sampling very sparsely the age-metallicity plane,
albeit with tests in clusters that span a metallicity range between [Fe/H]$\simeq$0.3, e.g. NGC6791 and [Fe/H]$\simeq$-1.1,  M4 (see \citealt{McKeever2019, Miglio2016, Valentini2019}).

There is also some ambiguity on how the average defining \deltanu\ is taken in the models and in the data. As shown in \citet{Handberg2017} and in Fig. 4 in \citet{Rodrigues2017}, if  individual radial-mode frequencies are available, then one can adopt a similar definition of \deltanu\ in the models and in the data. To test the effect of this, we determined individual radial-mode frequencies in the sub-sample of stars with [$\alpha$/Fe]>0.1 and compared the results with \deltanu\ measured using the method in \citet{Mosser2009} and \citet{Mosser2011} (see Table \ref{tab:runs} and \citealt{Khan2019}).

When inferring stellar properties of red-giant stars (here primarily mass and age) from a combination of seismic indices and photospheric constraints,  one should remember that such properties  depend on the observational constraints via power laws, that is mass scales as  \deltanu$^{-4}$ and age as \deltanu$^{\sim 14}$. It is thus inevitable to get higher resolution at younger ages, and a blurred view at older stellar ages; this is why we  adopt a logarithmic scale when discussing and showing age distributions.

While our theoretical understanding of \deltanu\, allows us to use model-predicted values, albeit with some still-standing issues related to the so-called surface effects and their  dependence on stellar properties \citep[e.g., see][]{Manchon2018}, we take the scaling relation of \numax\ at face value. Currently we lack a robust prediction from theory and the scaling relation is to be considered at this stage primarily an empirical relation (but see \citealt{Belkacem2011, Zhou2019}). We assume,
$$\numax = \frac{M/{\rm M_\odot}}{(R/{\rm R_\odot})^2 \sqrt{T_{\rm eff}/{\rm T_{\rm eff_\odot}}}} \,\numax_\odot\rm ,$$
where $\numax_\odot=3090\, \mu \rm{Hz}$ \citep[see][]{Handberg2017, Huber2011}.
The effect of a systematic bias of the \numax\ scaling relation on our results is investigated in one of our modelling runs (R9, see Table \ref{tab:runs}).

As mentioned earlier, we are aware that biases, for example in the \teff\ and metallicity scales, both in the models and in the constraints, can lead to significant systematic effects, hence we explore whether our findings are robust against those (see Sec. \ref{sec:systematics} and Table \ref{tab:runs}).

Finally,  we set a uniform prior on the age from 0 to 40 Gyr, which is intentionally largely uninformative. This is to avoid  setting an artificial lower limit to stellar mass, hence effectively introducing a  bias related to the prescription of the mass loss efficiency during the RGB which would also fail to reproduce stars that have likely lost significant mass  (see e.g. \citealt{Handberg2017}  for the case of the solar-metallicity 0.8 M$_{\odot}$ star in NGC6819).

\begin{figure*}
\hspace{-2cm}
     \resizebox{1.2\hsize}{!}{\includegraphics{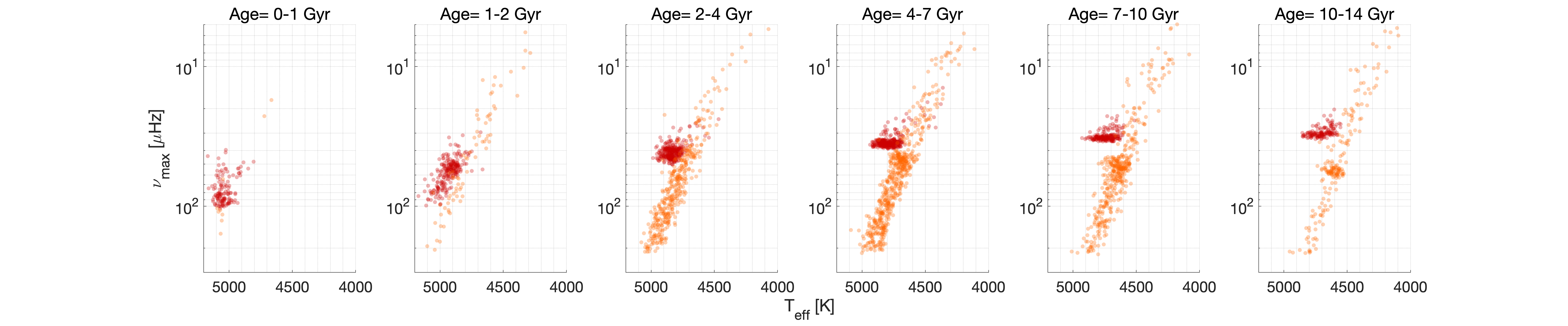}}
\caption{Observational properties ($T_{\rm eff}$ and \numax, where  $\numax\propto g /\sqrt{{T_{\rm eff}}}$, see Sec. \ref{sec:method}) of stars with $-0.15 \le {\rm [Fe/H]} \le 0.2$ in different age bins. Stars in the core-He burning phase are depicted in red. Age increases from left to right, where one notices how young stars populate almost exclusively the secondary clump. Age is inferred using the method described in Sec. \ref{sec:method}.}
 \label{fig:monoage_highmet}
\end{figure*}

\begin{figure*}
\hspace{-2cm}
     \resizebox{1.2\hsize}{!}{\includegraphics{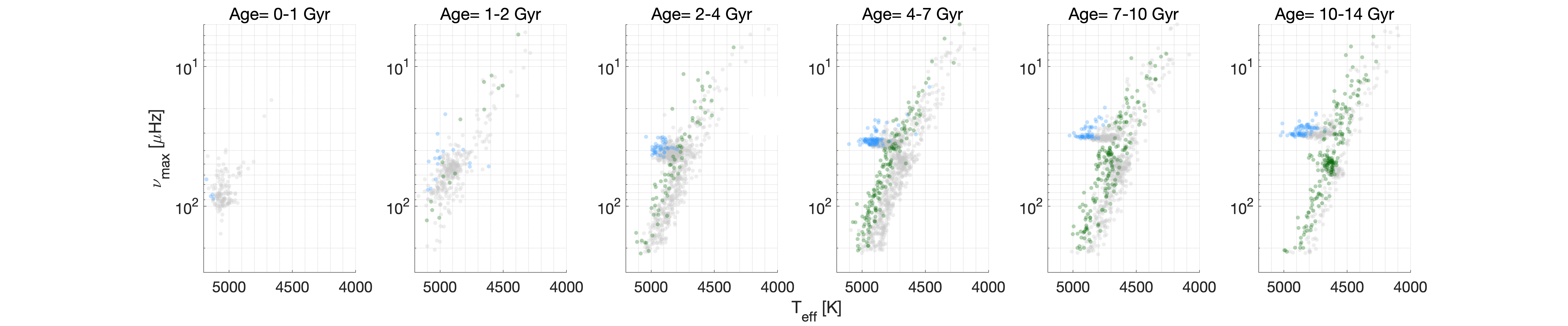}}
\caption{As in Fig. \ref{fig:monoage_highmet} but considering stars with lower metallicity, that is, $-0.5 \le {\rm [Fe/H]} \le -0.2$. Stars of higher metallicity ( $-0.15 \le {\rm [Fe/H]} \le 0.2$) are represented by grey dots in the background. Already from this plot one sees that in this sample young metal poor stars are rare while old stars are present in both metallicity bins. Stars in the core-He burning phase are shown light blue.}
 \label{fig:monoage_lowmet}
\end{figure*}
\section{Age dissection of the MW discs}
\label{sec:results}
This work builds upon and is a natural continuation of previous studies of the Milky Way using stars with asteroseismic constraints. Initially these approaches were limited to studying the distribution of average seismic parameters only \citep{Miglio2009}, then to reporting distributions of masses \citep{Chaplin2011, Miglio2013a} and eventually to dissecting the stellar population in age  intervals, as tests of the robustness of the seismically inferred properties became available 
\citep[e.g., see][]{Casagrande2016, Anders2017, Anders2017a, SilvaAguirre2018, Rendle2019b, Ciuca2020}. 
We now aim at an age dissection of the MW discs at higher precision and accuracy, benefiting from a larger dataset and the inclusion in the analysis of discussion and testing of the main systematic uncertainties affecting our method. 
\label{sec:context}

As shown by Figs. \ref{fig:monoage_highmet} and  \ref{fig:monoage_lowmet}, the data considered in this study, coupled to the modelling described in Sec. \ref{sec:method}, enable us to dissect the composite population in our sample into `stellar-cluster-like'  populations, in terms of age and chemical composition. 
From the distribution of stars in these two figures, one can already see both the effects of stellar evolution and the chemical evolution of the Galaxy at work. 
Recall that the population of stars explored by \Kepler\ is located  at a nearly constant galactocentric radius ($\langle R\rangle=7.7 \pm 0.1$ kpc), thus radial variations are minimised. 

With this in mind, some notable features of these two figures are: a) for a sample of red-giant stars, the youngest tend to concentrate in the secondary clump \citep{Girardi99}  which, as expected from stellar evolution, is populated by stars just massive enough to ignite He in non-degenerate conditions. These stars have a helium-core mass -- and hence luminosity -- lower than that of the main RC; b) young metal-poor stars are rare, while old stars cover a broad range of metallicity, just as expected from chemical evolution predictions for the solar neighbourhood; c) trends expected from basic stellar evolution predictions (e.g. \teff\ variations with mass, age, and metallicity) are evident in these plots, owing primarily to the precision and accuracy of the seismic and spectroscopic measurement available \citep[see also][]{Pinsonneault2018}. 

Going beyond this broad qualitative picture, additional considerations should be made if one wishes to define a sample of stars with ages less affected by systematic uncertainties. As mentioned earlier, the ages of stars in the red-giant phase are determined primarily by their initial mass. However, since stars are expected to experience mass loss while on the RGB, the age estimates of stars in the RC phase (which
constitute a large fraction of the red giants with detected oscillations) are plagued by our poor
understanding of RGB mass loss. Constraints on the efficiency of mass loss are therefore crucial to enable the accurate determination of ages of RC stars. Section \ref{sec:mloss} will be devoted to inferring an integrated mass-loss rate for the $\alpha$-rich population.

Here, we select stars with robust age estimates by removing stars in the RC with masses below 1.2 M$_\odot$, because their actual masses are expected to be more significantly affected by mass loss.
Mass loss from younger, more massive stars is expected to be negligible (from Reimers-like prescriptions, e.g. \citealt{Castellani2000}, and as inferred by asterosesimology,  e.g., \citealt{Miglio2012, Stello2016, Handberg2017}). The trends described below, however, are largely insensitive to this additional selection. 

Also, among the non core-He burning giants, we restrict the sample to stars with estimated radii smaller than 11 R$_{\odot}$. This avoids contamination by early-AGB stars, and removes stars with relatively low \numax, a domain where seismic inferences have not been extensively tested so far. This reduces our initial sample of $\sim5400$ stars to $\sim3300$. The median random uncertainty in mass of the stars in our complete sample is  $ 6 \%$, which translates to a $23\%$ median random uncertainty in age. 
In what follows we use this reduced sample to study the age-[$\alpha$/Fe] relation (Section~\ref{agealpha}), the old metal-rich stars in the solar vicinity (which have presumably undergone radial migration, see Section~\ref{oldrich}), the age gradients with distance from the mid-plane (Section~\ref{sec:orbits}),  and the age of the thick disc (Section~\ref{sec:alpha}). 

{ The resulting catalogue of stellar properties (available online) is presented in Appendix \ref{app:catalogue}.}

\begin{figure}
\centering
\vspace{-1.2cm}

  \resizebox{.99\hsize}{!}{\includegraphics{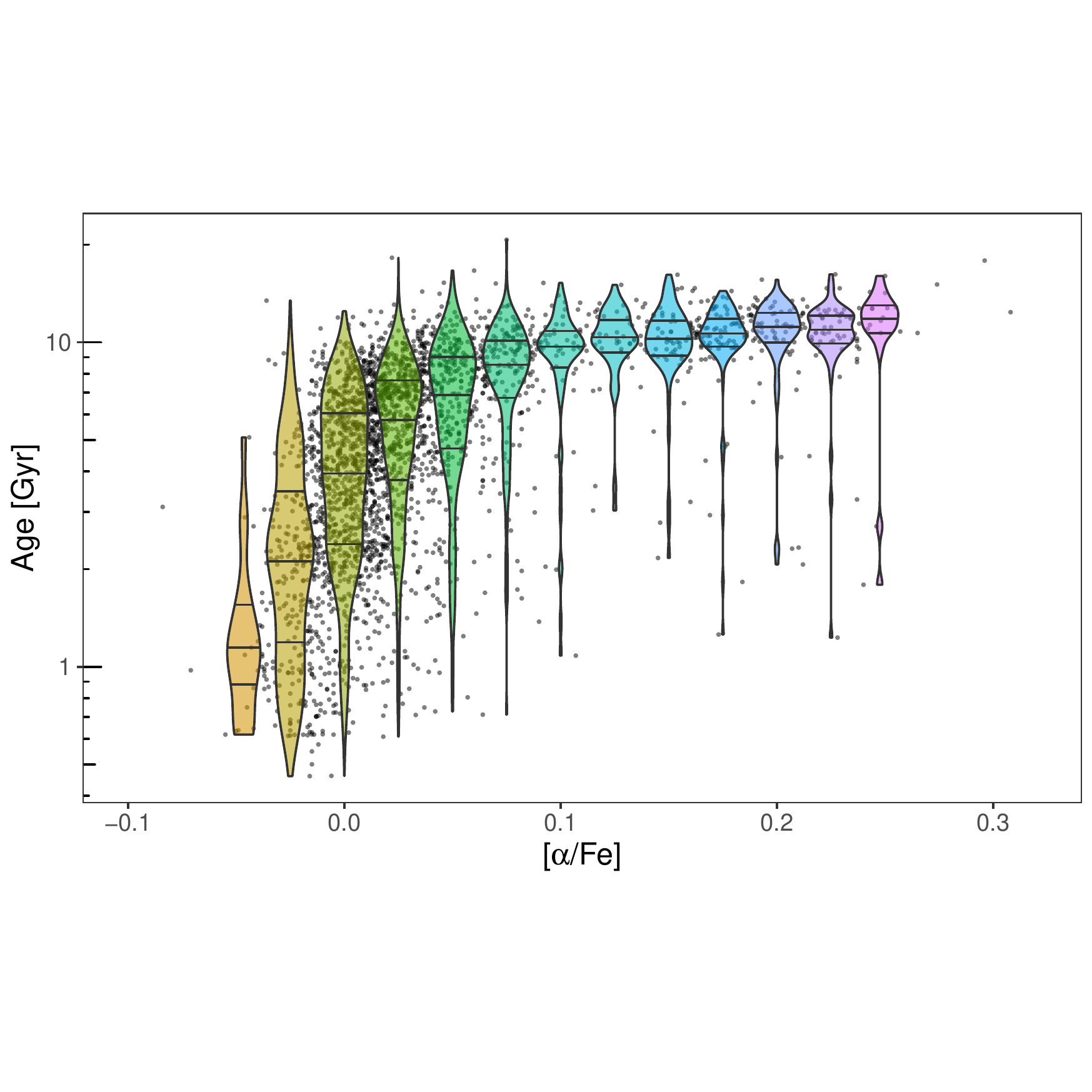}}
  \vspace{-1cm}

 \caption{Age as a function of [$\alpha$/Fe] of the stars in our sample. The age distributions of stars binned in [$\alpha$/Fe] are superposed on black dots representing stars in the sample. On each age distribution the three horizontal lines denote the 25th, 50th and 75th percentile. Long, thin tails extending to young ages are associated to the overmassive $\alpha$-rich stars (see Sec. \ref{sec:overmassive} for more details).}
\label{fig:agealphaR}
\end{figure}

\begin{figure*}
\centering
  \resizebox{.65\hsize}{!}{\includegraphics{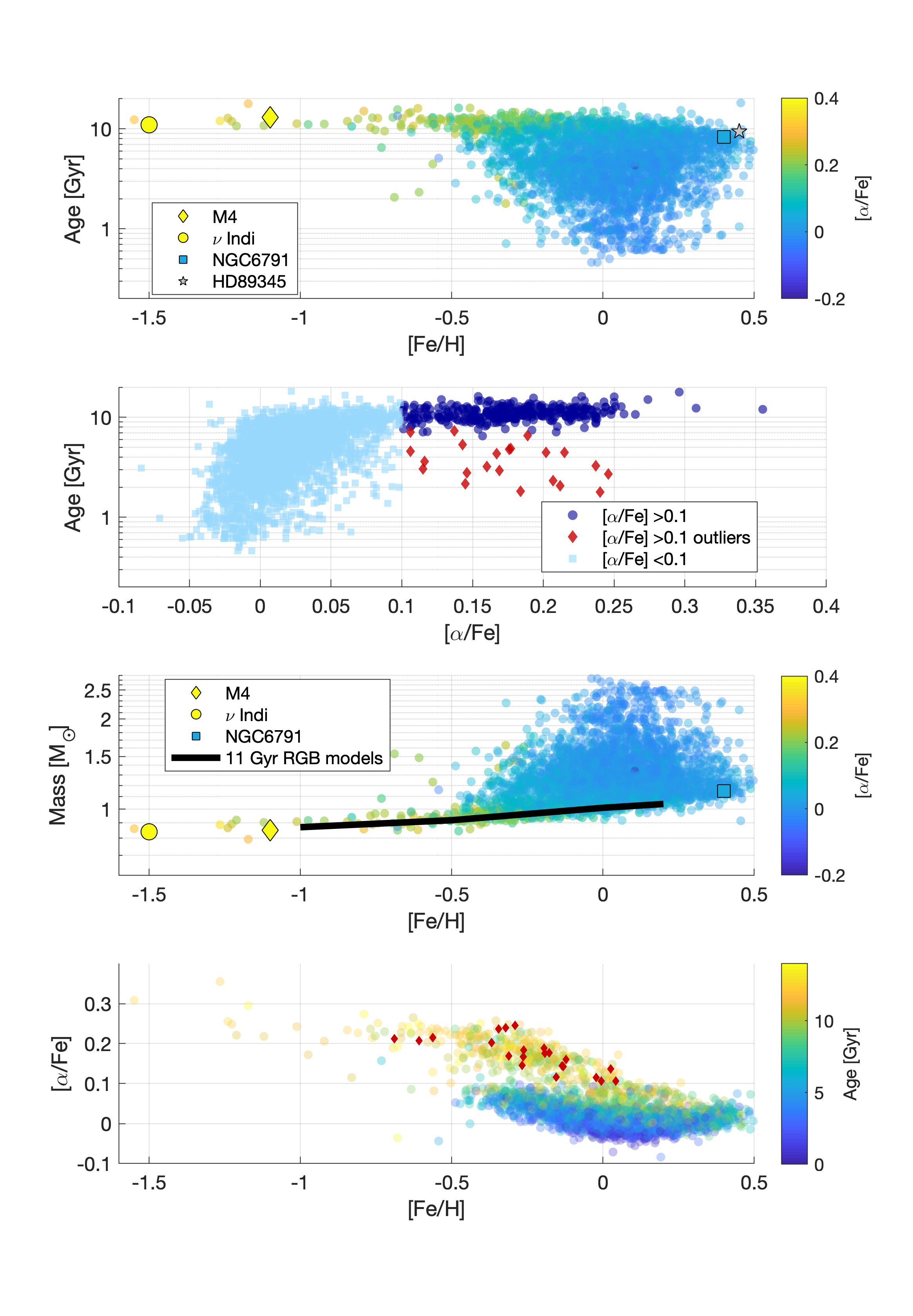}}

  \caption{Age-mass-chemical-composition scatter plots of stars in our sample ($R < 11$ R$_\odot$ and including RC stars with  $M > 1.2$ M$_\odot$, see main text for details). Black crosses indicate typical uncertainties on the relevant measured/inferred properties. {\it Top panel:} age vs. [Fe/H]. The colour represents [$\alpha$/Fe].   {\it Upper middle  panel:} age vs. [$\alpha$/Fe]. Red dots denote $\alpha$-rich stars that are considered outliers based on their mass or age (see Sec. \ref{sec:overmassive} for the criterion used). {\it Lower middle panel:}  Stellar mass versus [Fe/H]. { The mass of 11-Gyr-old RGB models of different metallicity is shown as a solid line.}  {\it Bottom panel:}  [$\alpha$/Fe] versus  [Fe/H], where the age is represented by colour. Red dots identify age/mass outliers among the $\alpha$-rich population.
  In the top and lower middle panels we also show the mass, chemical composition and inferred age for eclipsing binaries in the old-open cluster NGC6791 \citep{Brogaard2012}, the old metal-rich subgiant HD89345 (\citealt{VanEylen2018}, $\mathrm{[\alpha/Fe]}$ is not available for this object), and in well-studied metal-poor objects: RGB stars in the globular cluster M4 \citep{Kaluzny2013, Miglio2016} and the nearby subgiant $\nu$ Indi \citep{Chaplin2020}.}
\label{fig:agemasscomplete}
\end{figure*}
\subsection{Age-[$\alpha$/Fe] in the Solar circle}
\label{agealpha}
The lack of precise ages has been one of the main reasons to resort to more indirect ways of inferring broad ages in galactic evolution studies.  One can map galaxies in terms 
of their [$\alpha$/Fe] enhancement, a historical tracer of timescales \citep[e.g., see][]{Matteucci1990}.  In Fig.~\ref{fig:agealphaR} we show this relation in our sample. The  [$\alpha$/Fe]-rich  population (hereafter $\alpha$-rich) is composed primarily of very old objects, older than most of the [$\alpha$/Fe]-poor stars (hereafter, $\alpha$-poor). Although systematic uncertainties in absolute ages are still to be fully quantified, we find a  median age of $\sim 10\mhyphen12\, \rm Gyr$ which is in broad agreement with the ages of the thick disc stars in  \citet{Fuhrmann2011}, \citet{Haywood2013} and  \citet{Anders2018}, which were inferred from the HARPS-GTO sample of \citet{DelgadoMena2017}, and with the analysis of \citet{SilvaAguirre2018},  also based on \emph{Kepler} targets. However, in contrast to the latter, which was based on a smaller set of targets compared to ours, and on a combination of RGB and RC stars, we find a very tight age-[$\alpha$/Fe] relation in the [$\alpha$/Fe]-rich population, with important consequences for the thick-disc formation scenario. 

Because our ages are based on the assumption that the seismic masses are very close to the initial stellar masses, in what follows we discuss age-mass-chemistry plots. Figure~\ref{fig:agemasscomplete} focuses on the main trends which are robust against the systematic uncertainties tested in this study, and therefore presents our results based on the model grid we believe to be most reliable  (G2, run R1, see Appendix \ref{sec:systematics} for a discussion of systematic uncertainties).

Fig.~\ref{fig:agealphaR} and the upper-middle panel in Fig. ~\ref{fig:agemasscomplete} suggest that the chemical evolution of stars in the low-$\alpha$ sequence happened on much longer timescales compared to the high-$\alpha$ sequence. In Fig. \ref{fig:agemasscomplete},  we highlight the $\alpha$-rich stars. Here we selected stars with $\rm [\alpha/Fe]>0.1$, which is the value that seems to separate two regimes, namely: the very narrow age range of stars above this value, and the large age range for stars below [$\alpha$/Fe]=0.1. This separation is similar to that found by \citet{Anders2018} using a  dimensionality-reduction technique applied to a sample of around 500 stars from  HARPS-GTO, and hence covering a much smaller volume of $\sim$ 100 pc around the Sun. A  noticeable feature in the top panel of Fig.~\ref{fig:agemasscomplete} is the stark increase of the dispersion of [Fe/H] with age.
It is clear, however, that at sub-solar metallicities, at a given [Fe/H], stars in the high-$\alpha$ sequence are on average older than those in low-$\alpha$ sequence (Fig.~\ref{fig:agemasscomplete}, upper and lower panels).  One also notices that when the high- and low-$\alpha$ sequences intercept at $\mathrm{[Fe/H]\simeq 0}$, independent age information is crucial, as  $\alpha$ enrichment alone becomes an ineffective clock (see the lower panel of Fig.~\ref{fig:agemasscomplete}). The nearly coeval nature of the $\alpha$-rich RGB stars is also evinced from the  clear correlation of their mass with  metallicity (Fig.~\ref{fig:agemasscomplete}, lower middle panel), where the mass of 11-Gyr-old RGB models is shown as a solid line. The decline of stellar mass with decreasing metallicity follows closely what is expected for a coeval population, although a modest age increase with [$\alpha$/Fe] could be tentatively inferred from, for instance, Fig.~\ref{fig:agealphaR}. A detailed discussion of the age dispersion of stars in the high-$\alpha$ sequence is presented in Section~\ref{sec:alpha}. 
In Fig.~\ref{fig:agemasscomplete} additional objects with tight constraints on metallicity and age are shown to agree with the trend seen for the RG stars (the two metal-rich objects are discussed in  Section~\ref{oldrich}), namely: the globular cluster M4, and the metal-poor star $\nu$ Indi. Both objects have well determined ages \citep{Kaluzny2013,Miglio2016,Chaplin2020}, and show large [$\alpha$/Fe] ratios. Finally, both in the upper middle and bottom panels of Fig. ~\ref{fig:agemasscomplete} we highlight stars that, although being $\alpha$-rich, sample a broad age range, reaching ages as young as $\sim$ 2 Gyr. These are the so called young-$\alpha$-rich stars identified first in the two CoRoT fields by \citet{Chiappini2015a}, and in the \emph{Kepler} field \citep{Martig2015}. The large number of objects available in the present study allows us to identify these stars as outliers from a population of low-mass stars both in the RGB and in the core-He burning phase (see Section~\ref{sec:overmassive}).

\subsection{The old, metal-rich stars in the Solar neighbourhood: radial migration efficiency}
\label{oldrich}
We now turn our attention to the metal-rich part of Fig. \ref{fig:agemasscomplete} where we included HD89345, a subgiant with robust age estimates based on asteroseismic constraints \citep{VanEylen2018} and the old-open cluster NGC6791 \citep{Brogaard2012}.
The latter is a $\sim8$-Gyr-old high-metallicity cluster almost 1 kpc above the Galactic mid-plane (also observed by \Kepler, see e.g. \citealt{Stello2011, Brogaard2011, Miglio2012, Corsaro2012, McKeever2019}). Recent studies of NGC6791 \citep{Martinez-Medina2018, Villanova2018, Linden2017} strongly support  its origin being in the inner disc or in the bulge.  
NGC6791  was shown to also have a small but significant [$\alpha$/Fe] enhancement (see \citealt{Linden2017, Casamiquela2019}, and references therein). 

First, we notice a dearth of young, metal-rich ($\mathrm{[Fe/H] > 0.2)}$ stars (Fig.~\ref{fig:agemasscomplete}, upper panel).  In addition, we note the existence of a significant population of old, super-solar metallicity stars that are not significantly enriched in $\alpha$ elements. These are the so-called super-metal-rich stars, that is, stars whose metallicity exceeds that of the present-day ISM at the Solar radius  (see a discussion in \citealt{Chiappini2009}, \citealt{Asplund2009}, and \citealt{Chiappini2013}). These stars are too metal-rich to be a result of the star formation history of the solar vicinity, and cannot be explained by pure chemical evolution models which predict the maximum metallicity of the solar vicinity to be around $\mathrm{[Fe/H]\sim0.2}$ dex, once observational constraints are taken into account (such as the present day ISM composition, among others). Therefore, stars currently at the solar galactocentric distance, but with $\mathrm{[Fe/H] > 0.2}$ have, most probably, migrated from their birth positions towards the solar neighbourhood. These stars are then expected to be of intermediate-old ages and to have had time to travel from inner regions, where a star can reach larger metallicities in a shorter time due to the inside-out disc formation, to their current positions (see discussions in \citealt{Minchev2013,Minchev2014,Chiappini2015}).  

\begin{figure}
\centering
  \resizebox{.88\hsize}{!}{\includegraphics{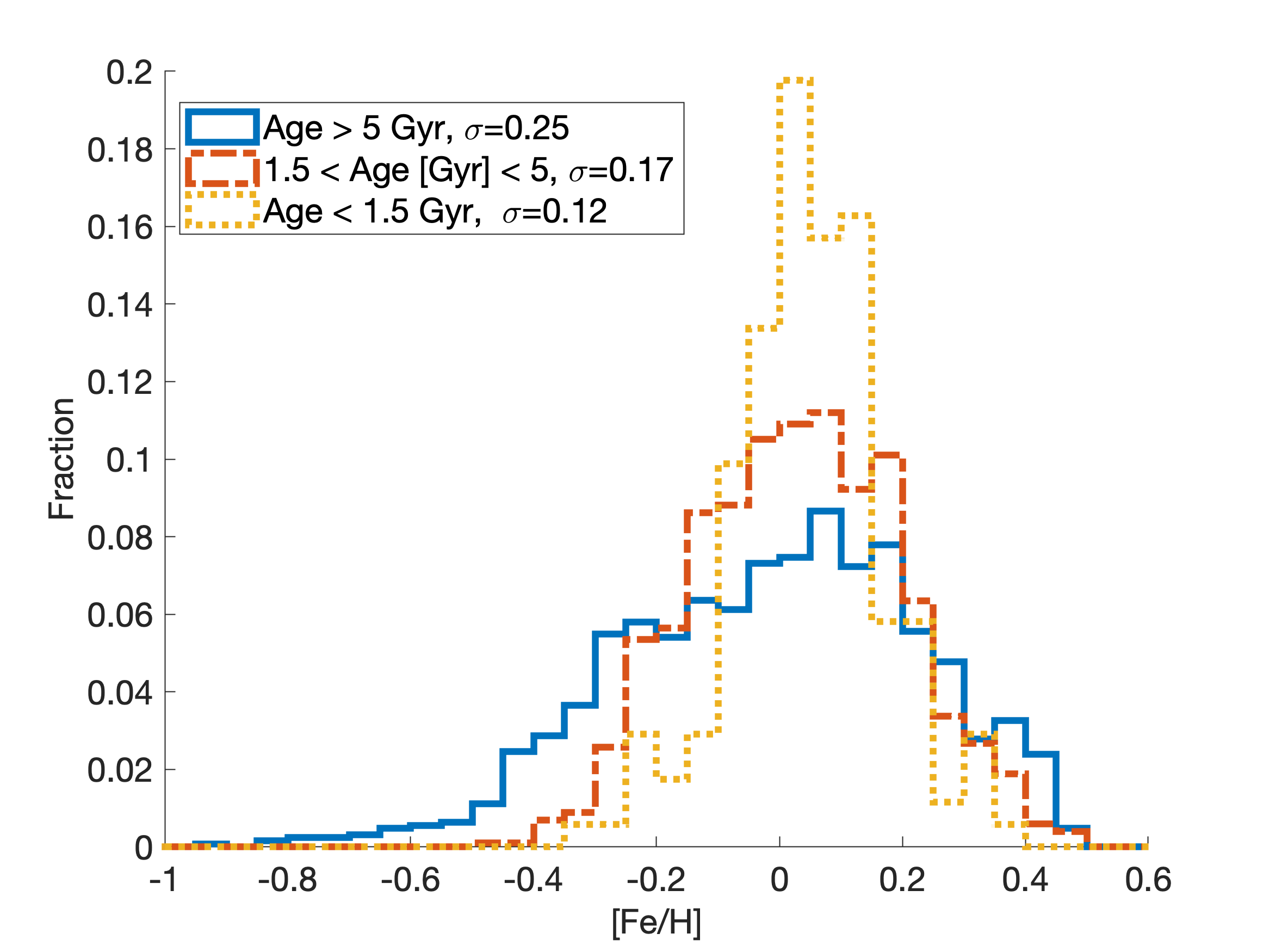}} 
\caption{{\it Upper panel:} Metallicity distribution function for stars in the sample with z$_{\rm max}$ $<$ 300 pc (to sample mostly thin-disc stars) divided into different age intervals. Stars having age $<$ 1 Gyr are shown in orange, 1 $\leq$ age/Gyr $<$ 5 in red and age $\geq$ 5 Gyr in blue. The standard deviation ($\sigma$) of the [Fe/H] distribution in each age bin is reported in the legend.}
\label{fig:luca}
\end{figure}

\begin{figure}
\centering
  \resizebox{.9\hsize}{!}{\includegraphics[angle=-90]{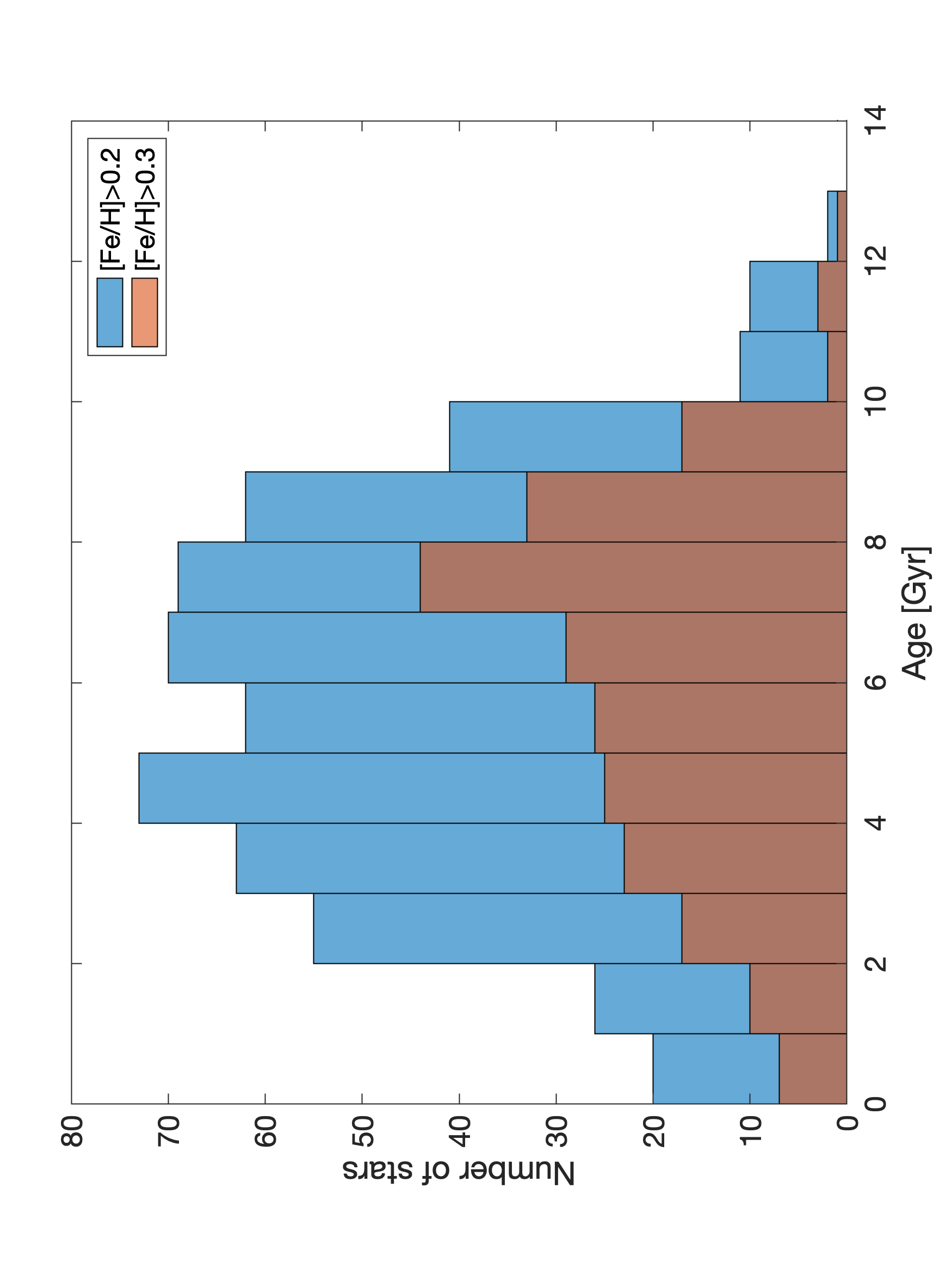}} 
\caption{Age distribution of super-metal-rich stars defined as those with $\mathrm{[Fe/H] > 0.2}$ or, being more conservative, $\mathrm{[Fe/H] > 0.3}$.}
\label{fig:metrich}
\end{figure}

The older ages of the super-metal-rich stars have been confirmed by \citet{Trevisan2011} and \citet{Casagrande2011} using isochrone fitting on the HR diagram for stars within the very small Hipparcos volume, but with the larger age uncertainties which are typical at old ages. 
Later, the existence of old metal-rich stars in the solar vicinity and longer galactocentric distances was confirmed by \citet{Anders2017} using the CoRoGEE sample. A hint of a trend showing a significant number of old metal-rich stars is also present in the first results from the SAGA survey -- see, for example, Fig. 13 in \citet{Casagrande2016}. Similar results were found by \citet{Grieves2018} who analysed a sample of subgiant stars from the MARVELS survey.

In summary, the population of intermediate-old, super-solar metallicity stars in very local samples has been interpreted as clear evidence of radial migration, as these stars do not share the common chemical evolution of the bulk of the local thin-disc stars (e.g. \citealt{Anders2018,Minchev2013}).  Other recent results, suggesting open clusters are affected by radial migration like field stars, are discussed in \citet{Anders2017}, \citet{Casamiquela2019}, and \citet{Donor2020}. Indeed, radial migration needs to be invoked to explain why, at the solar position, older open clusters are more metal-rich than the youngest. The interpretation suggested is that,  due to radial migration, older clusters can escape disruption, and appear at larger radii, although they formed in the more metal-rich inner regions of the Galactic disc. This also explains why the oldest clusters trace  steeper gradients than younger clusters, when the opposite is seen, for instance, in the CoRoGEE data. The results shown in Fig. \ref{fig:agemasscomplete} confirm this to be the case with more precise ages and a longer age baseline. 

\begin{figure*}
\centering
\resizebox{\hsize}{!}{\includegraphics[angle=-90]{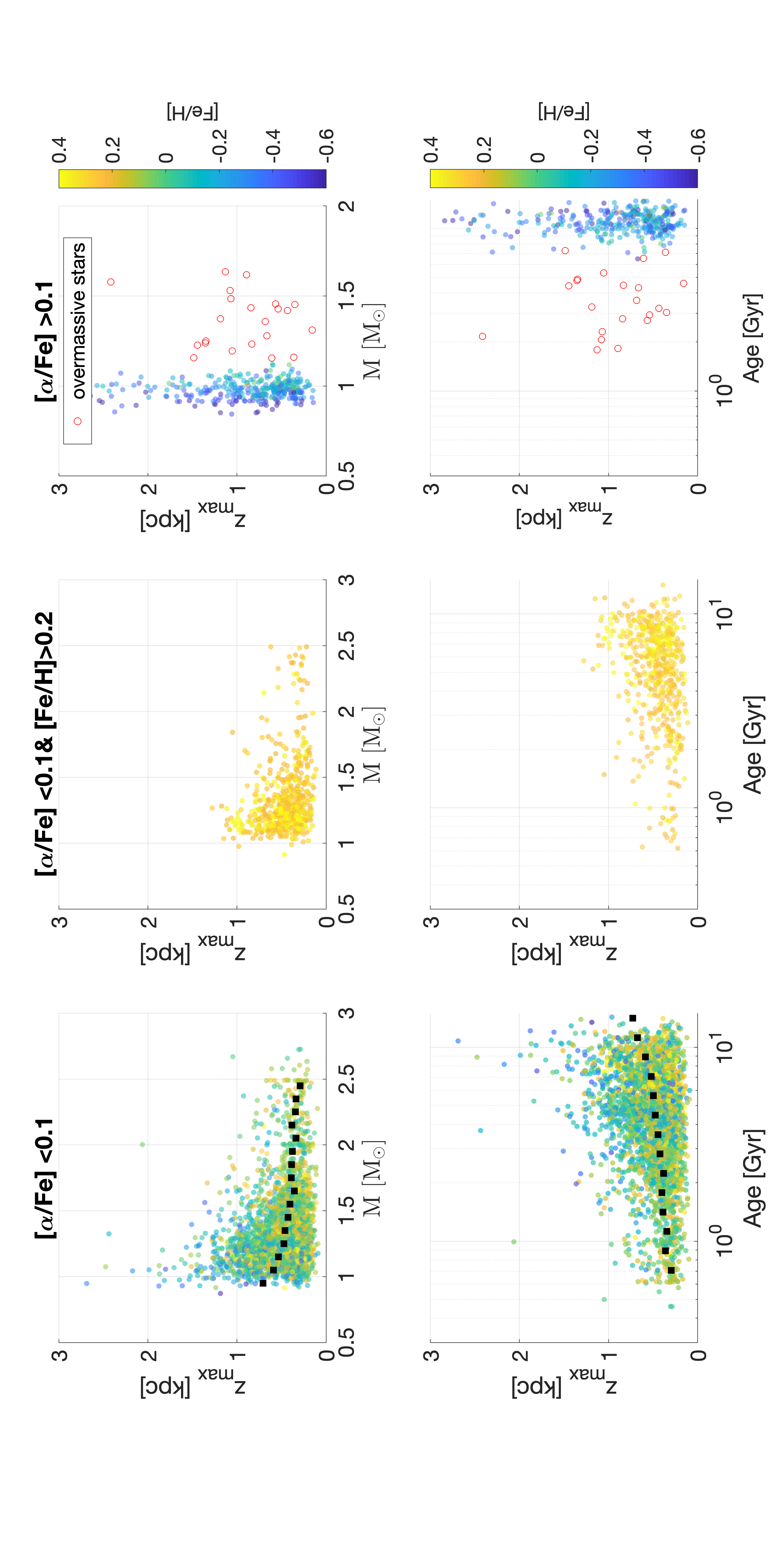}}

 \caption{{\it Upper (lower) panels:} Maximum height above the Galactic plane, $z_{\rm max}$, as a function of stellar mass (age) for stars with  $\mathrm{[\alpha/Fe]\le 0.1}$ ({\it left panel}), $\mathrm{[\alpha/Fe]\le 0.1}$ and $\mathrm{[Fe/H]\ge 0.2}$ ({\it middle panel}),  and $\mathrm{[\alpha/Fe]> 0.1}$ ({\it right panel}). Colour represents [Fe/H]. Stars among the $\alpha$-rich population identified as overmassive are indicated by red open circles.
 { Black squares in the upper(lower)-left panel represent the median $z_{\rm max}$ in different mass (age) bins.}
 }
\label{fig:vertical_zmax}
\end{figure*}

In Figure~\ref{fig:luca} we show the metallicity distribution of our sample divided into three age bins, similarly to  \citet{Casagrande2011}. We warn that, especially at super-solar metallicity, the trends of age versus metallicity strongly depend on the assumed relation between helium and metallicity which is, at this stage, a source of strong systematic uncertainty (Section~\ref{helio}). Stars  younger than 1.5 Gyr populate the $\mathrm{-0.2 < [Fe/H] < 0.2}$ range as expected from pure chemical evolution models (see, for instance, Figure 3 of \citealt{Minchev2013}). 

Estimating the age range of the super-metal-rich stars can  constrain the efficiency of radial migration. Indeed, according to the models of  \cite{Minchev2013}, while the low metallicity part of the metallicity distribution function in the solar vicinity is composed of stars with a wide range of birth radii, stars with [Fe/H] $>$ 0.25 (value used in that model) are mostly born in the $\mathrm{3-5~kpc}$ galactocentric regions and have later migrated to the solar neighbourhood. The authors argued that this corresponds to stars born just inside the bar co-rotation where, in the simulations, the strongest outward radial migration occurs. According to these simulations super-metal-rich stars are at most 6-7 Gyr old. Here, we have super-metal-rich stars as old as 9 Gyr, suggesting the efficiency of radial migration may be even larger than  in the simulation used by \cite{Minchev2013}.  \citet{Frankel2018}, using APOGEE RC stars with ages obtained by a data-driven approach \citep{Ness2016}, estimated the radial-migration efficiency to be such that a typical star  migrates  by around 3.6 kpc (i.e. the thin-disc scale length) over a timescale of 8 Gyr (age of the thin disc).  More recently \cite{Frankel2020} revised their model parameters, lowering the estimated efficiency of radial migration (mostly because they assume a flatter chemical abundance gradient in the innermost parts of the MW disc).

On the other hand, it seems that both the migration efficiency in the simulation of  \citet{Minchev2013} (MCM) and that estimated in \citet{Frankel2020} could still be lower limits. This was already suggested in \cite{Anders2017a} who carried out a comparison of models with data by mocking the MCM model according to the data selection in the two studied CoRoGEE fields. The conclusion was that the data implied stronger migration than in the MCM simulation. 

Figure~\ref{fig:metrich} shows the age distribution of the metal-rich, low-$\alpha$ ($\mathrm{[\alpha/Fe]<0.1]}$), stars. 
The age distribution is broad, and peaks at old ages, including many stars older than 6-7 Gyr. Another striking result is that many of these stars can be as young as 2 Gyr. Are these also overmassive stars similar to those found in the $\alpha$-rich population? Or are these stars just misclassified due to uncertainties in their metallicities? A more detailed investigation of these points, including a thorough analysis of the target selection function, is beyond the scope of the present work. Here, the main point to notice is that there is a significant fraction of stars of stars 8-9 Gyr old, as in NGC6791,  suggesting the migration efficiency is high, {and that more stars migrate out from the innermost regions of the Milky Way than in the MCM simulation}. Other studies also suggest large amounts of radial migration in the Milky Way disc to explain observations \citep[e.g.][]{Sellwood2014, Halle2015, Loebman2016, Frankel2018, Frankel2020}.

{ We also note that the age-metallicity trends we see in our sample are  broadly compatible with the findings reported in \citet{Feuillet2018,Feuillet2019}, if one restricts to the region in the disc representative of the \Kepler\ field. While \citet{Feuillet2019} can extend the analysis to different locations in the Milky Way, our higher age resolution allows using individual points to investigate age-chemical-composition trends (instead of using the metallicity binning as it is the case in \citealt{Feuillet2019}). The two approaches are thus complementary.}

\subsection{Adding kinematic constraints}
\label{sec:orbits}
Our sample concentrates on the solar galactocentric region, but  extends sufficiently above the Galactic plane ($z$) that one expects to see changes in the vertical structure (primarily of the $\alpha$-poor disc). Vertical trends in the population are presented in Fig. \ref{fig:vertical_zmax}. Although our aim is not to quantify such trends, which should be done by fully exploring, for instance, target selection biases, we notice that 
stars in the low-$\alpha$ sequence show evidence for an age-dependent  scale height, with the { median $z_{\rm max}$ increasing as age increases (or mass decreases)}. This is in line with tendencies reported by, e.g.,  \citet{Ting2018}  and \citet{Mackereth2017}, and from seismology by \citet{Casagrande2016, SilvaAguirre2018} and \citet{Rendle2019b}.  Another important point is that, as also noted by \citet{SilvaAguirre2018}, the $\alpha$-rich, overmassive stars show orbital parameters similar to the other high-$\alpha$ stars, which suggest they are part of the main high-$\alpha$ disc population, and not migrated from the inner disc (see the discussion about the two possibilities in \citealt{Chiappini2015a}). The middle panel of Figure \ref{fig:vertical_zmax} shows that metal-rich stars never reach the large $z_{\rm max}$ values of the $\alpha$-rich stars.

The increased number of targets, and the robustness of the inferred ages, allows constraints to be set on the age dependence of the vertical scale-height, hence ultimately on dynamical processes responsible for the vertical heating of the disc.
The high precision of the age constraints also allows a re-assessment of the age-velocity dispersion relationship (AVR) in the solar vicinity. The AVR is an important observational constraint for models of the formation and dynamical evolution of the MW disc. It is commonly fit by power-law relationships, such that $\sigma_{z}\propto\mathrm{age}^{\beta}$, with observational studies generally finding $\beta \sim 0.5$ (e.g \citealt{Wielen1977}, \citealt{Seabroke2007}, \citealt{Soubiran2008}, \citealt{Mackereth2019}). In particular, \citet{Minchev2013}, using a cosmological N-body zoom-in simulation of a MW-like galaxy fused with chemical evolution models, predicted an increase in the AVR at high age, indicative of a violent early origin for these old stars.

We fit the AVR using both a single power-law, as expressed before, and using a broken power-law, such that:
\begin{equation}
    \sigma_{z} \propto 
    \begin{cases}
      \mathrm{age}^{\beta_1}    & \quad \mathrm{age} < \mathrm{age_b} \\
     \mathrm{age_b}^{\beta_2-\beta_1}\mathrm{age}^{\beta_2}   & \quad \mathrm{age} \geq \mathrm{age_b}  
  \end{cases}\mathrm{,}
  \label{eq:vdisp}
\end{equation} where $\beta_{[1,2]}$ are the power-law indices either side of a break age $\mathrm{age_b}$. We determine the best model given the data by computing the Bayesian Information Criterion (BIC) for the best fit parameters of each model. In this way, if a significant or abrupt increase of $\sigma_z$ was preferred by the data it would be fit as such.

Initially, we fit both models to the entire data set, without selecting populations in element abundance space. In this case, the best fit model is the broken power law, with $\mathrm{age_b} = 7\pm1\ \mathrm{Gyr}$. The AVR is relatively flat before the break, with $\beta_1 = 0.24\pm0.03$, but becomes very steep afterwards with $\beta_2 = 1.2\pm0.2$. Since, as we have already discussed, the high and low $\mathrm{[\alpha/Fe]}$ population have very different age distributions, it is therefore likely that this break in the AVR is due to a superposition of these populations.

We divide the high- and low-$\mathrm{[\alpha/Fe]}$ populations, removing stars with $\mathrm{[Fe/H] < -0.7}$ (to ensure we avoid halo stars within our sample) and re-fit the AVR models above. Here, we adopt a more complex division in $\mathrm{[\alpha/Fe]}$, using a piecewise function to divide the populations { (see e.g. \citealt{Mikolaitis2014} and \citealt{Mackereth2019} for details on how to divide high- and low-$\mathrm{[\alpha/Fe]}$ populations)}:

\begin{equation}
    \mathrm{[\alpha/Fe]} = \begin{cases} 
    -0.2\ \mathrm{[Fe/H]} + 0.04 & \quad \mathrm{[Fe/H]} < 0 \\
    0.04 & \quad \mathrm{[Fe/H]} \geq 0
    \end{cases}\mathrm{.}
\end{equation} 

\begin{figure}
\centering
\resizebox{.8\hsize}{!}{\includegraphics{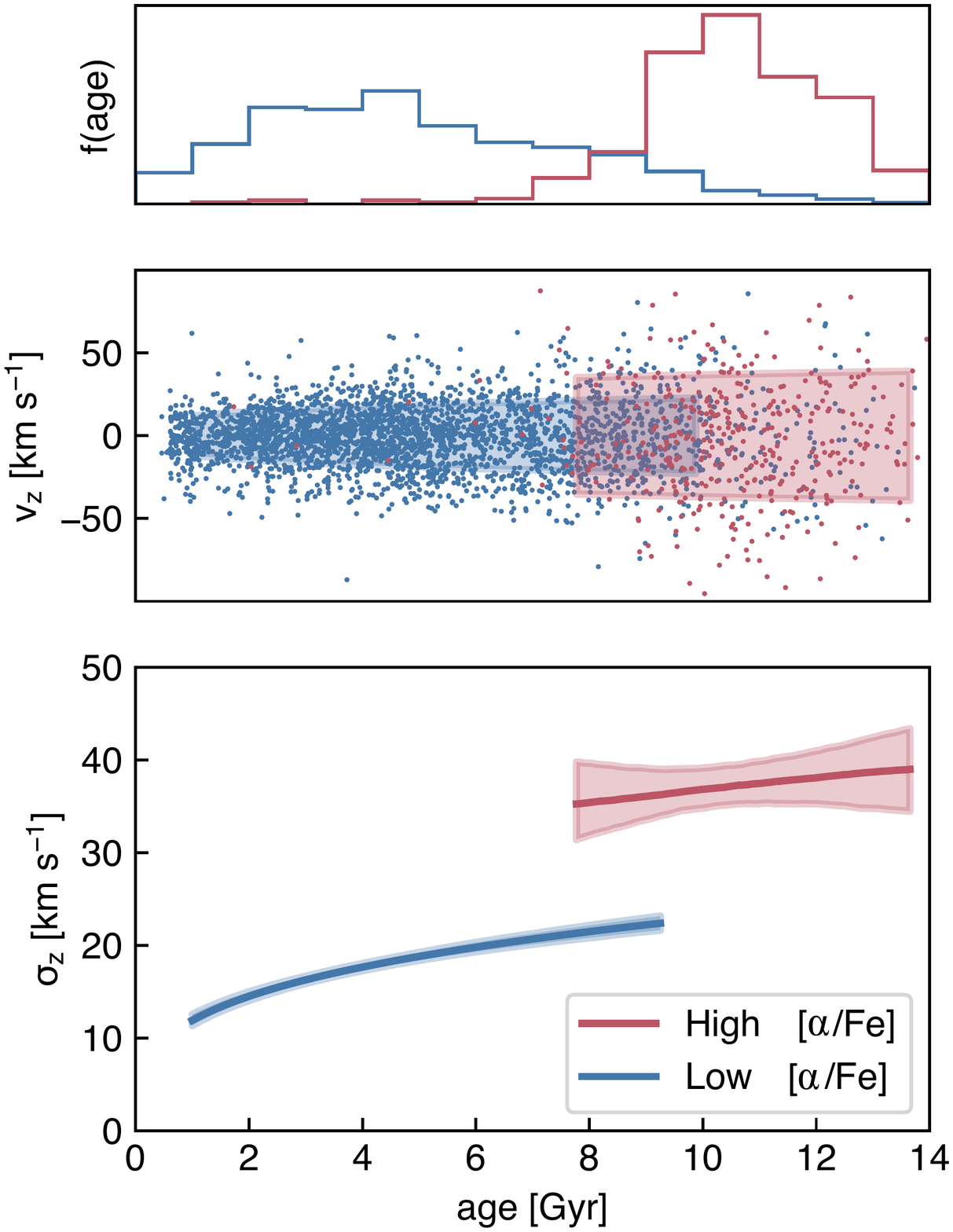}}
 \caption{Vertical ($\sigma_{z}$) velocity dispersion as a function of age. The middle panel shows the best-fit models for the AVR of the low- and high-$\mathrm{[Mg/Fe]}$ selection in age-$v_z$ space, compared with the data used in the fit. The lower panel shows the AVRs themselves, and the upper panel shows the age distribution of the $\mathrm{[\alpha/Fe]}$ selected populations. Blue and red lines represent the best fit power-law models for stars in the low- and high-$\alpha$ sequence, respectively. The coloured bands in the lower panel show the 5th to 95th percentile credible intervals of the inferred $\sigma_{z}$-age relation. The AVR for each population is only shown in the range of the 0.05 and 0.95 quantile of its ages.}
\label{fig:vel_disp}
\end{figure}

In both these populations, a single power law provides a marginally lower BIC. We adopt this model, noting that the broken power law which is fit is consistent with the single power law (such that the best fit parameters represent a single power law). The slope of the AVR is consistent between both the high and low $\mathrm{[Mg/Fe]}$ populations selected, such that $\beta_{\mathrm{low\ [Mg/Fe]}} = 0.29\pm0.02$ and $\beta_{\mathrm{high\ [Mg/Fe]}} = 0.4\pm0.2$ (note the much larger uncertainty in the high $\mathrm{[Mg/Fe]}$ population). The normalisation of $\sigma_z$ changes significantly between the two populations, such that $\sigma_z(10\ \mathrm{Gyr})_{\mathrm{low\ [Mg/Fe]}} = 22.6\pm0.6\ \mathrm{km\ s^{-1}}$ and $\sigma_z(10\ \mathrm{Gyr})_{\mathrm{high\ [Mg/Fe]}} = 36\pm2\ \mathrm{km\ s^{-1}}$. 
In Figure \ref{fig:vel_disp} we show the best-fit AVR model for the high- and low-$\mathrm{[Mg/Fe]}$ populations both in age-$v_z$ space (upper panel) and presenting the best-fit AVR relations in the regions representative of 95\% of the measured ages in each population (lower panel). 
{ Consistent results are obtained when excluding from the analysis metal-rich ($\mathrm{[Fe/H]> 0.25}$) stars, which have likely migrated from the inner disc.}

Our precise characterisation of the difference in kinematics between low- and high-$\mathrm{[Mg/Fe]}$ populations \citep[noted also in ][]{Fuhrmann2011,Adybekyan2013, Haywood2013, Hayden2018, Mackereth2019} indicates that the two populations likely had very different dynamical histories (as already suggested by the chemical discontinuity observed, for instance, in a [$\alpha$/Fe] vs. [Fe/H] diagram). The abrupt change at $\mathrm{\sim 10\, Gyr}$ (i.e. just before the beginning of the formation of the thin disc, similar to Figure 9 of \citealt{Minchev2013}), is an important observational constraint toward understanding the origin of this difference.
Indeed, as discussed in \citet{Martig2014}, the strong increase in the $\sigma_z$ at old ages is smoothed out when age errors are large. Moreover, the existence of a sudden increase in the velocity dispersion at old ages suggests the $\alpha$-rich disc was not formed by secular processes (such as radial migration), but either due to merger events or strong gas accretion (see \citealt{Brook2004}, \citealt{Minchev2013}, \citealt{Martig2014}, \citealt{Mackereth2018a} for theoretical suggestions in this line). {High-redshift galaxy observations and simulations suggest strong accretion to be a dominant process (e.g. \citealt{Lofthouse2017}, \citealt{Dekel2020})}.

\begin{figure}
           \resizebox{\hsize}{!}{\includegraphics{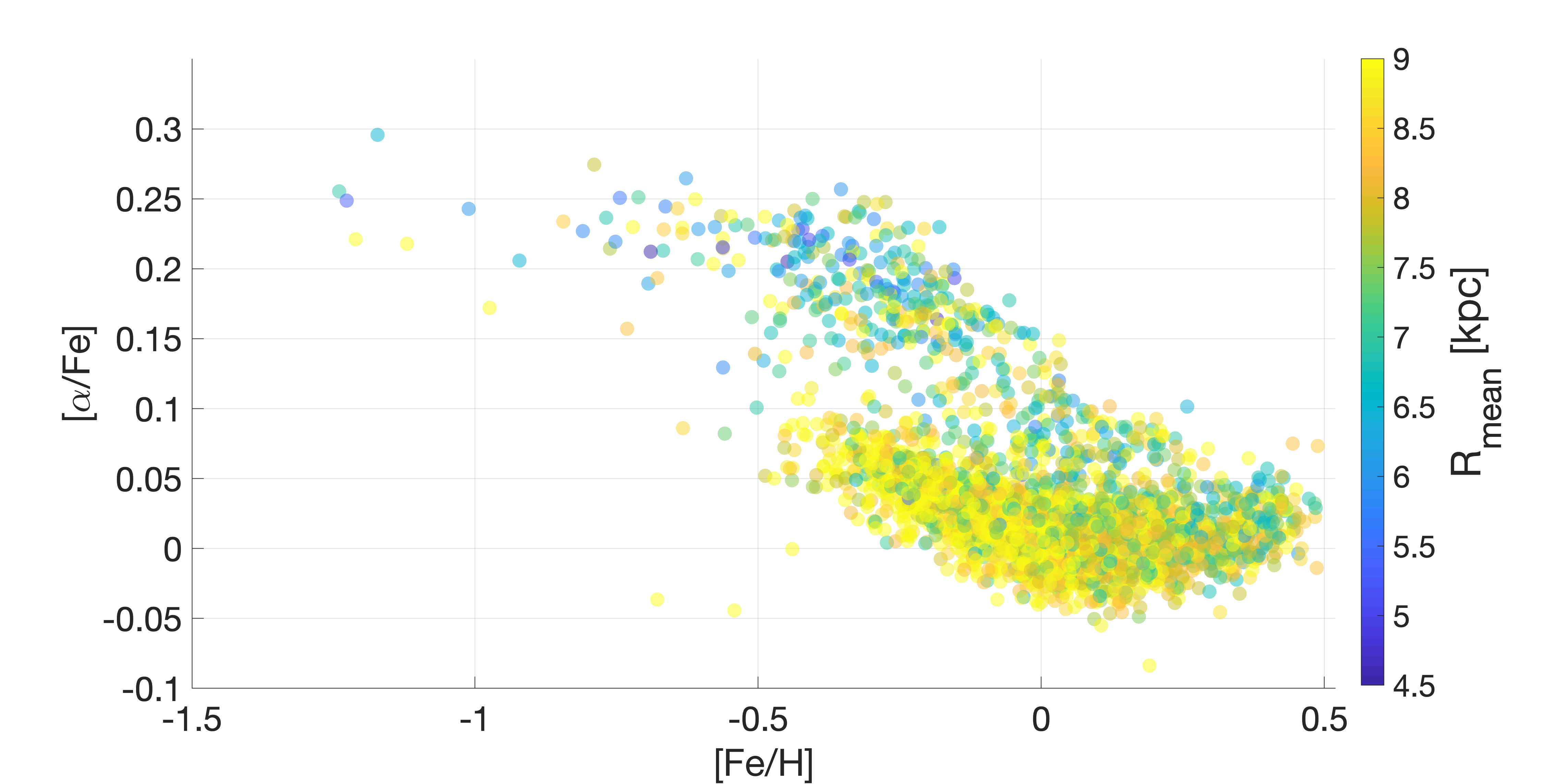}}
           \resizebox{\hsize}{!}{\includegraphics{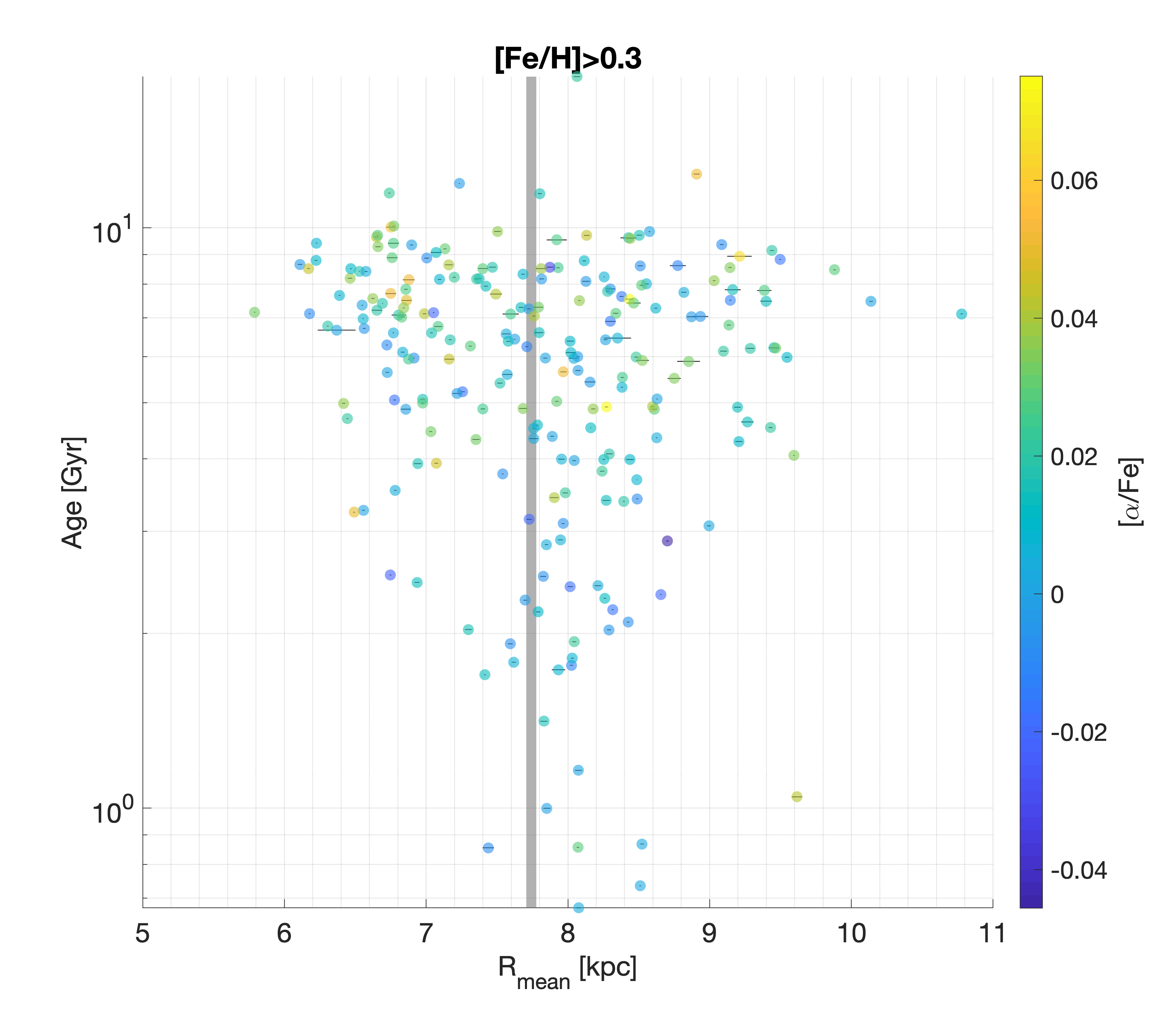}}.
           \caption{{\it Upper panel:}  [$\alpha$/Fe] vs. [Fe/H] of our sample colour-coded by the mean radius of the orbit; {\it Lower panel:}  age versus mean radius of stars more metal-rich than $\mathrm{[Fe/H] = 0.3\, dex}$, colour-coded by [$\alpha$/Fe] ratios. { The vertical line indicates the median galactocentric radius of the targets in the sample.}
           }
         \label{fig:orbits}
         \end{figure}

Finally, in Figure~\ref{fig:orbits} (upper panel) we show the [$\alpha$/Fe] vs. [Fe/H] diagram coloured by the  mean radius of the orbit. As already shown in several other APOGEE papers (e.g. \citealt{Anders2014}, \citealt{Nidever2014}, \citealt{Hayden2015}), the $\alpha$-rich stars have smaller mean radius. The high-$\alpha$ component, associated with the chemically-defined thick disc, is more concentrated in the inner regions of the Galaxy and becomes less important as one moves to the outer disc (e.g. \citealt{Bensby2011} and \citealt{Queiroz2020} for a more recent view with APOGEE DR16 data). 
In the low-$\alpha$, metal-rich population, one sees both young and old stars. In the bottom panel of the same figure, one can see the distribution of the $R_{\rm mean}$ for these super metal-rich stars  (here defined conservatively as [Fe/H] $>$ 0.3). While the youngest-metal-rich ones have  mean radii more concentrated around galactocentric distances of $\mathrm{8\,kpc}$, the $R_{\rm mean}$ distribution gets broader for progressively older stars (although still confined between { 6 and 10} kpc range, with many of them having $R_{\rm mean}$ near the solar neighbourhood).
This is important as it suggests that most of these super-metal rich stars cannot be explained by stars having inner $R_{\rm mean}$.  This is likely a signature of radial migration, that is, stars that have changed their angular momentum, and are now on a new orbit. Just like the other stars born in that orbit, the older they get, the hotter they become. The crucial point here is that  stars of such large metallicities are only common at much smaller galactocentric radii, irrespective of their age (see, for instance, Fig. 1 of \citealt{Anders2017a}). Instead, the bottom panel of Figure~\ref{fig:orbits} shows not only that all of them have $R_{\rm mean} > 6\, {\rm kpc}$, but also that this metal-rich population does not show any bias towards inner $R_{\rm mean}$ (being symmetrically distributed with respect to the central value defined by the youngest ones). 

Finally, we notice that while the majority of the super metal-rich sample is older than $\sim$4 Gyr (thus implying migration rates $\mathrm{<1-2\,kpc/Gyr}$, assuming the most probable birth radius of  stars with $\mathrm{[Fe/H] > 0.3}$ is  around $\mathrm{2-4~kpc}$ from the Galactic centre), { a few younger objects may imply either much more efficient migration rates, or an in-homogeneous enrichment of the interstellar medium (see discussion in \citealt{Magrini2015}, \citealt{Casamiquela2018}, \citealt{Quillen2018}, and \citealt{Frankel2020}), or may be stars that appear young due to a stellar merger or a mass-accretion event, potentially sharing the same origin with the overmassive stars identified in the high-$\alpha$ population (see Sec. \ref{sec:overmassive}).}

\subsection{The age of the chemically-defined thick disc}
\label{sec:alpha}

We now discuss the age and age spread of the thick disc component here defined as stars from the  $\alpha$-rich population discussed in the previous sections. We recall these stars were selected to be RGBs,  with a radius lower than 11 R$_\odot$, which are  the most robust tracers of age in the sample, as discussed in Sec. \ref{sec:context}. A comparison of  
the distribution of masses of RGB versus RC stars is presented in  Sections \ref{sec:mloss} and \ref{sec:overmassive}.

\begin{figure}
 \resizebox{1\hsize}{!}{\includegraphics{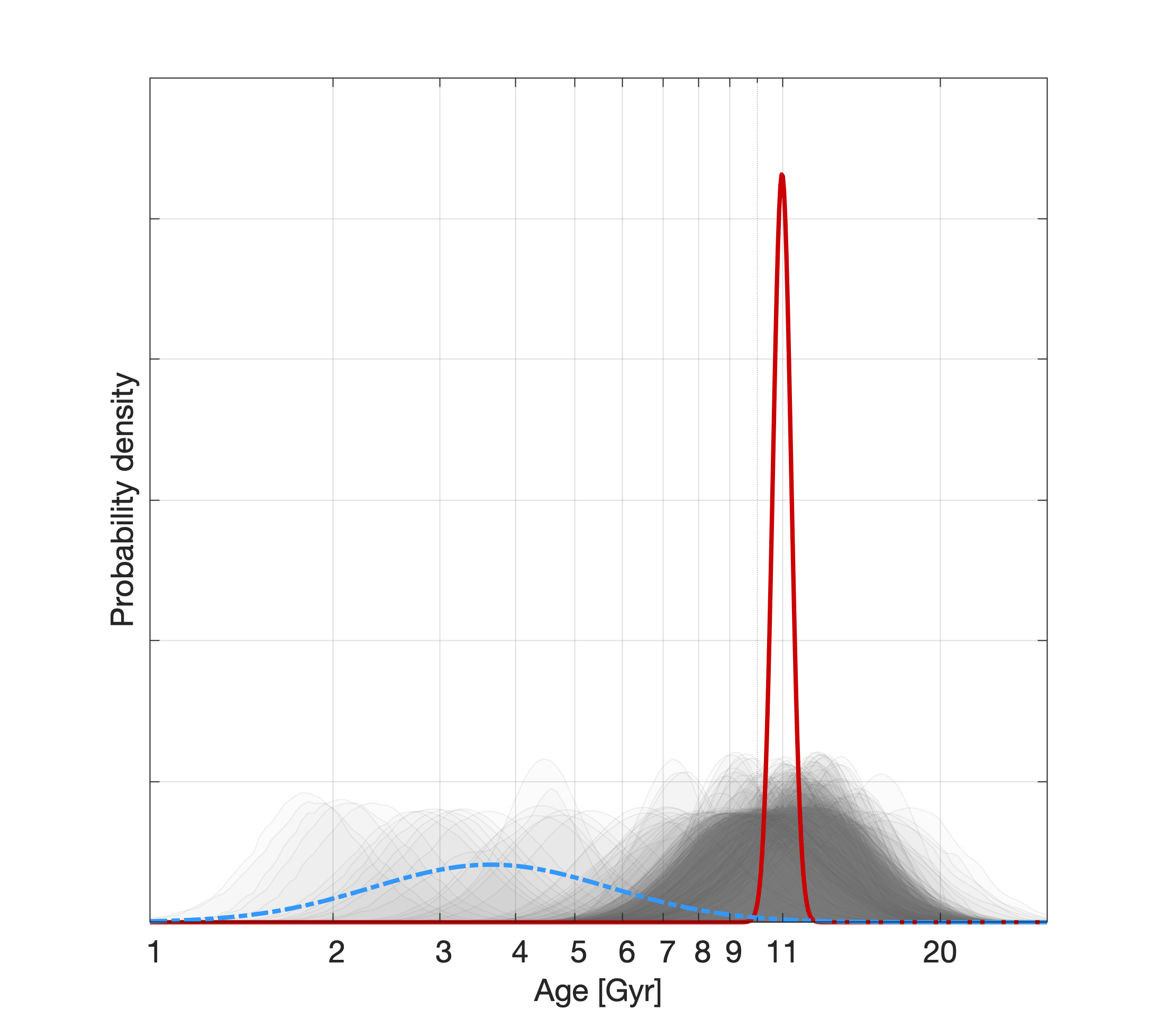}}
   \caption{Age posterior probability distribution functions of RGB stars with [$\alpha$/Fe]>0.1 (nearly 350 stars, see main text for details about the target selection). The red solid (dashed blue) line shows the intrinsic age distribution of the main population  (contaminants) inferred from the statistical model presented in  Sec. \ref{sec:rgbmass_age}. Results shown here refer for the modelling run  R1, see Table \ref{tab:runs}.}
\label{fig:age}
\end{figure}

The age posterior probability distribution functions for the stars in our sample are shown in Fig. \ref{fig:age}.
We now aim to  disentangle the intrinsic spread in mass and age of the population from that caused by observational uncertainties ($\sim$ 25-30\% in age).
We do this by fitting a hierarchical model to the stellar ages and masses  and assess the mean and the intrinsic spread of the high-$\alpha$ population (see Appendix \ref{sec:HBM}). We assume that the true age of each star is drawn from a normal distribution with a mean age $\mu$ and dispersion in $\log_{10}(\mathrm{age})$ $\sigma$, which is contaminated by a wider normal distribution at some fraction $\epsilon$ (such that the contribution of the targeted population is $1-\epsilon$) with  mean $\mu_c$ and spread $\sigma_c$ that captures the contribution of `over-massive' stars (see Sec. \ref{sec:overmassive}).
We then assume that the inferred ages are drawn from this true age distribution with a Gaussian uncertainty determined from the posterior probability given by {\sc param}.

\label{sec:rgbmass_age}

\begin{figure}
\centering
 \resizebox{\hsize}{!}{\includegraphics{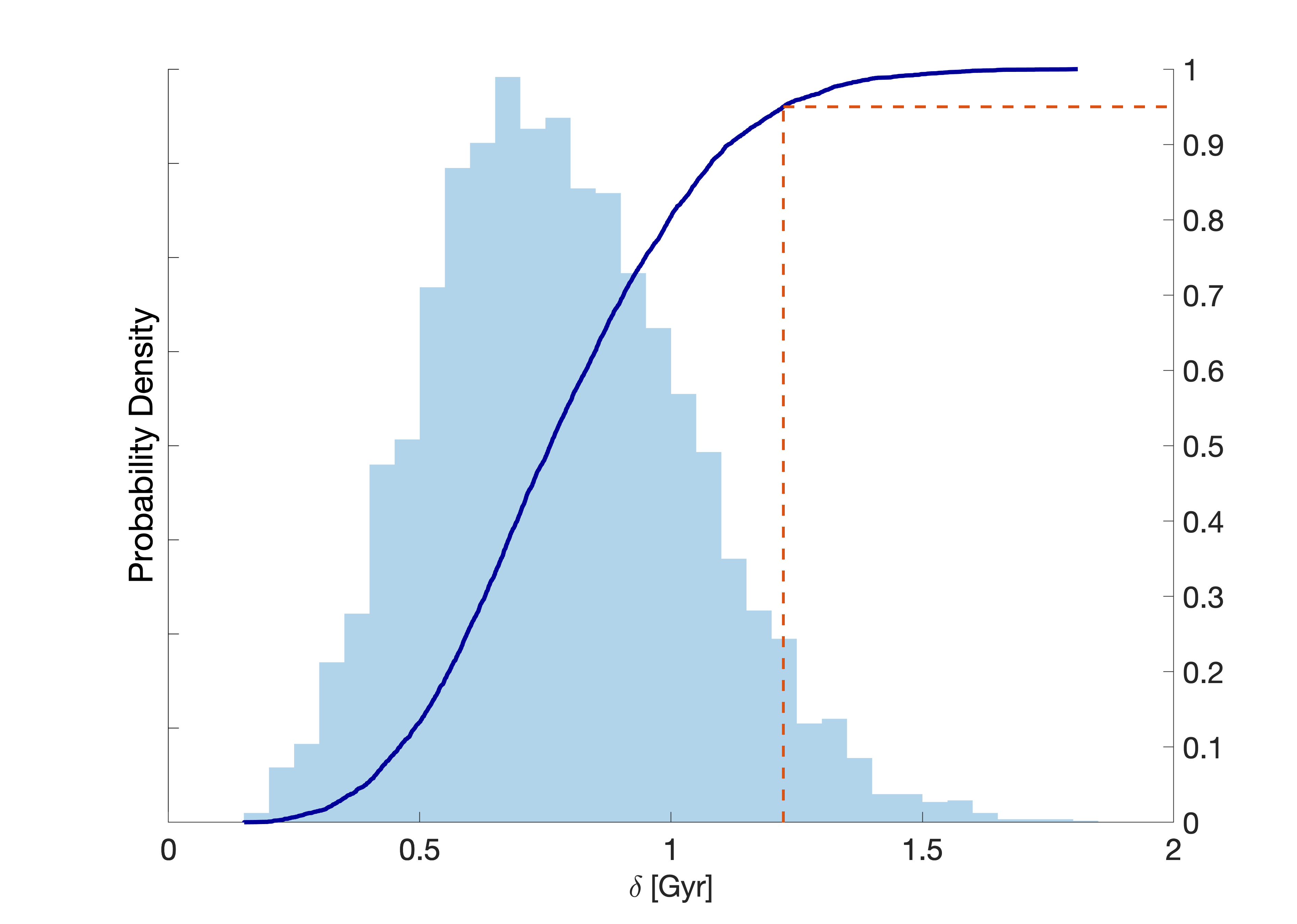}}
 \caption{Posterior probability distribution function of the age spread of the high-$\alpha$ population in the sample (R1, see Table \ref{tab:runs}), resulting from the statistical model described in Appendix \ref{sec:HBM}. The cumulative distribution function is shown as a solid line and indicates that the 95\% credible interval for the intrinsic age spread corresponds to $\delta \lesssim 1.25$ Gyr. Results from all the modelling runs are reported in Table \ref{tab:runs}.}
\label{fig:agespread}
\end{figure}

As reported in Table \ref{tab:runs}, we find a mean age of the high-$\alpha$ population in our sample of $\sim 11$ Gyr, with variations depending on the modelling run of the order of 1 Gyr, hence larger than the formal  uncertainties ($\sim 0.2$ Gyr), where the latter originate from the large number of stars in the populations.
Defining $\delta_{\rm Age}$ as the age range between $\mu-\sigma$ and $\mu+\sigma$, with $\mu$ and $\sigma$ the mean and standard deviation of the Gaussian in $\log_{10}(\mathrm{age})$ (see Appendix \ref{sec:HBM}),  we find that the age spread in the population is $\delta_{\rm Age}= 0.76^{+0.27}_{-0.23}~{\rm Gyr}$, with  variations depending on the modelling run which are well within the uncertainties.
We thus infer an upper limit to $\delta$ of 1.25 Gyr with 95\% confidence (see Fig. \ref{fig:agespread} for the reference run R1).
Alternatively, we can measure the age spread from the posterior samples, and infer that 95\% of the population was born within $1.52^{+0.54}_{-0.46}$ Gyr.

\subsection{Age distributions of stars in the thin and thick discs}
\label{sec:bias_age}

To understand how the observed population and age distribution is affected by the target selection, we compared it with a synthetic population generated by {\sc trilegal} \citep{Girardi2005} assuming constant star formation history (SFH) in the last 10 Gyr and a burst of star formation (between 11 and 12 Gyr) related, for instance, to the formation of the thick disc ($\alpha$-rich population  following our chemistry-based definition of the samples).

\begin{figure}
\centering
  \hspace{-.8cm}
  \resizebox{1.07\hsize}{!}{\includegraphics{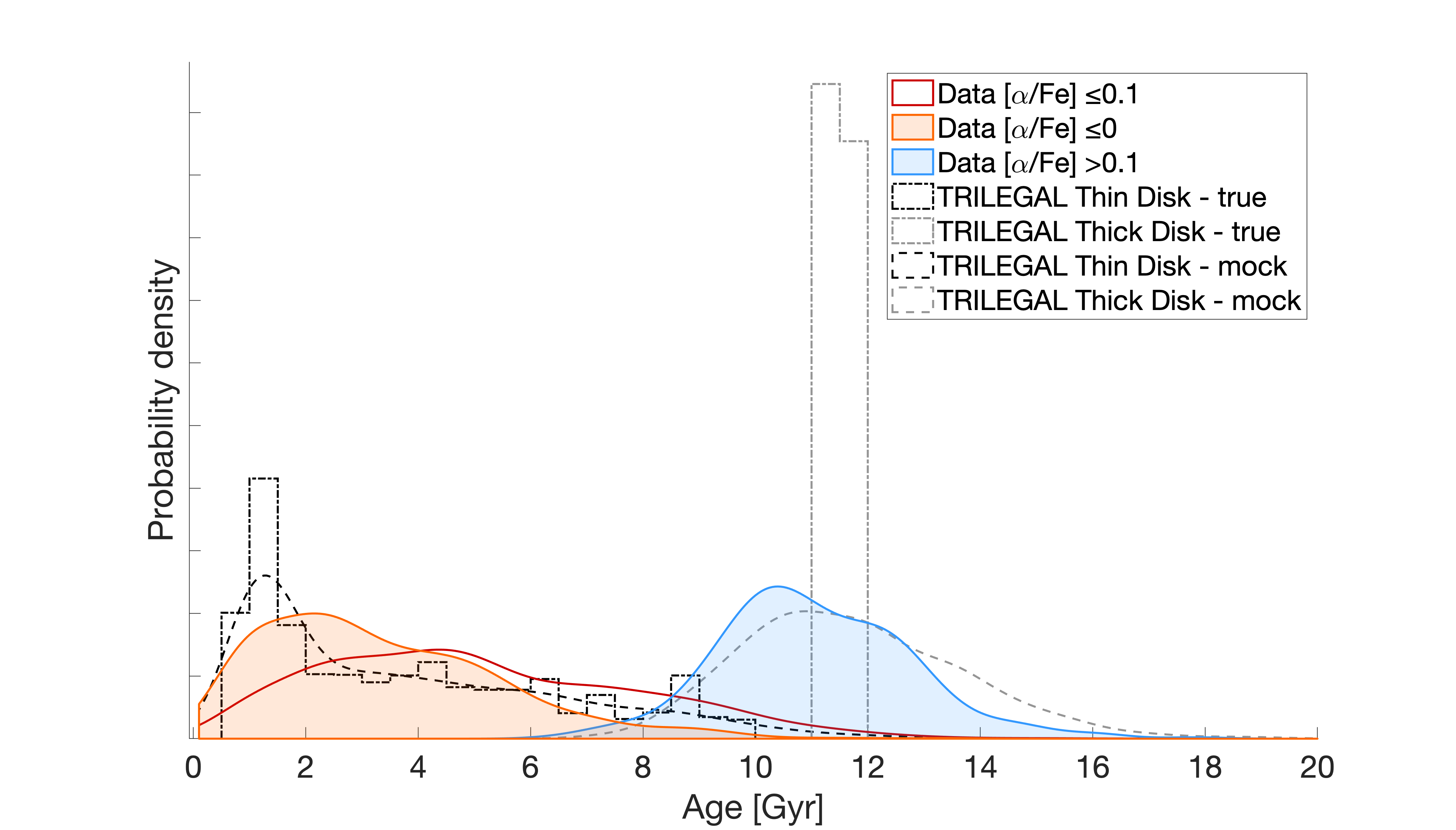}}
    \caption{Comparison between a reference {\sc trilegal} simulation, after a target selection similar to \Kepler's has been applied \citep[see][]{Miglio2014}, and ages estimated from observations. For the thick disc in the simulations we assume an age between 11 and 12 Gyr, while a uniform star formation history is assumed for the thin disc for the last 10 Gyr (dot-dashed lines). Ages are then perturbed assuming a 0.08 dex random uncertainty to reflect the current uncertainties in age ({ dashed lines}). The observed sample is divided into a high-$\alpha$ population  ($\mathrm{[\alpha/Fe] > 0.1}$,  { blue line and shaded area}) and a low-$\alpha$ population { defined using two thresholds: $\mathrm{[\alpha/Fe] \leq 0}$ (orange line and shaded area) and $\mathrm{[\alpha/Fe] \leq 0.1}$ (red line)}. }
\label{fig:trilegalSFH}
\end{figure}

The aim of such a comparison, presented in Fig. \ref{fig:trilegalSFH}, is not to infer the SFH, but to understand how one expects the observed population properties, for example the age distribution, to be affected by the target selection, which we have included in our synthetic population following  similar prescriptions as in \citet{Miglio2014}, with the additional criteria on mass, radius, and evolutionary state defined in Sec. \ref{sec:results} (see also \citet{Casagrande2016} for an alternative approach).
Such a simple comparison shows, for instance, that  a peak in the age distribution should not be interpreted necessarily as evidence of a burst in the star-formation history, as it may originate from the selection bias simulations where, for instance, stars in the secondary clump ($\sim 1$ Gyr-old) are over-represented (see also \citealt{Casagrande2016} and \citealt{Manning2017}). { Differences between the age distribution of the mock dataset and the observed sample need to be further investigated and may stem from a combination of several effects, including unaccounted target selection effects and limitations related to the simplistic, parameterised model used here.}

Also,  perturbing the age of the simulated stars by the typical uncertainties we have in our sample shows that the width of the observed age distribution of thick-disc stars is largely not due to an intrinsic age dispersion, but rather to the relatively large  uncertainties in ages, as discussed in Sec. \ref{sec:rgbmass_age}.
Fig. \ref{fig:trilegalSFH} also illustrates how age uncertainties can mask the evidence of a possible age gap between the two populations (see also \citealt{Rendle2019}), which is present in the simulated stars. While a quantitative assessment of the existence and width of such an age gap is beyond the scope of this work, the comparison with the simple synthetic population presented above shows that the epochs of star formation of the two populations are likely to be distinct. Not only age uncertainties, but also radial migration can contribute to blur a possible age-gap (see discussion in Chiappini et al. in prep.).

A star formation gap between the (chemical) thick and thin disc formation has been proposed as an explanation to the observed discontinuity in the 
$\mathrm{[\alpha/Fe]}$ versus [Fe/H] diagram \citep{Chiappini1997,Fuhrmann1998}. Reasons for such a gap are discussed in the recent literature \citep[e.g][]{Noguchi2018,Grand2018}.

\section{Evidence for mass loss on the red-giant branch and for products of mass exchange / coalescence}
\label{sec:stevo}
Thanks to asteroseismology we can not only measure the masses of red-giant stars, but also discriminate between
stars on the red giant branch  and in the red clump. These achievements are, however, insufficient to accurately determine the ages of stars in the RC \citep[e.g. see][]{Casagrande2016, Anders2017}. As mentioned earlier, the ages of stars in the red-giant phase are determined primarily by their initial mass. However, since stars on the RGB
are expected to experience mass loss, the age estimates of stars in the RC phase, which
constitute a large fraction of the red giants with detected oscillations, are plagued by our poor
understanding of RGB {mass loss}. Constraints on the efficiency of  mass loss is therefore crucial to enable the accurate determination of ages of RC stars.

In addition to enabling robust age estimates, setting constraints on the efficiency of the mass loss on the RGB has implications for our understanding of the dynamical evolution of planetary systems, including our own (e.g. see \citealt{Schroder2008}), for our understanding of the physical parameters shaping the horizontal branch (HB) in globular clusters \citep[e.g., ][and references therein]{DAntona2002, Milone2018}, and of the formation channel of, for instance, sdB stars with impacts on the origin of the UV excess in old stellar systems, like elliptical galaxies \citep{Han2002}.
So far, most of the constraints on the integrated mass loss during the RGB are provided by the morphology of the HB of globular clusters \citep[e.g., see ][for a recent analysis and description of the limitations]{Salaris2016}, by estimates of mass-loss rates based on the evidence for dust formation in infra-red photometry  \citep[e.g.][]{Origlia2014}, and by inferring an upper limit on the integrated RGB mass loss by measuring the mass segregation on the radial distribution of stars in different evolutionary states \citep{Heyl2015, Parada2016}. Estimates from these methods are in some cases in stark disagreement (e.g. of NGC 104, \citealt{Salaris2016}) and are mostly from globular clusters. 

 Asteroseismology has started to provide estimates of the integrated mass loss in the old-open clusters NGC6791, NGC6819 and M67 \citep{Miglio2012, Stello2016, Handberg2017}. These estimates  consistently suggest that mass loss, in the age and metallicity domain explored, is rather inefficient, translating to a Reimers parameter $\eta \lesssim 0.2$.
 Detailed asteroseismic studies of stars in the $\sim 2.5$-Gyr-old, solar-metallicity open cluster NGC6819 have also found evidence for a RC object (KIC 4937011, see \citealt{Handberg2017}) that most likely experienced higher-than-average mass loss \citep[for a possible mechanism to explain such enhanced mass loss, see the  companion-reinforced attrition process proposed by][]{Tout1988}. With a mass of $\simeq 0.8$ \msun\, this object would  appear to be significantly older than it is, older than the age of the Universe in this specific case. This highlights the caution that needs to be taken when age-dating RC stars, even when detailed asteroseismic constraint are available. 

Stars that underwent mass loss on the RGB are not the only complication to simple age-mass relations for
red-giant stars: so are {products of coalescence or mass exchange} in binary stars. On top of the well-studied case of blue stragglers (see e.g. \citealt{FusiPecci1992}, and references therein), evidence for objects that appear to have a mass larger than expected has been found, thanks to seismology, studying red-giant stars in clusters \citep{Brogaard2016, Leiner2016, Handberg2017} and, possibly, among
$\alpha$-rich stars, which are expected to have a small age spread \citep{Martig2015,Chiappini2015a,Jofre2016,Yong2016,Izzard2018, SilvaAguirre2018}.
Little is known about the frequency of these objects in the Galactic field also compared to clusters,
although recent studies \citep[e.g.][]{Santucci2015} suggest that these objects are more likely to exist in the field, which is exactly where they are most difficult to find.
It is thus fundamental for Galactic archeology
studies to be able to identify these overmassive and undermassive stars or, at least, to quantify their
occurrence which, complemented with a precise characterisation of similar objects on the main sequence \citep[e.g. see the recent works by][]{Fuhrmann2017b, Fuhrmann2018, Brogaard2018b}, promises to give us insights into their origin and into processes related to mass loss and mass transfer involving interactions with companions of stellar and planetary nature \citep[e.g. see][]{DeMarco2017}.

\subsection{Evidence for RGB mass loss}
\label{sec:mloss}
{ We consider the nearly coeval population of high-$\alpha$ stars and look for the signature of mass loss by comparing the mass distribution of stars in the RC with that of stars on the RGB.}
A clear trend which we  consistently find in all  sets of results is that the average mass of RC stars is smaller than that of RGB stars.

\begin{figure}
\centering
  \resizebox{.9\hsize}{!}{\includegraphics{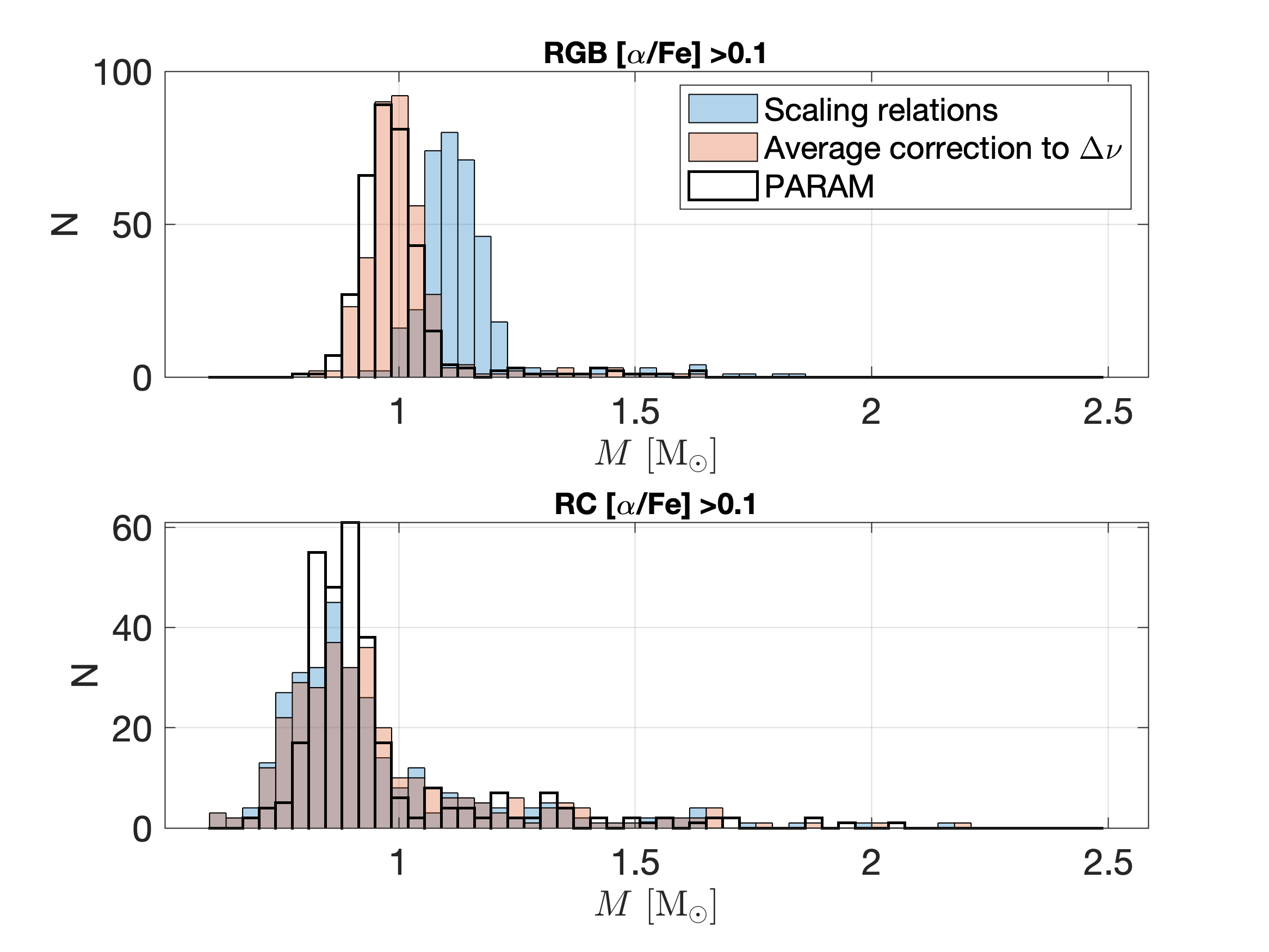}}
  \caption{Mass distribution of stars with [$\alpha$/Fe]>0.1 obtained using asteroseismic scaling relations with and without average model-suggested corrections to the \deltanu\ scaling relation \citep[e.g., see ][]{Miglio2016, Brogaard2018}. For comparison, the distribution of masses obtained using {\sc param} (R1) is also shown (unfilled bars).}
\label{fig:massSC_met}
\end{figure}

To check whether the mass difference is present irrespective of using {\sc param}, which may introduce biases from stellar models, in Fig. \ref{fig:massSC_met} we show the distribution of masses as estimated from a combination of the \deltanu\ and \numax\ scaling relations at face value, and after having applied an average correction to \deltanu. The latter is inferred by comparing \deltanu\ calculated from models' radial-mode frequencies to the assumed scaling of \deltanu\ with the square root of the star's mean density. As already shown in many papers \citep{White2011,Miglio2013, Sharma2016, Guggenberger2016, Miglio2016, Handberg2017, Rodrigues2017} corrections of the order of a few percent are expected, especially for low-mass RGB stars ($\sim 3\%$ assuming that $\alpha$-rich stars have a mass of 0.8-1.0 M$_\odot$, see \citealt{Rodrigues2017,Miglio2016}). If no correction is applied, we find that that masses of RGB stars would be overestimated by $\sim 12\%$ and we find a mean mass difference between RGB and RC stars of about $0.25\ {\rm M}_\odot$. Adding the theoretically motivated corrections to \deltanu, even approximately, shows that although the  mass difference is still present, it is significantly reduced. 

Similar results are obtained with {\sc param}, as reported in Table \ref{tab:runs}. We infer \DM\, from the distribution of the difference between the mean of the intrinsic mass distribution of RGB stars ($\mu_{\rm M, RGB}$) and that of RC stars ($\mu_{\rm M, RC}$), using  the statistical  model introduced in Sec. \ref{sec:alpha} and Appendix \ref{sec:HBM}. In our reference modelling run R1 we find  $\DM=0.10 \pm 0.01$ M$_\odot$. 
Such a small uncertainty stems from the large number of stars in the populations and is likely smaller than the systematic component to the uncertainty, which we estimate in what follows. 

In addition to being motivated by stellar evolution models, model-suggested corrections to the \deltanu\ scaling relation (or better,  using \deltanu\ computed from radial-mode frequencies instead of assuming a scaling relation) significantly reduces the discrepancies between asteroseismically inferred distances and radii and those from \Gaia\ DR2, as presented extensively in \citet{Khan2019}.
In particular, as reported in  in Sec. \ref{sec:gaia}, the comparison of seismically-determined parallaxes with those from \Gaia\ DR2 suggests that \Gaia's parallax zero-point offset does not significantly depend on the evolutionary state, which lends confidence to our inferred relative (RC versus RGB) mass. At this stage, however, we cannot exclude systematic effects on the mass difference of the order of 0.02 M$_\odot$ (see Sec. \ref{sec:gaia}), which we decide to adopt as a conservative uncertainty on our best estimate for integrated mass loss.

To test the robustness of this finding we also perform several runs of {\sc param} and we recover the results within the estimated uncertainties. Among the results presented in Table \ref{tab:runs}, the only cases where the estimated \DM\, is significantly different are R10 and R11, that is, if we adopt in the observational constraints an average large separation defined differently (see Sec. \ref{sec:asterodata}). 
%
The small, yet systematic ($\sim +1\%$), difference between a global \deltanu\ as determined by \citet{Mosser2011} and from individual radial-mode frequencies, or by \citet{Yu2018},  leads to a  $\sim 4\%$ reduction in the estimated mean mass of RGB stars when using \citet{Mosser2011}'s \deltanu, hence to a smaller inferred integrated mass loss ($\mathrm{\DM=0.05\mhyphen0.06\, M_\odot}$). However, as discussed in Sec. \ref{sec:asterodata} and based on the comparisons with \Gaia\ DR2 parallaxes (see Sec. \ref{sec:gaia} and \citealt{Khan2019}), we have reasons to consider the results of these particular runs (R10 and R11) as less accurate.

\begin{figure}
\centering
  \resizebox{1\hsize}{!}{\includegraphics{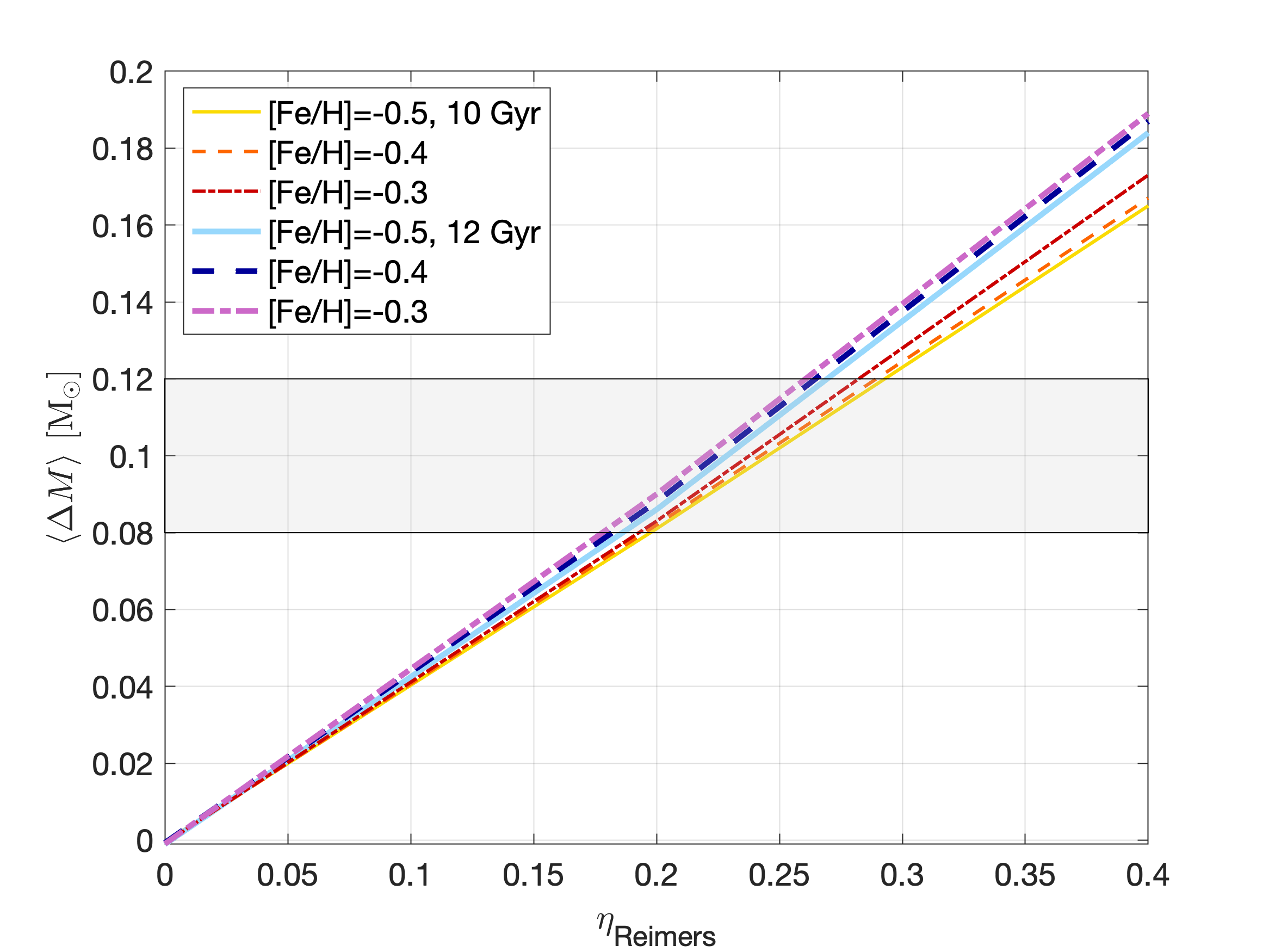}}
  \caption{Difference between the average mass of RGB stars and RC stars as a function of Reimers' $\eta$ parameter for 10 and 12-Gyr {\sc parsec} isochrones with metallicities representative of our sample. The grey area represent the 1-$\sigma$ region of the observed \DM\,. We adopt a conservative uncertainty on \DM\,, see discussion in Sec. \ref{sec:mloss}.}
\label{fig:DM}
\end{figure}

Since we are considering a composite population,  \DM\ is representative of an average integrated mass loss only. With this caveat in mind,  we compare the observed  value of \DM\ against the expected \DM\, based on the widely-used  parameterisation of mass loss along the RGB by \citet{Reimers1975a}. In Fig. \ref{fig:DM} we show \DM\, as predicted from {\sc parsec} \citep{Bressan2012} isochrones of 10 and 12 Gyr and $\mathrm{[Fe/H]=[-0.3, -0.4, -0.5]}$ using different $\eta$ values in Reimers' prescription for mass loss, as implemented in {\sc parsec}. Our findings are compatible with a mass-loss efficiency parameter $\eta_{\rm Reimers}\sim 0.25$. As mentioned earlier, we adopt a conservative uncertainty of 0.02 M$_\odot$. The availability in the near future of more precise and accurate parallaxes from \Gaia\ will provide more stringent tests of asteroseismically inferred radii and masses, and a significant reduction of the uncertainty on \DM.

{We also notice that, among the stars considered in this work, there are 7 lower-RGB  ($R< 11\, \mathrm{R_\odot}$) and 9 RC stars belonging to the old-open cluster NGC6791. In our reference modelling run R1, we find a  median mass of the RGB stars to be 1.15 \msun\ with a standard deviation of  0.04 \msun\,  (compatible with results based on turnoff EBs, see \citealt{Brogaard2012}, and with the detailed modelling in \citealt{McKeever2019}). On the other hand, we obtain a median mass of RC stars of 1.06 \msun\ with a standard deviation of 0.03 \msun\,  leading to an estimated \DM\ $\sim 0.09$ \msun, consistent with the value reported in \citet{Miglio2012}.}

As a final point, we discuss whether the sample of stars we have mass estimates for may be biased against stars that had lost a larger fraction of their mass. 
\subsubsection*{An observational bias against low-mass core-He-burning stars?}
\label{helio}

\begin{figure}
\centering
  \resizebox{\hsize}{!}{\includegraphics{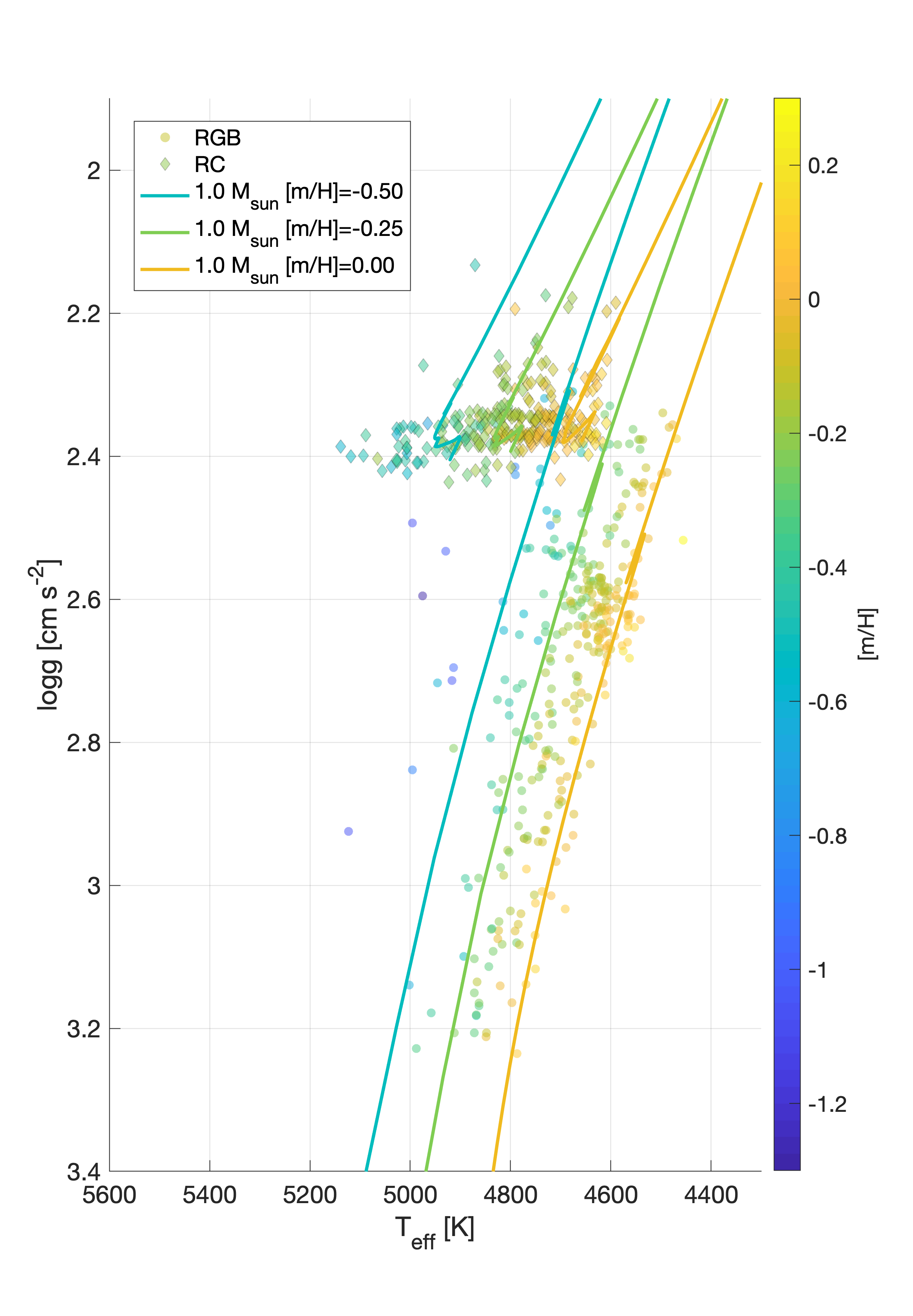}}
  \caption{Kiel diagram based on asteroseismically-determined \logg\  (stellar-evolutionary-track independent), \teff\ and metallicity from APOGEE DR14. From this plot one can evince that the spread in \teff\ is compatible with a spread in metallicity and not necessarily a large spread in mass. The overdensity of points at \logg$\sim 2.7$ is associated with the RGB bump \citep{Khan2018}.}
\label{fig:kiel}
\end{figure}

Low-mass, low-metallicity, core-He-burning stars are expected to be significantly hotter than the main clump, hence potentially excluded by \Kepler's target selection, and -- when sufficiently hot -- to not show solar-like oscillations due to the proximity to the red-edge of the classical instability strip. 
To check for potential biases against low-mass stars in the core-He burning phase we look at ratios of RGB to RC stars in the sample. We use the fraction of RC/RGB (in a restricted \logg\ domain, between 3.1 and 2.4, see Fig. \ref{fig:kiel}) as an indicator of whether the clump stars in the sample are representative of the underlying  population of core-He burning stars.  
We find $\rm N_{\rm RC}/N_{\rm RGB}=0.7-0.85$ depending on whether we consider also overmassive stars, which have higher occurrence rates among RC stars compared to RGB stars (see Sec. \ref{sec:overmassive}).\\

A similar exercise looking at 1.0 and 0.8 M$_\odot$ stellar evolutionary models { with $-0.5\leq \rm [Fe/H]\leq 0$}  gives $\rm N_{\rm RC}/N_{\rm RGB}$ in the range 0.6-0.8. Of course one should be  careful to give too much weight to this test, given the uncertainties in the duration of the RC phase itself. However,  the evolutionary tracks we are using predict  R2 parameters and a distribution of period spacing in good agreement with observations (e.g. \citealt{Bossini2015}, \citealt{Bossini2017}). 

Moreover, the median metallicity of stars in the high-$\alpha$ sequence is relatively high and, for example, for models with $\mathrm{[Fe/H]=-0.4}$, only masses below $\approx$ 0.7 M$_\odot$ are expected to become sufficiently hot to approach the RR Lyr instability strip (e.g. see also core-He burning tracks by \citealt{Bressan2012}). We therefore expect that the average integrated mass loss estimate given here is not significantly affected by such a potential bias.

\subsection{Overmassive stars}
\label{sec:overmassive}
As introduced earlier, among stars with $\mathrm{[\alpha/Fe] > 0.1}$ we find evidence for stars whose mass is higher than expected in an old population.
We use the statistical mixture model presented in Sec. \ref{sec:alpha} to quantify the fraction of outliers ($\epsilon$) and its uncertainty. The occurrence rate of overmassive stars among RGB and RC stars is presented in Table \ref{tab:runs}, for various modelling runs.
Comparable fractions are obtained by selecting  outliers by defining the width, $\sigma$, of the distribution as the mass difference between the median and the 15.8th percentile and by identifying overmassive stars that have masses  3-$\sigma$ above the median. 

Additional tests on the robustness of our mass estimates using parallaxes measured by \Gaia\ (see Sec. \ref{sec:gaia}) and the behaviour of other seismic indicators expected to be mass-dependent (see Sec. \ref{sec:smallsep}) give results consistent with the estimates provided by {\sc param}.

Our results indicate that the fraction of such `over-massive' stars is lower ($\sim 5-6\%$) in the portion of the RGB explored by our targets (\logg\ between 3.1 and 2.4) than among core-He-burning stars ($\sim 18\%$). 
While quantitative comparisons between predictions from interacting binary evolution is beyond the scope of this paper \citep[e.g. see][]{Izzard2018,Abate2018}, we notice that if these stars have undergone a merger or significant mass accretion event, one would expect the latter to occur with higher probability near the RGB tip, hence to find a higher fraction of these overmassive stars in the core-He-burning phase compared to the low-luminosity RGB. This is supported also by the work by \citet{Badenes2018} and \citet{PriceWhelan2020}, who find evidence for a reduced fraction of binary stars between the low-luminosity RGB and the RC, which would suggest that a higher number of binary systems had undergone interaction when observed in the RC compared to the low-luminosity RGB. 

Part of the increased fraction of overmassive stars in the RC is, however, expected simply from mass-dependent stellar evolutionary timescales. It is well known that the duration of the core-Helium burning compared to that of the RGB phase increases as a function of stellar mass. This is also evident observationally comparing the ratio of RC to RGB stars in clusters of different age, hence corresponding stellar mass in the giant branches. Here we restrict ourselves to the  \logg\ domain between 3.1 and 2.4 and consider tracks of 1 \msun\ and 1.4 \msun\ stars.  For a 1 \msun\ model we expect N$_{\rm RC}$/N$_{\rm RGB}$ to be of the order of 0.7 (see also Sec. \ref{sec:mloss}). For more massive stars,  1.4 \msun, representative of overmassive stars, one expects an increased N$_{\rm RC}$/N$_{\rm RGB} \simeq 1$. A mass-dependent N$_{\rm RC}$/N$_{\rm RGB}$ can only partially account for the evolutionary-state dependent observed occurrence rates, strongly suggesting that a significant fraction of these stars underwent a merger or a mass accretion event during the high-luminosity RGB phase.

\begin{figure}
    \resizebox{\hsize}{!}{\includegraphics{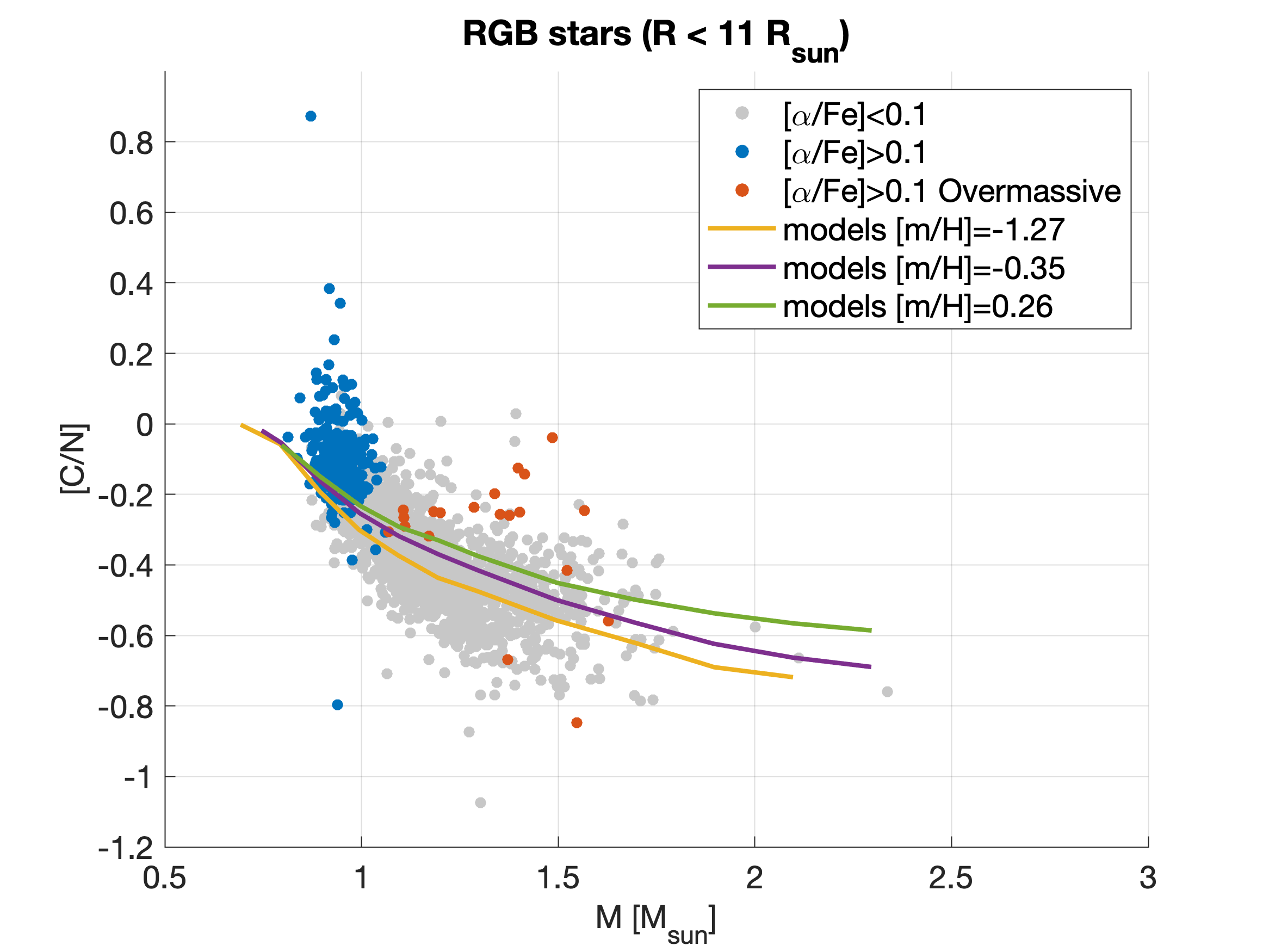}}
    \resizebox{\hsize}{!}{\includegraphics{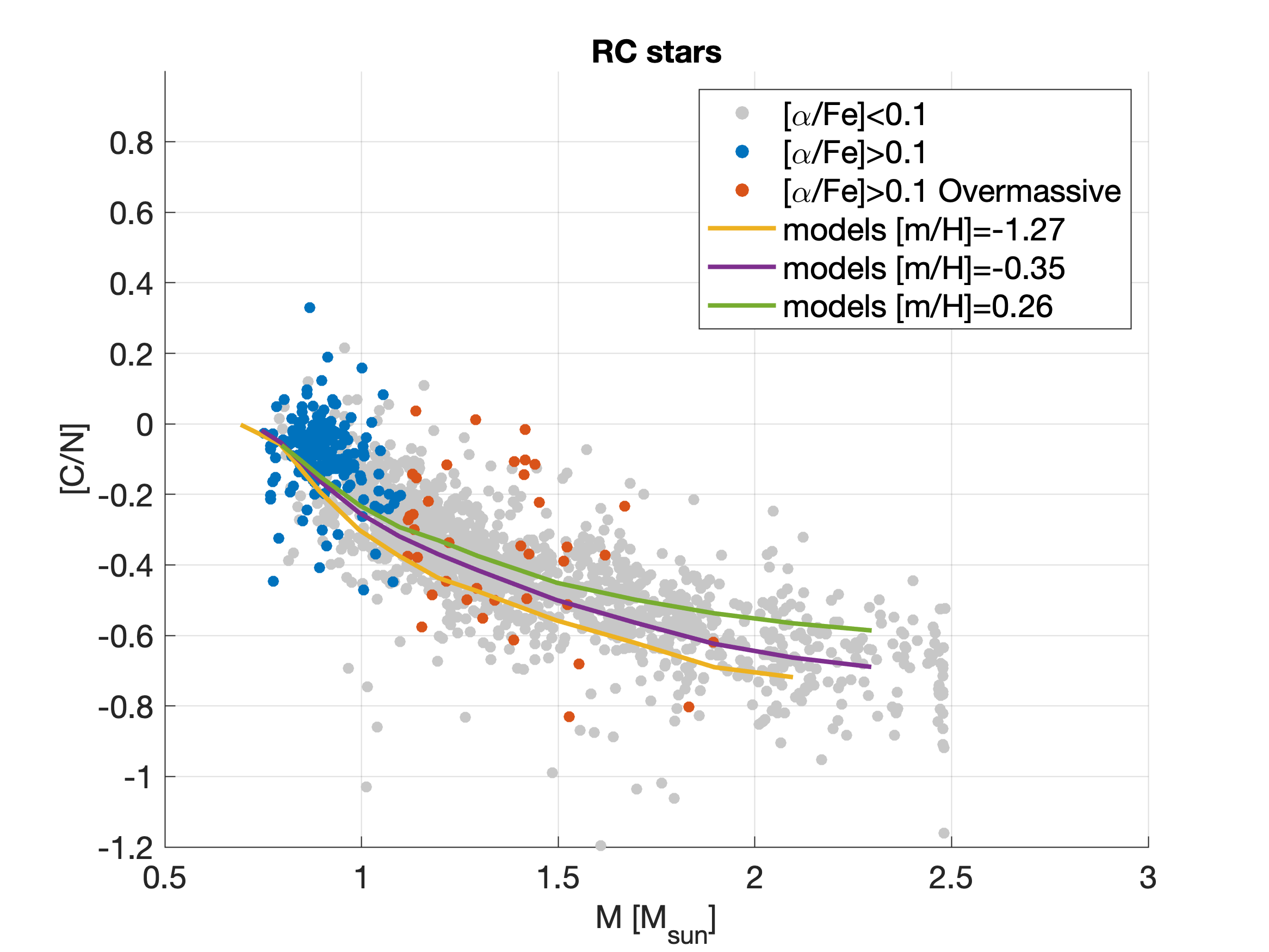}}
  \caption{[C/N] ratios of the stars in the sample as a function of their estimated mass. Red dots represent stars identified as overmassive among the population with [$\alpha$/Fe]>0.1. Solid lines represent the predicted [C/N] at the first dredge up for stellar models of different metallicities \citep{Salaris2018}.}
\label{fig:cn}
\end{figure}

Additional information on the nature of these stars may be gathered from photospheric abundances, in particular from [C/N] ratio, which is available from APOGEE DR14. In Fig. \ref{fig:cn} we show [C/N] of the stars in the sample (see also \citealt{Hekker2019}). Overplotted are predictions from \citet{Salaris2018} showing [C/N] at the first dredge up for stellar models of different metallicities. Among the overmassive stars, while some show [C/N] typical of lower-mass stars, others have values more in-line with their higher mass. 
{ This can potentially provide evidence for such stars being the result of a merger happening after or before the first dredge up. However, a robust inference about the nature of these stars will only be possible  by quantitative comparisons with predictions from binary evolution.}

{ For completeness, we note that overmassive stars  have orbital parameters similar to the rest of the $\alpha$-rich population (see e.g. Fig. \ref{fig:vertical_zmax}), as also discussed in \citet{Martig2015} and \citet{SilvaAguirre2018}. Quantitative inferences are however limited by the low number of overmassive stars in the sample, and, for instance, by the potential effects of their binary origin on their estimated orbital parameters.}

\section{Summary and conclusions}
  \label{sec:summary}
We use a combination of spectroscopic and asteroseismic constraints (described in Sec. \ref{sec:data_models}) to infer masses and ages of about 5400\ red-giant stars observed by {\it Kepler} and APOGEE. Crucially, we explore some of the systematics that may affect the accuracy of the inferred properties, both related to biases in the data (e.g. \Teff\ and metallicity scales, definition of seismic average parameters) and from the grid of stellar models adopted in the code (e.g. changing the assumed chemical enrichment relation $\Delta Y/\Delta Z$). By performing several runs of the code {\sc param} (see Sec. \ref{sec:method}, Table \ref{tab:runs}, and \citealt{Rodrigues2017}) we present patterns in mass, evolutionary state, age, chemical abundance, and orbital parameters that we deem robust against the systematic uncertainties explored. 

By selecting stars with robust age estimates (see Sec. \ref{sec:results}), we obtain a sample of $\sim3300$ stars with a median random uncertainty in mass of  $ 6 \%$, which translates to a $23\%$ median random uncertainty in age.
\subsection{Galactic archaeology}

The use of robust ages led to the following main results on Galactic archaeology:
\begin{itemize}
\item[1] the identification of a nearly coeval $\alpha$-rich population of old stars, which we identify with the chemical-thick disc, and the confirmation of a much larger age spread for the low-$\alpha$ population, implying a longer timescale for the formation of the thin disc;
\item[2] we find evidence that radial migration has brought old stars born in the {metal-rich, inner (2-4~kpc)} regions of the disc/bulge into the solar vicinity { ;}
\item[3] we find the thick disc to be as old as $\mathrm{z\sim2}$, and presumably formed at the same epoch as the $\mathrm{z\sim2-3}$ gas-rich, thick discs observed in faint high-redshift spectroscopic surveys (e.g. \citealt{Genzel2017}). It is important to note, however, that very likely the MW at $\mathrm{z\sim2}$, which would consist of the thick disc and the bulge, would be too faint to be observable with the current spectroscopic instrumentation;
\item[4] the chemical discontinuity in the $\mathrm{[\alpha/Fe]}$ versus [Fe/H] diagram seems to correspond to an abrupt change in the velocity dispersion at old ages.  Moreover, the age distributions of these two populations confirm the different  chemo-dynamical histories of the
chemical-thick and thin discs, and is also suggestive of a halt in the star formation (quenching) after the formation of the chemical-thick disc. 
\end{itemize}

\noindent More specifically, we investigated:

\subsubsection*{\it The age-[$\alpha$/Fe] relation in the solar circle and signatures of radial-migration in the disc} As shown in Figs. \ref{fig:agealphaR} and \ref{fig:agemasscomplete}, we find that the chemical evolution of the $\alpha$-rich population happened on a significantly shorter timescale compared to that of the low-$\alpha$ sequence, as expected from chemical evolution models based on the \emph{two-infall} scenario.

We  note both a dearth of young, metal-rich ($\mathrm{[Fe/H] > 0.2)}$ stars (upper panel of Fig.~\ref{fig:agemasscomplete}  and Fig.~\ref{fig:metrich}), and the existence of a significant population of old (8-9 Gyr), super-solar metallicity stars, reminiscent of the age and metallicity of the well-studied open cluster NGC 6791. 
 These stars, currently at the solar galactocentric distance, but with $\mathrm{[Fe/H] > 0.2}$, have likely migrated from their birth positions towards the solar neighbourhood, being too metal-rich to be a result of the star formation history of the solar vicinity. 
While a full comparison with chemodynamical models is needed to quantify the efficiency of radial migration, our results give indications of a higher efficiency than in the simulations and/or previous studies (see also the section below). 

\subsubsection*{\it Chemo-kinematic constraints}
 In Sec. \ref{sec:orbits} we consider additional information provided by the orbital parameters inferred using the  exquisite constraints from \Gaia\ DR2 (see Sec. \ref{sec:orbitalparam}).
Looking at stars in the low-$\alpha$ sequence, we find evidence for an age-dependent vertical scale height, which can be used to set stringent constraints on vertical disc heating (e.g. see \citealt{Casagrande2016, Ting2018, Mackereth2019, Rendle2019}, and references therein).
 
The high precision of the age constraints also allows a re-assessment of the age-velocity dispersion relationship (AVR) at the solar radius. 
Initially we consider the entire data set and fit the AVR with both a single and a broken power law ($\sigma_{z}\propto\mathrm{age}^{\beta}$). We find the broken power law to be the best model, with a break age $\mathrm{age_b} = 7\pm1\ \mathrm{Gyr}$ (see Eq. \ref{eq:vdisp}). However, if we divide the data set into low- and high-$\mathrm{[\alpha/Fe]}$ samples and apply the same approach, we find the single power law to be a better model. The slope of the AVR is consistent between the two populations (see Fig. \ref{fig:vel_disp}), however, we find a significantly different normalisation, such that at $\mathrm{\sim10\,Gyr}$  $\sigma_{z, {\mathrm{low\ [\alpha/Fe]}}} = 22.6\pm0.6\ \mathrm{km\ s^{-1}}$ and $\sigma_{z, {\mathrm{high\ [Mg/Fe]}}} = 36\pm2\ \mathrm{km\ s^{-1}}$.
The clear difference in kinematics between low and high $\mathrm{[\alpha/Fe]}$ populations  indicates that they likely had very different dynamical histories. The abrupt change in normalisation at $\sim 10$ Gyr, which we have constrained here, is an important observational constraint toward understanding the origin of this difference, suggesting that the thick disc was not formed by secular processes, but either due to merger events or strong gas accretion (see e.g. \citealt{Brook2004}, \citealt{Minchev2013}, \citealt{Martig2014}).

Finally, we note that $\alpha$-rich stars have smaller mean galactocentric radius, as expected from their likely origin in a centrally concentrated thick disc, as also shown by, for instance,  \citealt{Anders2014}, \citealt{Nidever2014}, \citealt{Hayden2015}, and \citealt{Queiroz2020}. 
On the other hand, the low-$\alpha$ stars show a wide distribution of mean radius, also for stars with super-solar metallicity, suggesting that most of these super-metal rich stars cannot be explained by stars having inner $R_{\rm mean}$. This is likely a signature of radial migration, that is, stars that have changed their angular momentum, and are now on a new, near circular, orbit at a different radius.

\subsubsection*{\it The age of the high-$\alpha$ population}

 In Sec. \ref{sec:rgbmass_age} we focus on the ages of RGB stars in the high-[$\alpha$/Fe] sequence. 
 We find the ages and masses of the nearly 400 $\alpha$-rich RGB stars to be compatible with those of an old population.  As reported in Table \ref{tab:runs}, we find a mean age of $\sim 11$ Gyr, with variations depending on the modelling run of the order of 1 Gyr, hence larger than the formal  uncertainties ($\sim 0.2$ Gyr), where the latter originate from the large number of stars in the populations. 

The width of the observed age distribution is dominated by the random uncertainties on age. Using a statistical model (see Appendix \ref{sec:HBM}) we infer the spread of the intrinsic age distribution $\delta_{\rm Age}$ to be of the order of $\rm 0.75-1.0~Gyr$, with  variations depending on the modelling run which are within the uncertainties.
Considering R1 as the reference run, we therefore find that 95\% of the population was born within $1.52^{+0.54}_{-0.46}$ Gyr.

 Our precise asteroseismic ages  suggest that the age of the (chemically-defined) thick disc component is comparable to that of the double sequences dated by \citet{Gallart2019}, and is not significantly younger. Moreover, the small age spread in the thick disc component, at least in the Solar vicinity, suggests the thick disc was already formed and in place by the time the Gaia-Enceladus / sausage merger event  happened, in agreement with \citet{Montalban2020}. This supports the picture of a first peak in the star-formation history at $z\sim 2$ (corresponding to look-back times of $\sim$ 10-12 Gyr) in line with the peak of the cosmic star formation rate (e.g. \citealt{Madau2014}) and subsequent works, which show that the red sequence in the halo population is actually the thick disc (\citealt{diMatteo2019}, \citealt{Sahlholdt2019}, \citealt{Belokurov2020}).

  \subsection{Stellar evolution}
In Sections \ref{sec:mloss} and \ref{sec:overmassive} we exploit the almost coeval $\alpha$-rich population to gain insight into processes that may have altered the mass of a star along its evolution. These inferences are key to improving the mapping of the current, observed, stellar mass to the initial mass and thus to age. The main results concerning stellar evolution are:
\begin{itemize}
\item[1] a quantitative inference of the mean integrated mass loss along the red-giant branch:  $\DM\ = 0.10\pm 0.02$ M$_\odot$;
\item[2] constraints on the fraction of massive $\alpha$-rich stars, which we find to be of the order of 5\% on the RGB, and significantly higher in the RC, supporting the scenario in which most of these objects had undergone interaction with a companion.
\end{itemize}

\subsubsection*{\it Mass loss} Comparing the mass distribution of stars on the lower RGB ($R< 11\, \mathrm{R_\odot}$) with those in the RC, we find evidence for a mean integrated RGB mass loss $\DM\ = 0.10\pm 0.02$ M$_\odot$ (see Sec. \ref{sec:mloss}), an estimate which we find robust against the systematic effects explored in this work.

If the inferred \DM\ was to be mapped into a  Reimers' parameter \citep{Reimers1975a} describing the mass loss rate on the RGB, then it would correspond to $\eta\sim0.25$ (see Fig. \ref{fig:DM}). 
However, we discourage from simply adopting the estimated $\eta$ parameter in models. First, the mapping from \DM\ to $\eta$ depends on details of stellar models themselves due, for example, to variations in the predicted luminosity of the RGB tip where, following Reimers' prescription, most of the mass loss occurs \citep[e.g.][]{Castellani2000, Serenelli2017}. Moreover, in this study we are considering a composite population (in mass and metallicity), hence \DM\ is representative of an average integrated mass loss only. Finally, our results combined with additional inferences on \DM\ in old-open clusters \citep[e.g.][]{Miglio2012,Stello2016,Handberg2017},   indicate that \DM\ has a stronger stellar mass dependence than given by  Reimers' parameterisation.

These results enter in the lively debate about mass loss efficiency, with contradicting evidence from HB (e.g. \citealt{Salaris2016}) and other works based on cluster dynamics (\citealt{Heyl2015a, Heyl2015}, see \citealt{Salaris2016} for a summary), and provide strong,  independent observational constraints to scenarios for the physical origin and efficiency of mass loss along the RGB.

\subsubsection*{\it Over-massive / rejuvenated stars}
 Finally, we find that the occurrence of massive $\alpha$-rich stars \citep[e.g., see][]{Chiappini2015a, Martig2015, Jofre2016, SilvaAguirre2018} is of the order of 5\%  on the RGB, and significantly higher  in the RC. As discussed in Sec. \ref{sec:overmassive}, this supports the scenario in which most of these stars had undergone interaction with a companion when observed in the RC, compared to the low-luminosity RGB. Our findings are also in agreement with evidence \citep[e.g., see][]{Badenes2018, PriceWhelan2020} of a reduced fraction of binary stars as we move from the low-luminosity RGB to the RC, which would imply a higher fraction of products of binary interaction in the RC.

 Moreover, there is recent evidence for an increased intrinsic fraction of close binaries in metal poor environments \citep[e.g., see][]{Moe2019,ElBadry2019}, which is expected to lead to a higher rate of binary interactions. 
 This is also supported by \citet{Fuhrmann2017a} where there is evidence for an increased fraction ($\simeq 10\%$) of blue stragglers among old, metal-poor stars.
 A significant fraction of likely products of binary evolution is also found in \citet{Brogaard2016} and \citet{Handberg2017}, where they estimate an occurrence rate  of overmassive red-giant stars of about $10\%$ and $5-10  \%$ in the old-open clusters NGC6819 and NGC6791.

 Comparisons with binary population synthesis models, expanding for example those in \citet{Izzard2018}, including target selection effects, would be beneficial to interpret the occurrence rates and surface abundances of these over-massive stars, and to make predictions about their internal structures, which could then potentially be tested against detailed asteroseismic inferences.

\subsection{Caveats and prospects}
The combination of APOGEE, \Gaia, and \Kepler\ data represents a treasure trove from which one can pick and combine elemental abundances, orbital parameters, stellar masses and ages for thousands of stars to recast and address long-standing questions in the evolution of the Galaxy (Sec. \ref{sec:results}) and of stars (Sec. \ref{sec:stevo}). 

As presented here, however, the precision enabled by asteroseismic constraints, even when using average seismic parameters, is such that one needs to consider sources of systematic uncertainties in stellar models. 
 The latter have direct impact not only on absolute ages, but on age trends as well. We have shown, for instance, that -- as expected -- the helium enrichment relation that one assumes in the models determines the age trend for the oldest stars as a function of metallicity.
 Here, we have attempted to quantify some of the uncertainties, and to use such tests to select trends that we consider robust.

Moreover, while initial-mass to age relations for giants are quite robust, one has to keep in mind that stars may have gained (lost) significant amounts of mass hence appearing younger (older) than they are. While there is clear evidence for these objects in clusters, it is harder to flag them in a composite population, and to firmly distinguish --based on observational evidence  alone--  their origin. Finally, given the nature of how age depends on the observational constraints, the broadening of age distributions at old ages is inevitable, however, one can attempt at inferring the intrinsic age spread using statistical models, as presented in Sec. \ref{sec:rgbmass_age}. 

There is certainly room for improvement, and to mitigate some of the systematic effects. Stars and stellar systems with independent determinations of masses and radii are  fundamental  in strengthening the foundations of asteroseismically-inferred stellar properties (see e.g. the encouraging results obtained with detached eclipsing binaries and clusters, and the comparison with \Gaia\ DR2 astrometric constraints, also discussed in Sec. \ref{sec:gaia}). Moreover, one can transition from using average seismic parameters to individual mode frequencies (e.g. \citealt{Buldgen2019, Rendle2019b, Montalban2020}) or at least, as presented here for stars in the high-$\alpha$ sequence, to using average seismic parameters defined from individual mode frequencies, hence removing the ambiguity in the definition of the global seismic quantity \deltanu.

 Improved stellar models, and modelling techniques, promise to deliver more precise and - crucially - more accurate ages. 
 Delivering accurate seismic ages is evermore relevant, given that stars with asteroseismic constraints are used increasingly often as training sets to data-driven techniques which, while on the one hand enable us to infer ages of millions of stars \citep[e.g., see][]{Ness2016, Das2019, Ting2018, Mackereth2019, Ciuca2020}, on the other hand carry the potential risk of propagating age biases from few thousands to millions of stars.

\appendix

\section{Additional tests}
\label{sec:systematics}

\subsection{Comparison with \Gaia\ parallaxes}
\label{sec:gaia}
The  analysis presented in this paper includes as key constraints global properties of acoustic modes, hence one expects inferences on the stellar mass to be strongly correlated with those on radius (as is obvious from the simple, approximated expressions relating \deltanu\ and \numax\ to global stellar properties). The seismically inferred radius, coupled with effective temperature, can be used to infer the luminosity of a star. The latter,  combined with apparent magnitude, bolometric correction, and an estimate of interstellar extinction, enables the determination of the distance to that star \citep[e.g. see][]{Miglio2013a, Casagrande2014, Rodrigues2014}.  To assess whether significant biases are present in the seismically-determined radii, one can compare seismically inferred parallaxes to \Gaia\ DR2  \citep{GaiaCollaboration2018}.

The presence of a zero-point offset in the parallaxes published in \Gaia\ DR2 has been established both in papers describing \Gaia\ data \citep{Lindegren2018} and by comparison with independent distance determinations \cite[e.g., see][]{ Riess2018, Stassun2018, Khan2019, Hall2019, Zinn2019, Schoenrich2019, Chan2019}. Such an offset is largely due to a degeneracy between the parallax and the basic-angle variation of the \Gaia\ satellite. Its amplitude varies with location on the sky, magnitude and colour, and is on average  $\mathrm{-29 \mu as}$ (as inferred using quasars as fiducial objects at zero parallax).
As already presented in \citet{Hall2019,Khan2019,Zinn2019} an offset of  few tens of $\mu$as is also present when comparing \Gaia\ DR2 parallaxes to those inferred by seismology in a sample of red giants in the \Kepler\ field.

Here we focus on stars in the $\alpha$-rich sample, and check against possible trends of the parallax offset with evolutionary state and mass. This is relevant as systematic effects related to those two properties would potentially impact also the relative mass scale, with consequences on the estimates of mass loss and the identification of overmassive stars.

\begin{figure}
\centering
\resizebox{.8\hsize}{!}{\includegraphics{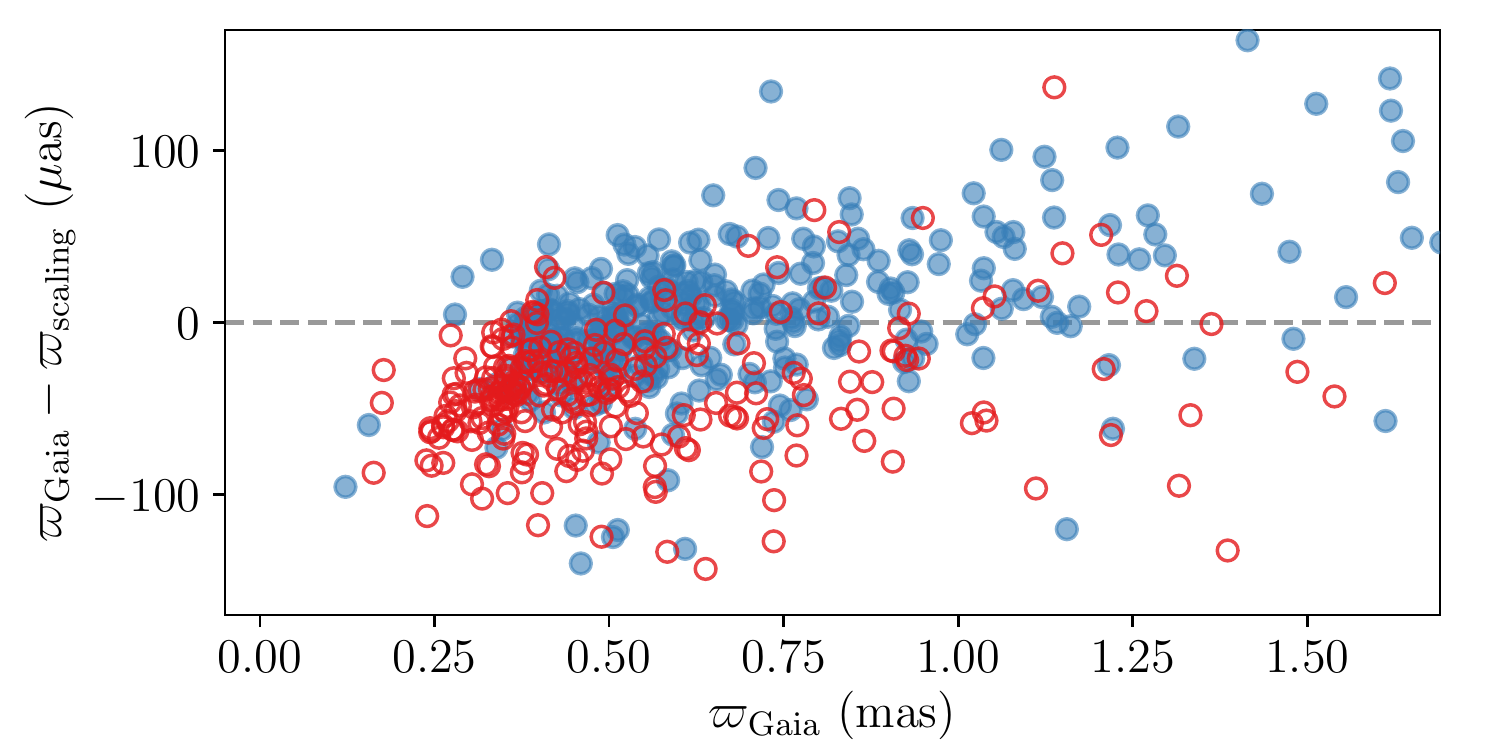}}
\resizebox{.8\hsize}{!}{\includegraphics{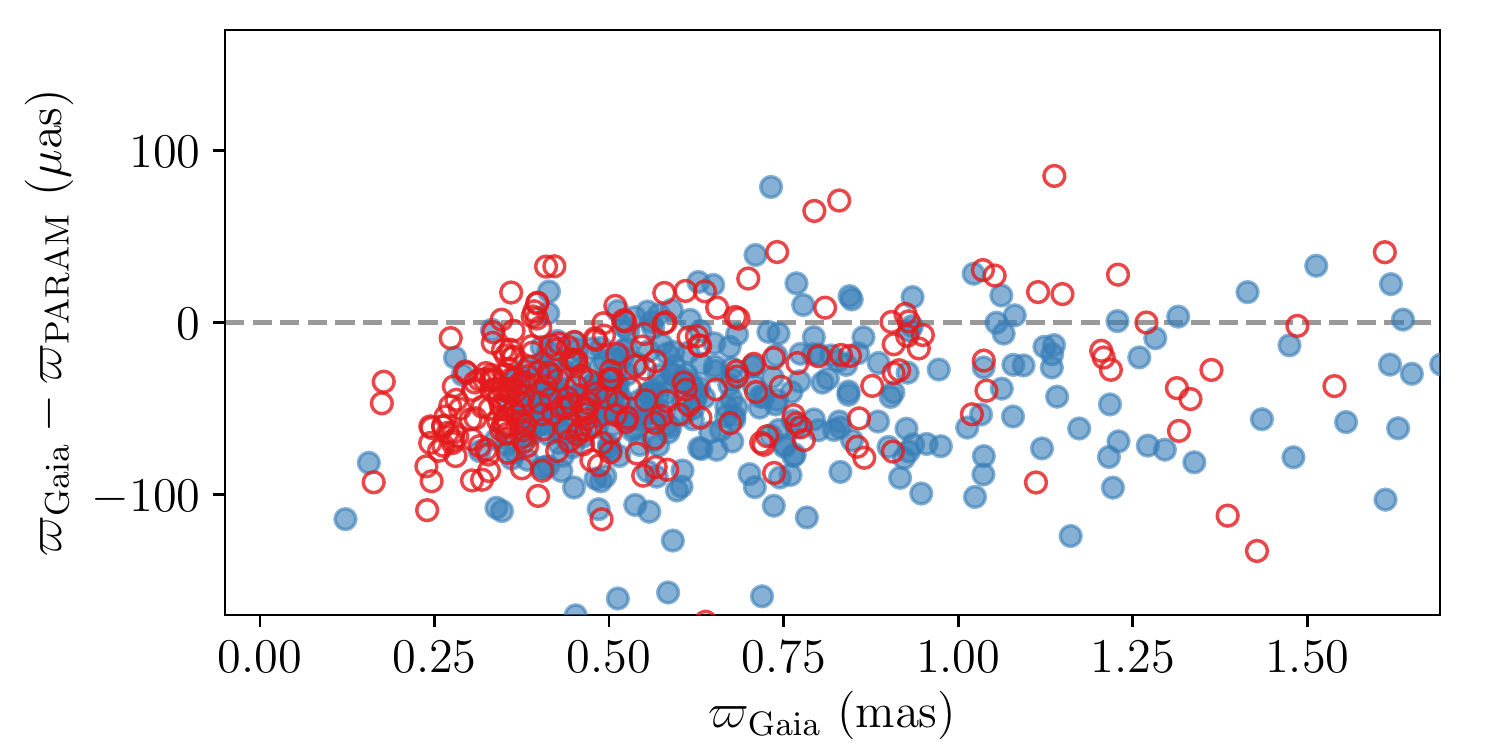}}
  \caption{Difference between \Gaia\ DR2 parallax and the parallax determined using asteroseismic constraints coupled with APOGEE DR14 spectroscopic constraints, as a function of \Gaia\ DR2 parallax. The asteroseismic parallax is computed either using scaling relations at face value ({\it upper panel}) or using {\sc param} ({\it lower panel}). Red (blue) dots denote RC (RGB) stars. The scatter in the plot is dominated by random uncertainties on \Gaia\ DR2 parallaxes \citep[see][for details]{Khan2019}.}
\label{fig:gaia}
\end{figure}

\begin{figure}
\centering
\resizebox{.8\hsize}{!}{\includegraphics{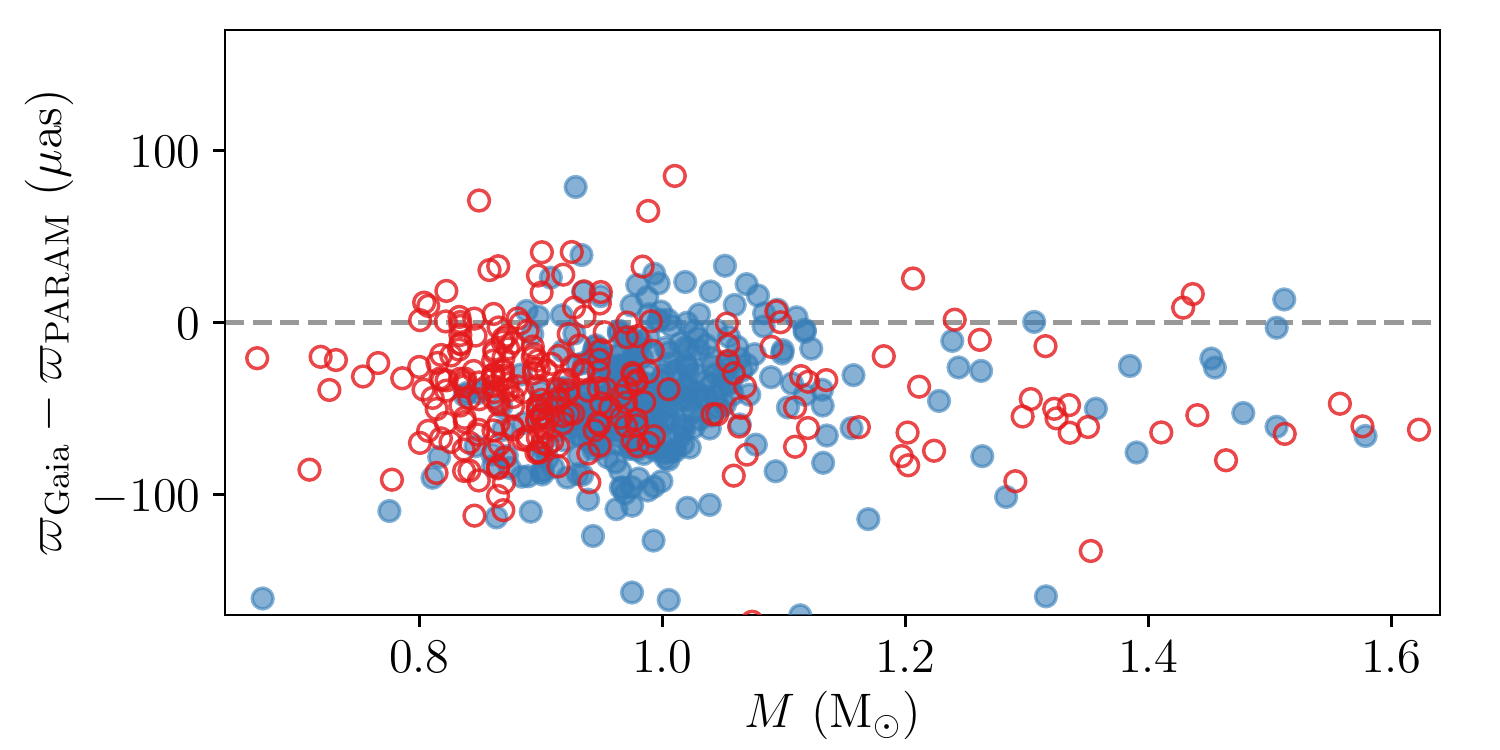}}
  \caption{Difference between \Gaia\ DR2 parallax and the parallax determined using {\sc param}, as function of the estimated stellar mass. Red (blue) dots denote RC (RGB) stars.}
\label{fig:gaia_mass}
\end{figure}

When using DR14 effective temperatures and scaling relations at face value, we note that the weighted mean between \Gaia\ DR2 and asteroseismic parallaxes ($\delta\varpi=\varpi_{\rm Gaia}-\varpi_{\rm seismo}$) is strongly dependent on the evolutionary state, as shown in the upper panel of Fig. \ref{fig:gaia} ($\delta\varpi_{\rm RGB}=-13$ $\mu$as, while $\delta\varpi_{\rm RC}=-37 $ $\mu$as). 
The different parallax offset between RGB and RC stars suggests that distances, hence radii, of RGB stars are overestimated compared to those of RC stars (or radii of RC stars are underestimated compared to those of RGB stars).
This would then imply that, if one were to use the \deltanu\ scaling relation at face value, one would overestimate the mass difference between RC and RGB stars \citep[e.g. see][]{Miglio2012}, given the tight correlation between seismically-inferred mass and radius.

We now look at comparisons with \Gaia\ DR2 parallaxes when using distances inferred from {\sc param}, coupled to our reference grid of models (G2, run R1). Using model-predicted  \deltanu, the relative parallax offset between RC and RGB stars is significantly reduced, as shown in the lower panel of Fig. \ref{fig:gaia} ($\delta\varpi_{\rm RGB}=-46$ $\mu$as, while $\delta\varpi_{\rm RC}-41$ $\mu$as). As mentioned earlier, details about how sensitive the average offset is on the \Teff\ scale, extinction, and on the model grid are given in \citet{Khan2019}. Here we simply take the decreased differential offset as an additional indication that using \deltanu\ from model-predicted frequencies is a definite improvement while, based on this comparison alone, we cannot make any strong conclusions on the absolute scale unless we associate the offset to \Gaia\ DR2 parallaxes alone. 

Moreover, we would like to note that at this stage we cannot exclude that seismic radii of RC and RGB suffer from different biases at the few $\%$ level, and this can have an impact on their average mass difference. For instance, if we were to consider the difference $\delta\varpi_{\rm RGB}-\delta\varpi_{\rm RC} \simeq-5$ $\mu$as, this would imply a $\sim 1\%$ relative offset in distance, taking $0.5~\mu$as as a representative parallax for the stars in the sample. If we associate that difference to a bias in \deltanu, we would conclude that, for instance, the distances, hence radii, of RGB stars are underestimated with respect to RC stars by $\sim 1\%$, and we may expect masses to be underestimated about twice as much, that is, $\sim 2\%$ \citep[e.g., see Eq. 1 and 2 in][]{Miglio2012}.
This may lead to an underestimation of the average mass difference between RGB and RC stars of about $0.02$ M$_{\odot}$.
While we think it is premature to adopt such a correction, this comparison gives us an estimate of the potential biases affecting the determination of the average mass difference between RGB and RC stars.

A similarly encouraging comparison holds for the parallax difference as a function of estimated mass (Fig. \ref{fig:gaia_mass}). We do not see a trend with mass, suggesting that the estimated mass of stars which were identified as overmassive is not related to an overestimation of their radius and distance. 

\subsection{Overmassive stars: small frequency separations}
\label{sec:smallsep}
As shown by \citet{Montalban2010}, the average separation between frequencies of radial and quadrupolar acoustic modes ($\langle d_{02}\rangle$) in red-giant branch stars is known to correlate strongly with mass. Although on a star-by-star basis the uncertainties on the average small separation are typically too large for a precise mass inference, such a trend is also clearly found in giants belonging to open clusters \citep{Corsaro2012, Handberg2017}. 

\begin{figure}
    \resizebox{\hsize}{!}{\includegraphics{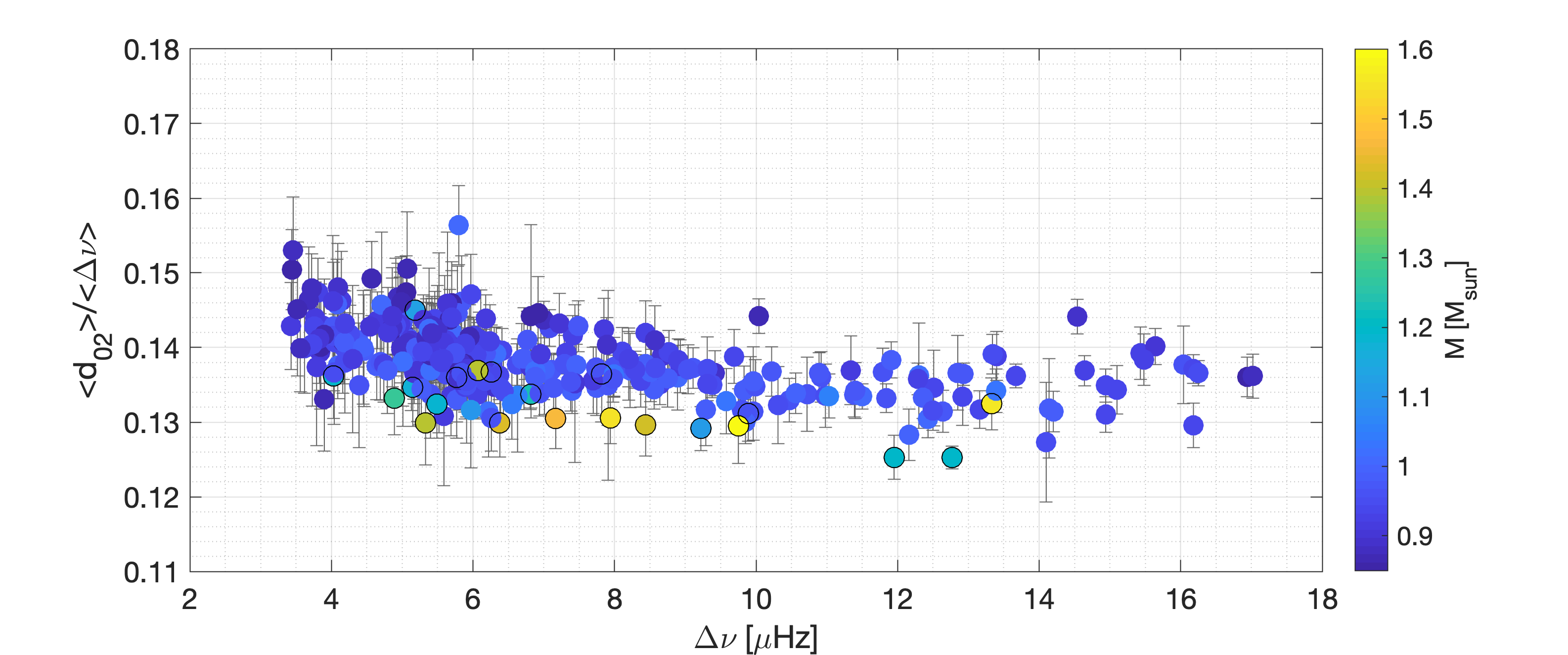}}
    \resizebox{\hsize}{!}{\includegraphics{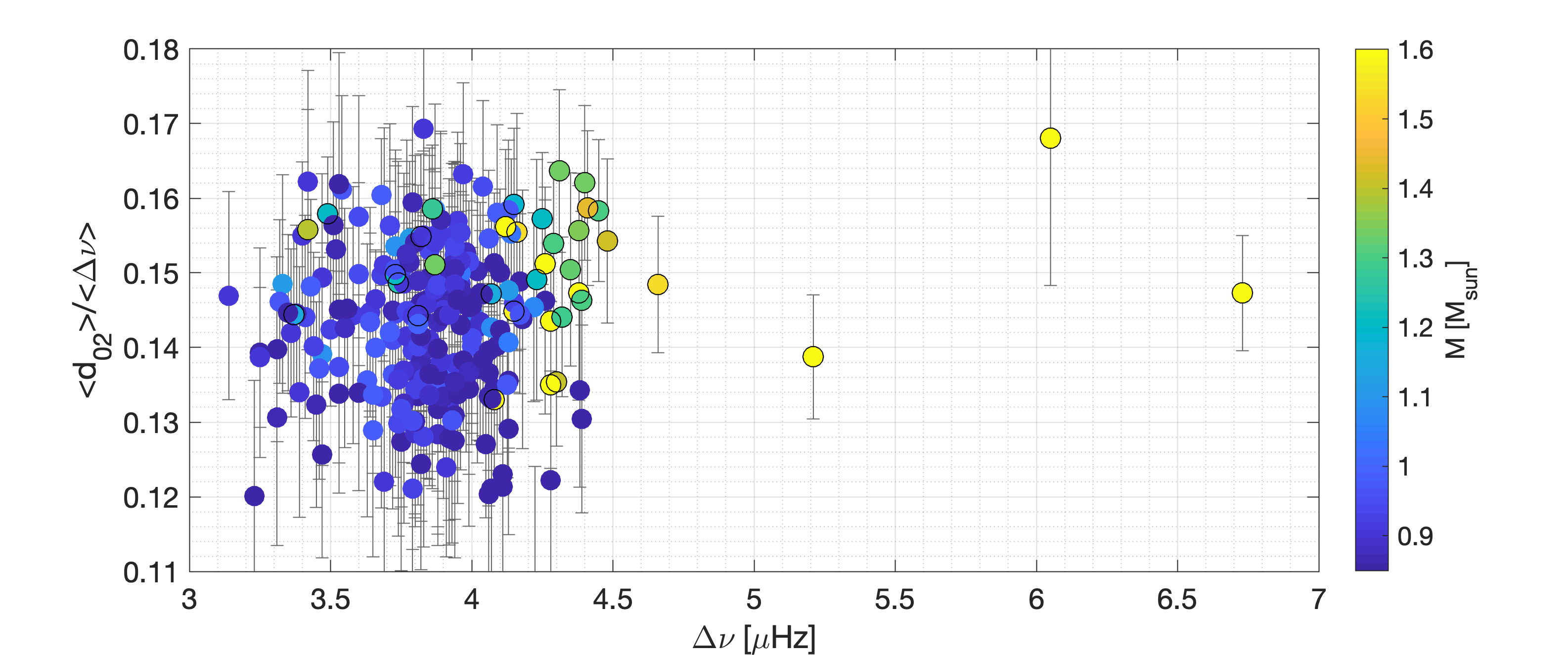}}
  \caption{{\it Upper panel:} Ratio of the average-small $\langle \rm d_{02}\rangle$  to average-large $\langle \Delta\nu \rangle$ frequency separation in RGB stars with $\rm [\alpha/Fe]>0.1$. The estimated stellar mass (R1) is colour coded. {\it Lower panel:} Same as the upper panel, but for stars in the RC.}
\label{fig:dnu02}
\end{figure}

Here we look at the ratio between the average small and large separations for the stars belonging to the high-$\alpha$ population ($\rm [\alpha/Fe]>0.1$). As shown in Fig. \ref{fig:dnu02}, in the RGB phase we note that stars flagged as overmassive in our previous analysis also tend to occupy the lower band of the $\langle d_{02}\rangle/\langle \Delta\nu \rangle$ relation, lending additional evidence that these stars are genuinely more massive.
For stars in the clump, the situation is less clear, but compatible with the trend noticed in \citet{Corsaro2012} and \citet{Handberg2017}, that is, that lower-mass stars tend to have lower ratios, albeit this does not follow the model predictions \citep{Montalban2012}.

\subsection{Impact of helium enrichment relations on ages}
\label{sec:he_enrich}
The helium mass fraction assumed in stellar models has direct impact on the inferred ages, also when asteroseismic constraints are included \citep[e.g. see also][]{Lebreton2014}.

If one restricts the analysis to RGB stars, and considers evolutionary tracks of stars of a given mass and metallicity, the impact of changes of the initial helium abundance on age are of the order of a  15\% decrease if $Y$ is increased by 0.02 (see e.g. Fig. \ref{fig:tracks_Y}). This is also expected from simple scalings of luminosity with mass and chemical composition. From mass-luminosity-chemical composition scalings (here under the crude assumption of energy being transported by radiation only), one can estimate that $L\propto \mu^{15/2}$ \citep{Schwarzschild1958, Kippenhahn2013} hence a $\sim$ 10-15\% decrease in duration of the MS if $Y$ is increased by 0.02, and one assumes that age $\propto M/L$, which is roughly compatible with what one finds from detailed stellar models.

\begin{figure}
\centering
  \resizebox{\hsize}{!}{\includegraphics{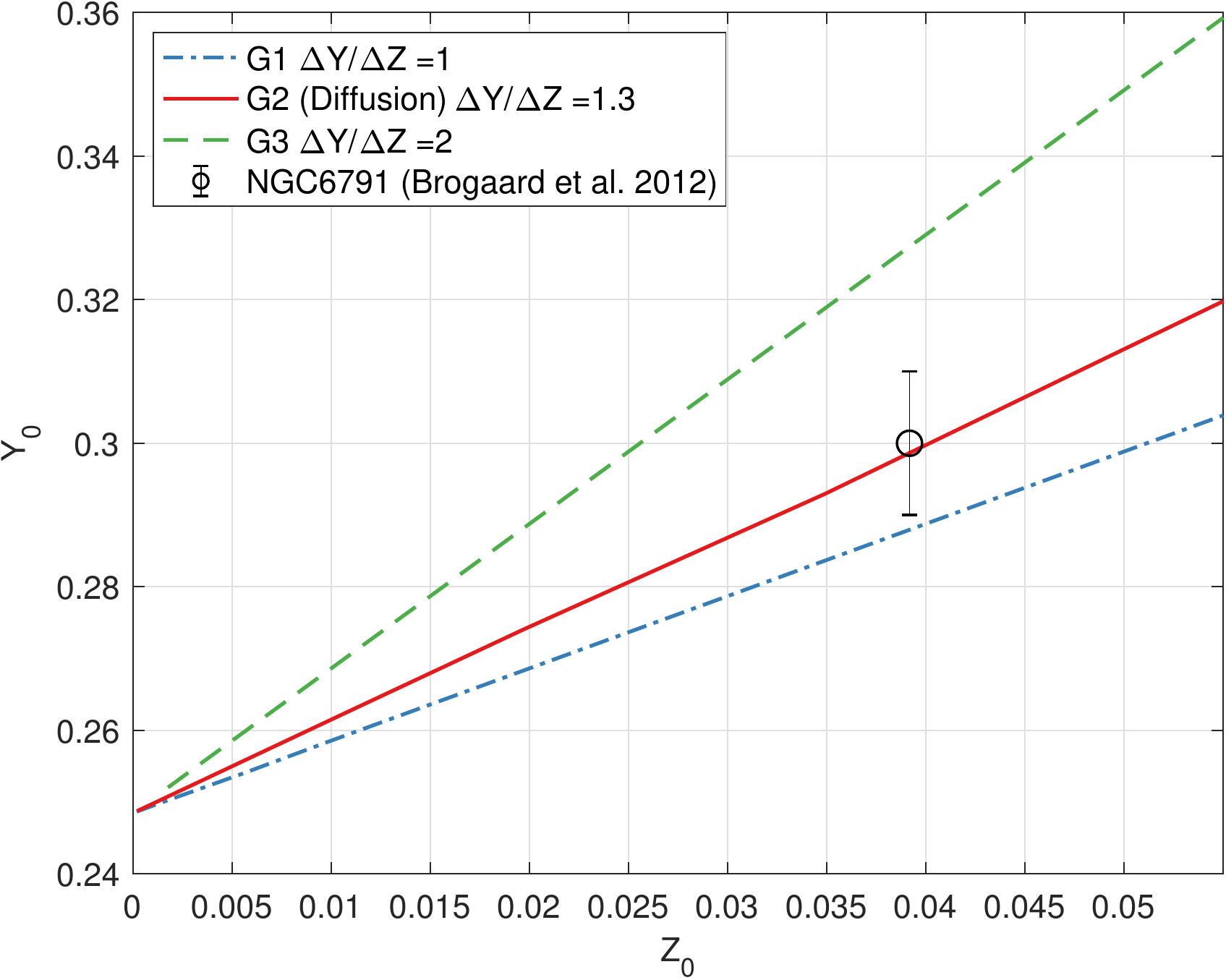}}
  \caption{Initial Helium-to-metallicty (Y$_0$ to Z$_0$) relations assumed in the grids of stellar evolution models. The black circle with error bars represents the helium mass fraction inferred for the old, metal-rich cluster NGC6791 \citep[see][]{Brogaard2012}.}
\label{fig:DYDZ}
\end{figure}

The impact of changes in $Y$ is, in reality, more complex, as one needs to consider changes in the complete set of predictions that are then used in grid-based modelling. For instance, from the tracks presented in Fig. \ref{fig:tracks_Y}, if one considers models with the same mass and metallicity, a change of 0.01 in $Y$ leads to a  negligible variation of the model-motivated corrections to the  \deltanu\ scaling. On the other hand, by changing $Y$, one displaces the predicted \teff\ and, to a larger extent, luminosities hence radii, and this is particularly relevant for stars in the RC. In the latter case, as well known from stellar evolution, one expects models of higher luminosity / radius when $Y$ is increased \citep[e.g. see][ for a recent review of the effect on the luminosity of RR Lyrae stars]{Marconi2018}. When using a grid-based modelling approach, if we include the evolutionary state as a constraint, then the inferences on radii of RC stars are strongly biased by the evolutionary tracks (which occupy a very well confined area in radius).

Even more complex is to interpret the effects of using a grid of models computed with diffusion. While the change in the initial helium abundance plays a prominent role, diffusion itself on the RGB and RC has direct effects, for instance, on the mass of the helium core   \citep{Michaud2007, Michaud2010, Michaud2011}. Moreover, models computed with and without diffusion differ also in terms of the mixing-length parameter obtained in the calibration of a solar model, which then translate into effective temperature shifts.

 As far as helium enrichment is concerned, we consider the grid computed with diffusion (G2) our preferred choice, given that it is compatible with constraints on $Y$  in NGC6791 \citep{Brogaard2012,McKeever2019}, is in better agreement with the initial solar $Y$ inferred indirectly from helioseismology (coupled with stellar models including atomic diffusion), and supported by comparisons with astrometric constraints from \Gaia\ DR2 (see Sec. \ref{sec:gaia}).
While we consider results based on grid G2 to be our reference results, throughout the paper we explore whether the our main conclusions are stable against changes in the model grids.

\begin{figure*}
\centering
  \resizebox{.33\hsize}{!}{\includegraphics {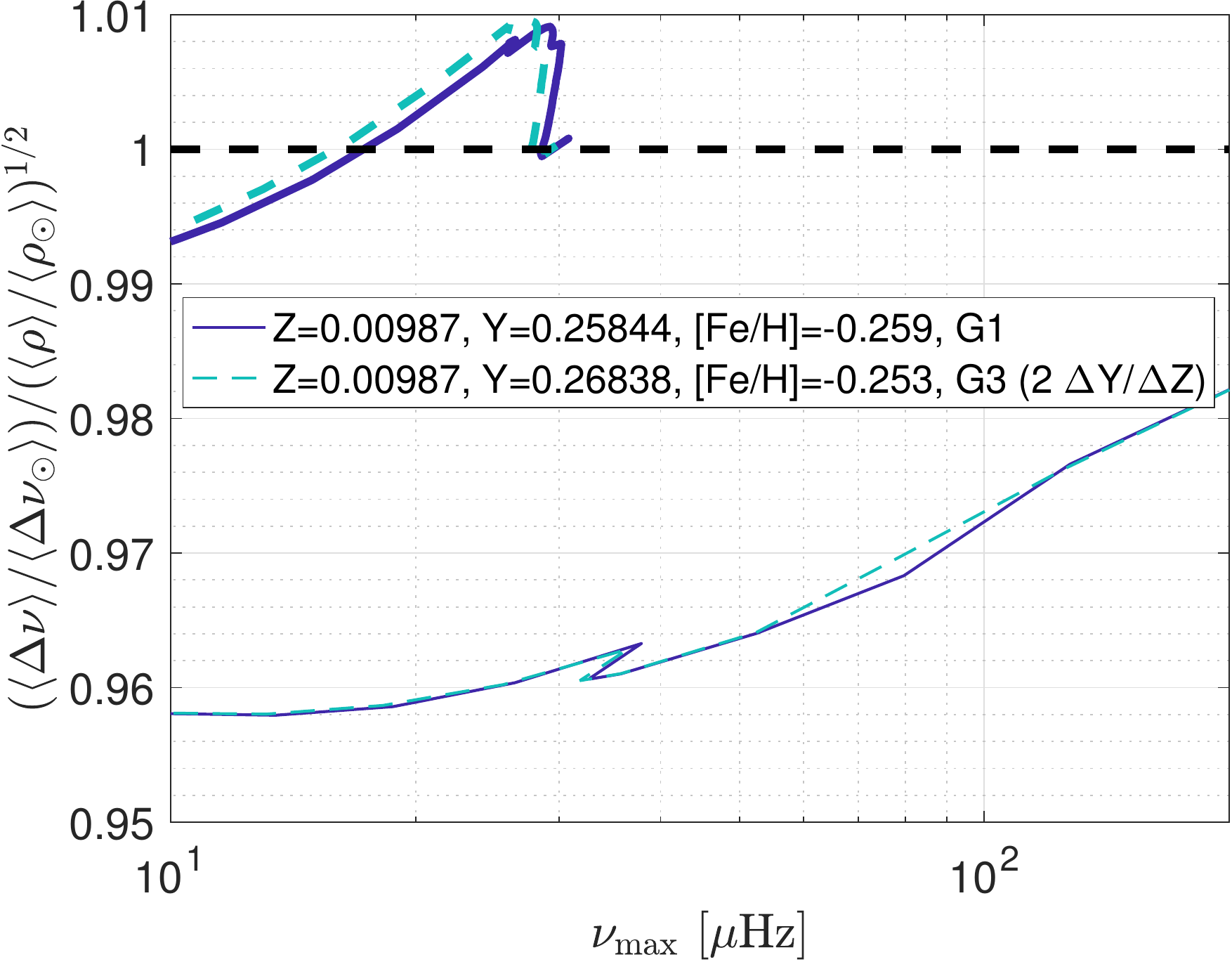}}
  \resizebox{.33\hsize}{!}{\includegraphics {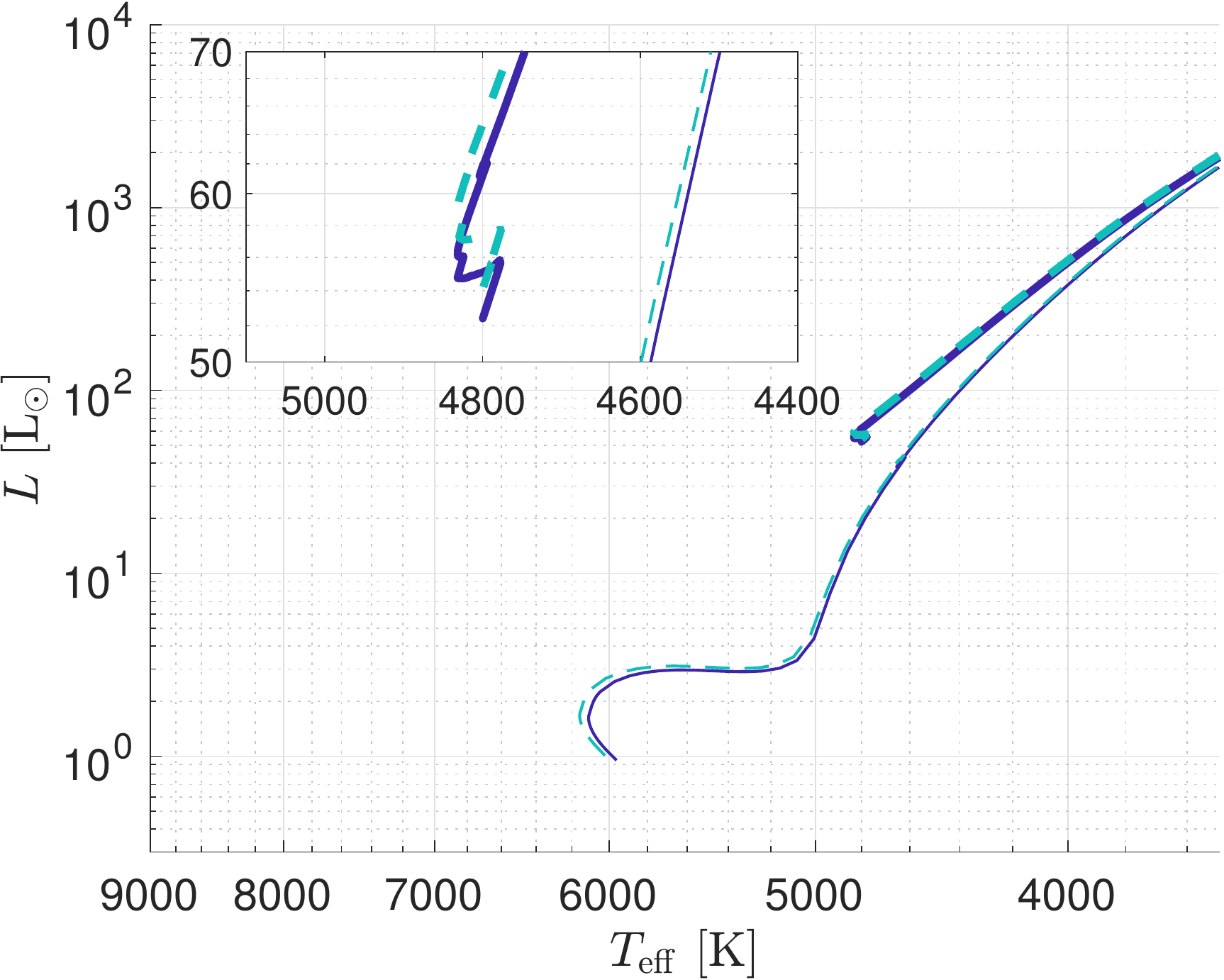}}
  \resizebox{.33\hsize}{!}{\includegraphics {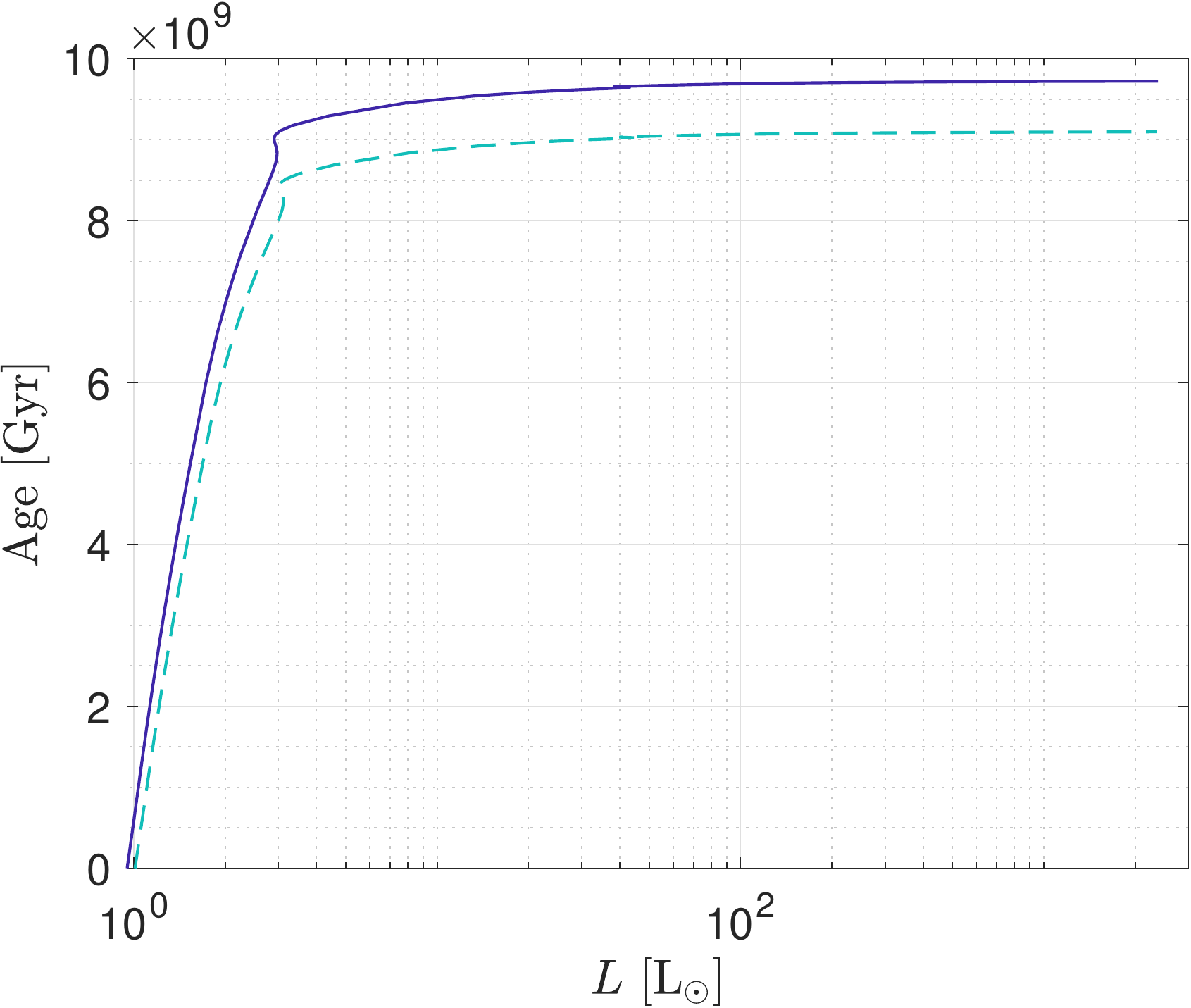}}
  \caption{Effects of changing the initial helium mass fraction by 0.01, in a model of a 1 M$_{\odot}$, $Z=0.00987$ star. While modest changes affect the \deltanu-average density scaling relation ({\it left panel}) and the HR-diagram, except the luminosity increase of the RC, ({\it middle panel}), the impact on the duration of the main sequence, hence on the age of the models on the RGB is $\sim 7\%$ ({\it right panel}).}
\label{fig:tracks_Y}
\end{figure*}

The uncertainties in the assumed helium-metallicity relation have therefore non-negligible effects on the age, in particular on the age-chemical composition trends in our sample. For instance, we notice that higher increases of $Y$ at solar and super-solar metallicities lead to younger age determinations.
Taking the grid with the lowest $\Delta Y/ \Delta Z$ as a reference, at $\rm [Fe/H]=0.25$ the other helium-enrichment relations described in Sec. \ref{sec:models} lead to differences in $Y$ of 0.01 or 0.03 and associated changes in age (at the same mass and metallicity) of 8\% or 23\%.
At solar metallicity the difference is of 0.005 or 0.017 in $Y$ leading to 4\% or 12\% in age. The effect itself is relatively small but significant if one considers uncertainties to be random only and takes advantage of the large number of targets to look for trends in the age-chemistry relations.

\section{Modelling the age and mass distributions of the high-[$\alpha$/Fe] sample}
\label{sec:HBM}

We model the intrinsic age and mass distribution of the high-$\alpha$ population  using a hierarchical Bayesian model (see Sec. \ref{sec:alpha}).  We assume that age/mass measurements of stars in a given population are drawn from a normal distribution with a mean $\mu$ and intrinsic spread $\sigma$, with some measurement error $\sigma_x$. In the case of the age distribution, we choose to fit in $\log_{10}(\mathrm{age})$, as the uncertainties are symmetric and roughly constant in log-space.  We include an outlier term in our model, assuming that  there is an overdensity at younger age / higher mass  as explained in Sec. \ref{sec:overmassive}. We assume that these outliers are also distributed normally with a mean $\log_{10}(\mathrm{age})$/mass $\mu_c$, a spread $\sigma_c$ and contributing some fraction $\epsilon$. When modelling the intrinsic mass distribution of RGB and RC stars, we use the distribution of the difference between $\mu_{M, RGB}$ and $\mu_{M, RC}$  to infer the integrated mass loss \DM and its related uncertainty. 

\begin{figure}
    \centering
    \includegraphics{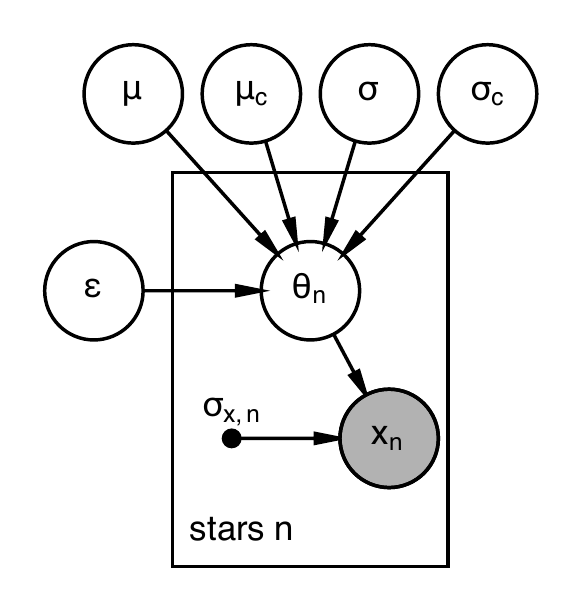}
    \caption{Probabilistic graphical model used to fit the mean age / mass and intrinsic age / mass spread. We assume the measured ages / masses ($X$) are drawn from an underlying true $\theta$ distribution that is Gaussian with a mean $\mu$ and standard deviation $\sigma$. We assume that the true age (mass) distribution is contaminated by stars whose mass is higher than expected. We model these contaminants as also being drawn from another normal distribution with a mean $\mu_c$ and spread $\sigma_c$ which has a fractional contribution $\epsilon$ to the total age distribution (hence the main population contributes $1-\epsilon$).}
    \label{fig:hbm}
\end{figure}

Our model can be summarised via the likelihood function:
\begin{multline}
\mathcal{L}(\vec{\theta} | \vec{x}, \sigma_{\vec{x}})\ =\ \sum_{i} [(1-\epsilon), \epsilon]\mathcal{N}([\mu, \mu_c], [\sigma^{2}, \sigma_{c}^2]+\sigma_{\vec{x}}^{2})
\end{multline}
where $\vec{\theta}$ is the vector of parameters described above $[\mu, \mu_c, \sigma, \sigma_c, \epsilon]$, $\mathcal{N}$ is the normal distribution, which is evaluated at each observed data point $\vec{x}$, which itself is drawn from a normal distribution centreed on the true data, with a width proportional to the observational uncertainty on $\vec{x}$, $\sigma_{\vec{x}}$. The likelihood is a sum over each of the components $i$ of the Gaussian mixture, determined by the parameter vectors in square brackets. In practice, we model the effect of the observational uncertainties by simply convolving the intrinsic distribution with the uncertainty to find the likelihood, as indicated in the equation above. The model is also shown as a graphical model in Figure \ref{fig:hbm}.

We adopt broad priors on all the parameters, which we outline in Table \ref{tab:HBMpriors}.
We have ensured that the priors are sufficiently broad, so that the posterior does not rail against artificial prior boundaries.  At the same time, we find that the posterior distribution is significantly different from the prior distribution, meaning that the posterior is ultimately informed by the data through the likelihood, rather than by the choice of prior.

\begin{table}[]
\caption{Priors adopted in the modelling of mass and age distributions.}
\begin{tabular}{|c|c|c|}
\hline
Model                 & Parameter  & Prior                                      \\ \hline
\multirow{5}{*}{age [Gyr]}  & $\mu$      & $\mathcal{N}(10\ \mathrm{Gyr}, 16\ \mathrm{Gyr}^{2})$ \\ \cline{2-3} 
                      & $\mu_c$    & $\mathcal{N}(4\ \mathrm{Gyr}, 16\ \mathrm{Gyr}^{2})$  \\ \cline{2-3} 
                      & $\sigma$   & $\mathrm{Lognormal}(\ln(0.01), 1)$         \\ \cline{2-3} 
                      & $\sigma_c$ & $\mathrm{Lognormal}(\ln(0.15), 1)$         \\ \cline{2-3} 
                      & $\epsilon$ & $\beta(2, 5)$                              \\ \hline
\multirow{5}{*}{mass [\msun]} & $\mu$      & $\mathcal{N}(1, 0.4^{2})$                  \\ \cline{2-3} 
                      & $\mu_c$    & $\mathcal{N}(1.4, 0.4^{2})$                \\ \cline{2-3} 
                      & $\sigma$   & $\mathrm{Lognormal}(\ln(1.), 0.5)$         \\ \cline{2-3} 
                      & $\sigma_c$ & $\mathrm{Lognormal}(\ln(1.2), 0.5)$        \\ \cline{2-3} 
                      & $\epsilon$ & $\beta(2, 5)$                              \\ \hline
\end{tabular}
\label{tab:HBMpriors}
\end{table}

We sample the posterior probability distribution given the data using \texttt{pymc3}. We make use of the No-U-Turn-Sampler (NUTS), a variant of Hamiltonian Monte Carlo, which uses the gradients of the likelihood function to facilitate rapid convergence and sampling of the posteriors over many parameters. For each population, we take 1000 samples of the posterior over 4 independent chains after allowing 1000 burn-in steps, for a total of 4000 samples. 

\begin{figure}
    \centering
    \includegraphics[width=\columnwidth]{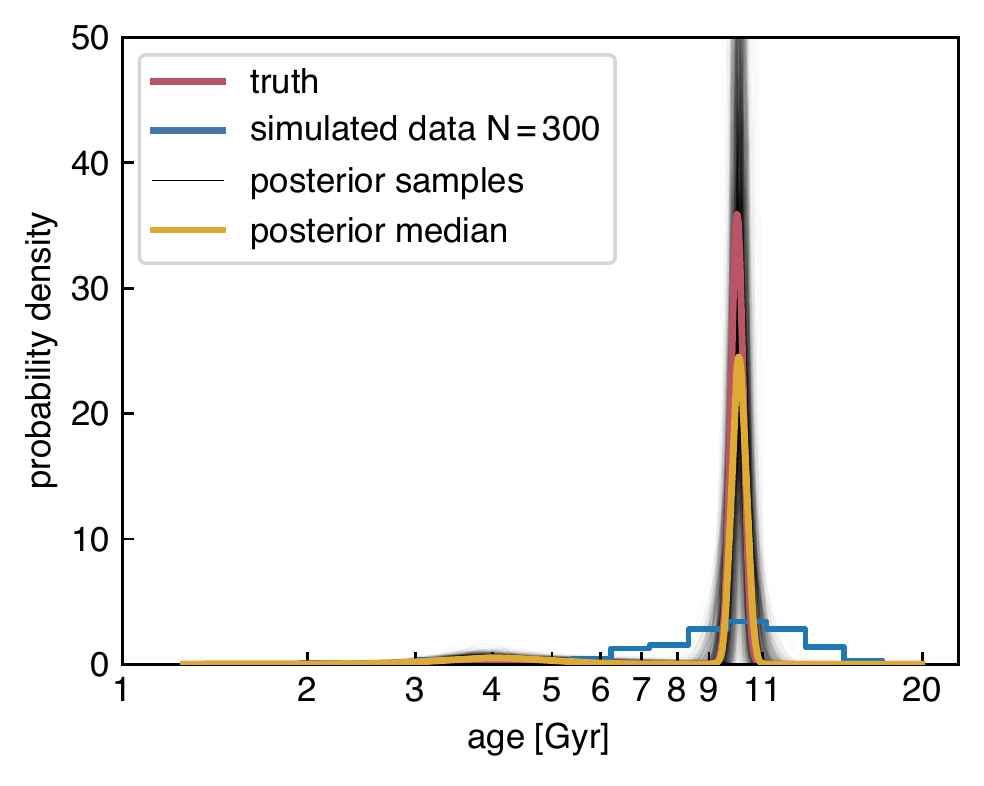}
    \caption{Mixture model for the age distribution of high $\mathrm{[\alpha/Fe]}$ stars (which includes outliers at young age due to over-massive stars), fit to simulated data generated from a set of known parameters and with simulated uncertainty on age representative of that in the observational data set (i.e. $\sim 25\%$).}
    \label{fig:mock}
\end{figure}

We use simulated data to test the ability of this model to fit the true age distribution of our sample, which has considerable uncertainties on the observed ages. We fix the parameters of the \texttt{pymc3} model and sample a set of simulated ages. We then include a simulated observational uncertainty (which may be over or underestimated) by addition of a random variable sampled from a normal distribution with 
$\sigma = 0.1$ on the logarithmic age, that is, a $\sim 25\%$ uncertainty on the absolute age. When fitting the simulated data, we assume that all uncertainties are exactly $25\%$. We then fit the simulated data, recovering all the input parameters to within $1\sigma$. We demonstrate the result of this test in Figure \ref{fig:mock}. It is clear that the width of the age distribution in the simulated data (blue histogram) is inflated relative to the truth distribution (red curve) by the uncertainties. However, our procedure can recover to good accuracy the intrinsic width of the distribution by accounting for these uncertainties (black and yellow curves). The relative height of the distributions is somewhat over-estimated, but the width and position of the peaks is well predicted by the median of the posterior parameter samples.

{
\section{Catalogue of seismic, photometric, spectroscopic and kinematic properties}

\label{app:catalogue}
We provide here a catalogue of stellar properties for the sample defined in Sec. \ref{sec:results}, which we use to present trends in age-kinematics-chemical abundances. Ages, masses, and radii in the catalogue result from the reference modelling run (R1, see Table \ref{tab:runs} for details). No information on the age is provided for $\alpha$-rich stars identified as overmassive.
The columns included in the catalogue are described in Table  \ref{tab:datamodel}.

\begin{table*}
\caption{The data relative to the sample of stars defined in Sec. \ref{sec:results}. Photometric and spectroscopic constraints are taken from 2MASS \citep{Skrutskie2006} and APOGEE DR14 \citep{DR14}. Orbital parameters are computed as described in Sec. \ref{sec:orbitalparam}. Ages, masses, radii, distances and extinction are inferred using {\sc param}'s modelling run R1 (see Table \ref{tab:runs}). All relevant columns are accompanied by an associated uncertainty, defined either as the standard deviation or the 16th and 84th percentiles. Uncertainties are found in accompanying columns labelled with the suffix `\texttt{err}'. \label{tab:datamodel}}
\begin{tabular}{llc}
\hline
Column Identifier & Description & Units \\
\hline
\texttt{APOGEE\_ID} & APOGEE ID in DR14 & $\mathrm{None}$ \\
\texttt{KIC\_ID} & {\it Kepler} Input Catalogue ID & $\mathrm{None}$ \\
\texttt{jmag} & 2MASS $J$-band magnitude & $\mathrm{mag}$ \\
\texttt{hmag} & 2MASS $H$-band magnitude & $\mathrm{mag}$ \\
\texttt{kmag} & 2MASS $K_{S}$-band magnitude & $\mathrm{mag}$ \\
\texttt{ra} & Right Ascension & $\mathrm{deg}$ \\
\texttt{dec} & Declination & $\mathrm{deg}$ \\
\texttt{FE\_H\_APOGEE} & APOGEE $\mathrm{[Fe/H]}$ (DR14) & $\mathrm{None}$ \\
\texttt{ALPHA\_M\_APOGEE} & APOGEE {[$\alpha$/M]} (DR14) & $\mathrm{None}$ \\
\texttt{age} & Age from {\sc param} (R1) & $\mathrm{Gyr}$ \\
\texttt{mass} & Mass from {\sc param} (R1) & $\mathrm{M_{\odot}}$ \\
\texttt{rad} & Radius from {\sc param} (R1) & $\mathrm{R_{\odot}}$ \\
\texttt{dist} & Distance from {\sc param} (R1)& $\mathrm{kpc}$ \\
\texttt{Av} & Extinction from {\sc param} (R1) & $\mathrm{None}$ \\
\texttt{evstate} & Evolutionary state from Yu et al. (2018). RGB = 1, core-He-burning = 2 & $\mathrm{None}$ \\
\texttt{zmax} & Maximum vertical excursion in \texttt{MWPotential2014} & $\mathrm{kpc}$ \\
\texttt{ecc} & Orbit eccentricity in \texttt{MWPotential2014} & $\mathrm{None}$ \\
\texttt{rperi} & Pericentre radius in \texttt{MWPotential2014} & $\mathrm{kpc}$ \\
\texttt{rap} & Apocentre radius in \texttt{MWPotential2014} & $\mathrm{kpc}$ \\
\texttt{GALR} & Position (Galactocentric cylindrical coordinates): Galactocentric radius & $\mathrm{kpc}$ \\
\texttt{GALPHI} & Position: azimuth & $\mathrm{rad}$ \\
\texttt{GALZ} & Position: height above the Galactic midplane & $\mathrm{kpc}$ \\
\texttt{vR} &  Galactocentric radial velocity & $\mathrm{km/s}$ \\
\texttt{vT} &  Galactocentric tangential velocity & $\mathrm{km/s}$ \\
\texttt{vz} &  Galactocentric vertical velocity & $\mathrm{km/s}$ \\
\hline
\end{tabular}
\end{table*}
}
\begin{acknowledgements}
AM, JTM, JM, and FV acknowledge support from the ERC Consolidator Grant funding scheme (project ASTEROCHRONOMETRY, \url{https://www.asterochronometry.eu}, G.A. n. 772293).
 GRD has received funding from the European Research Council (ERC) under the European Union’s Horizon 2020 research and innovation programme (CartographY GA. 804752).
FV acknowledges the support of a Fellowship from the Center for Cosmology and AstroParticle Physics at The Ohio State University.

AM, GRD, and WJC  acknowledge the support of the UK Science and Technology Facilities Council (STFC).  We are extremely grateful to the International Space Science Institute (ISSI) for support provided to the \mbox{asterosSTEP} ISSI International Team (\url{http://www.issibern.ch/teams/asterostep/}).
DB is supported in the form of work contract FCT/MCTES through national funds and by FEDER through COMPETE2020 
in connection to these grants: UID/FIS/04434/2019; PTDC/FIS-AST/30389/2017 \& POCI-01-0145-FEDER-030389.
LC is the recipient of the ARC Future Fellowship FT160100402.
Parts of this research were conducted by the ARC Centre of Excellence ASTRO 3D, through project number CE170100013. IM is a recipient of the Australian Research Council Future Fellowship FT190100574.
CC acknowledges partial support from DFG Grant CH1188/2-1 and from the ChETEC COST Action (CA16117), supported by COST (European Cooperation in Science and Technology).

We thank Maurizio Salaris for many useful discussions and Will Farr for providing suggestions on the statistical models. \\

This work has made use of data from the European Space Agency (ESA) mission
{\it Gaia} (\url{https://www.cosmos.esa.int/gaia}), processed by the {\it Gaia}
Data Processing and Analysis Consortium (DPAC, \url{https://www.cosmos.esa.int/web/gaia/dpac/consortium}). Funding for the DPAC
has been provided by national institutions, in particular the institutions
participating in the {\it Gaia} Multilateral Agreement. 

This paper includes data collected by the \Kepler\ mission. Funding for the \Kepler\ mission is provided by the NASA Science Mission directorate.

Funding for the Sloan Digital Sky Survey IV has been provided by the Alfred P. Sloan Foundation, the U.S. Department of Energy Office of Science, and the Participating Institutions. SDSS acknowledges support and resources from the Center for High-Performance Computing at the University of Utah. The SDSS web site is www.sdss.org.

SDSS is managed by the Astrophysical Research Consortium for the Participating Institutions of the SDSS Collaboration including the Brazilian Participation Group, the Carnegie Institution for Science, Carnegie Mellon University, the Chilean Participation Group, the French Participation Group, Harvard-Smithsonian Center for Astrophysics, Instituto de Astrofísica de Canarias, The Johns Hopkins University, Kavli Institute for the Physics and Mathematics of the Universe (IPMU) / University of Tokyo, the Korean Participation Group, Lawrence Berkeley National Laboratory, Leibniz Institut für Astrophysik Potsdam (AIP), Max-Planck-Institut für Astronomie (MPIA Heidelberg), Max-Planck-Institut für Astrophysik (MPA Garching), Max-Planck-Institut für Extraterrestrische Physik (MPE), National Astronomical Observatories of China, New Mexico State University, New York University, University of Notre Dame, Observatório Nacional / MCTI, The Ohio State University, Pennsylvania State University, Shanghai Astronomical Observatory, United Kingdom Participation Group, Universidad Nacional Autónoma de México, University of Arizona, University of Colorado Boulder, University of Oxford, University of Portsmouth, University of Utah, University of Virginia, University of Washington, University of Wisconsin, Vanderbilt University, and Yale University.

The computations described in this paper were performed using the University of Birmingham's BlueBEAR HPC service, which provides a High Performance Computing service to the University's research community. See \url{http://www.birmingham.ac.uk/bear} for more details.

\end{acknowledgements}

%
%
\bibliographystyle{aa} 
\bibliography{andrea_m} 
%
%
%
%
%
%
%
%
%
%
%
\end{document}